\documentclass[letterpaper,twocolumn,english,noeprint,nodoi,longbibliography]{revtex4-1}
\usepackage[T1]{fontenc}
\usepackage[latin9]{inputenc}
\usepackage{verbatim}
\usepackage{enumitem}
\usepackage{bm}
\usepackage{amsmath}
\usepackage{amssymb}
\usepackage{graphicx}

\makeatletter

\pdfpageheight\paperheight
\pdfpagewidth\paperwidth

\usepackage{braket}

\makeatother

\usepackage{babel}
\begin{document}
\title{Quantum mechanics of composite fermions}
\author{Junren Shi}
\email{junrenshi@pku.edu.cn}

\affiliation{International Center for Quantum Materials, Peking University, Beijing
100871, China}
\affiliation{Collaborative Innovation Center of Quantum Matter, Beijing 100871,
China}
\begin{abstract}
We establish the quantum mechanics of composite fermions based on
the dipole picture initially proposed by Read. It comprises three
complimentary components: a wave equation for determining the wave
functions of a composite fermion in ideal fractional quantum Hall
states and when subjected to external perturbations, a wave function
ansatz for mapping a many-body wave function of composite fermions
to a physical wave function of electrons, and a microscopic approach
for determining the effective Hamiltonian of the composite fermion.
The wave equation resembles the ordinary Schr\"{o}dinger equation
but has drift velocity corrections which are not present in the Halperin-Lee-Read
theory. The wave-function ansatz constructs a physical wave function
of electrons by projecting a state of composite fermions onto a half-filled
bosonic Laughlin state of vortices. Remarkably, Jain's wave function
ansatz can be reinterpreted as the new ansatz in an alternative wave-function
representation of composite fermions. The dipole picture and the effective
Hamiltonian can be derived from the microscopic model of interacting
electrons confined in a Landau level, with all parameters determined.
In this framework, we can construct the physical wave function of
a fractional quantum Hall state deductively by solving the wave equation
and applying the wave function ansatz, based on the effective Hamiltonian
derived from first principles, rather than relying on intuition or
educated guesses. For ideal fractional quantum Hall states in the
lowest Landau level, the approach reproduces the well-established
results of the standard theory of composite fermions. We further demonstrate
that the reformulated theory of composite fermions can be easily generalized
for flat Chern bands.
\end{abstract}
\maketitle

\section{Introduction}

Exotic correlated states of electrons emerge in fractional quantum
Hall systems, where strong magnetic fields completely quench the kinetic
energy of electrons, rendering conventional many-body techniques inadequate
in addressing the effects of correlations between electrons~\citep{tsui1982}.
The theory of composite fermions, proposed by Jain in 1989, offers
a comprehensive framework for understanding these exotic states~\citep{jain2007}.
It introduces a new paradigm, interpreting correlated states of electrons
as non-correlated or weakly-correlated states of fictitious particles
called composite fermions, which are assumed to be the bound states
of electrons and quantum vortices. Based on the insight, the theory
prescribes an ansatz for constructing many-body wave functions that
achieve nearly perfect overlaps with those determined by exact diagonalizations
for various fractional quantum Hall states in the lowest Landau level~\citep{jain1997}.
On the other hand, for predicting the responses of these states to
external perturbations, one usually employs the effective theory proposed
by Halperin, Lee and Read (HLR)~\citep{halperin1993}, which has
been shown to make predictions that align well with experimental observations~\citep{heinonen1997}.
The two components of the theory, namely the wave function ansatz
and the effective theory, complement each other, forming a versatile
framework for understanding the rich physics of the fractional quantum
Hall systems.

Despite the remarkable success, the theory still lacks a concrete
foundation. On the one hand, one usually relies on intuition or educated
guesses when constructing composite fermion wave functions. This is
in sharp contrast to the deductive approach typically employed for
ordinary particles like electrons, for which one can confidently write
down a Hamiltonian for a given physical circumstance, and obtain wave
functions by solving the Schrödinger equation. On the other hand,
the conjecture of the HLR theory - a composite fermion also obeys
the ordinary Schrödinger equation - is only justified heuristically.
Lopez-Fradkin's theory~\citep{lopez1991} is often cited as the rationale
behind both the wave function ansatz and the effective theory~\citep{jain2007}.
However, the theory can only be viewed as a tentative argument rather
than a rigorous foundation for the theory of composite fermions due
to two obvious issues. 

Firstly, Jain's ansatz prescribes wave-functions of electrons in the
form~\citep{jain1997}
\begin{equation}
\Psi\left(\left\{ z_{i}\right\} \right)=\hat{P}_{\mathrm{LLL}}J\left(\left\{ z_{i}\right\} \right)\tilde{\Psi}_{\mathrm{CF}}\left(\left\{ \bm{z}_{i}\right\} \right),\label{eq:Psistd}
\end{equation}
which differs from the form suggested by Lopez-Fradkin's theory based
on the singular Chern-Simons (CS) transformation~\citep{lopez1991}
\begin{equation}
\Psi\left(\left\{ \bm{z}_{i}\right\} \right)=\frac{J\left(\left\{ z_{i}\right\} \right)}{\left|J\left(\left\{ z_{i}\right\} \right)\right|}\tilde{\Psi}_{\mathrm{CF}}\left(\left\{ \bm{z}_{i}\right\} \right),\label{eq:CFCS}
\end{equation}
where $\Psi$ and $\tilde{\Psi}_{\mathrm{CF}}$ represent the wave
function of electrons and composite fermions, respectively, 
\begin{equation}
J\left(\left\{ z_{i}\right\} \right)\equiv\prod_{i<j}\left(z_{i}-z_{j}\right)^{2}\label{eq:Jastrow}
\end{equation}
is the Bijl-Jastrow factor which presumably attaches a vortex with
two flux quanta to each electron, $\hat{P}_{\mathrm{LLL}}$ is the
projection operator to the lowest-Landau-level, and $\{\bm{z}_{i}\equiv(x_{i},y_{i})\}$
and $\{z_{i}\equiv x_{i}+\mathrm{i}y_{i}\}$ denote the coordinates
of electrons in the vector and complex forms, respectively. Equation
(\ref{eq:CFCS}) is formulated in the full Hilbert space of free electrons
while Eq.~(\ref{eq:Psistd}) is defined in the restricted Hilbert
space of a single Landau level. Reconciling the two is non-trivial~\citep{murthy2003}.

Secondly, the picture of composite fermions implied by Lopez-Fradkin's
theory, which is also inherited by the HLR theory, differs from that
obtained by directly inspecting the ansatz wave function Eq.~(\ref{eq:Psistd}).
For the latter, Read's analysis indicates that the electron and the
vortex in a composite fermion are spatially separated~\citep{read1994}.
The finding contradicts the picture implied by Eq.~(\ref{eq:CFCS}),
which suggests that a composite fermion is a point particle, consisting
of an electron and $\delta$-function flux tubes. More recently, Son
points out that the HLR theory lacks the particle-hole symmetry~\citep{son2015},
whereas the ansatz wave function is shown to preserve the symmetry
well~\citep{balram2016}. These observations raise doubts about whether
the HLR theory accurately describes the composite fermions implied
by the ansatz wave function. It prompts the need for an alternative
effective theory, or ideally, an alternative foundation from which
both the wave-function ansatz and the effective theory can be inferred.

The dipole picture of composite fermions, initially proposed by Read,
offers an alternative picture for the fictitious particle~\citep{read1994,rezayi1994}.
The picture differs from the HLR view in two fundamental ways. Firstly,
instead of being a point particle, the composite fermion has a dipole
structure with the electron and vortex spatially separated. Secondly,
the electron and the vortex are confined in two separated Landau levels
created by the physical magnetic field and an emergent CS magnetic
field, respectively, as opposed to moving in a free space~\citep{shi2017,shi2018}.
The dipole picture is shown to yield low-energy and long-wavelength
electromagnetic responses that are identical to those predicted by
the Dirac theory of composite fermions~\citep{ji2021}, indicating
that it satisfies the general requirements of particle-hole symmetry.
The feature, along with the fact that the picture is inferred directly
from microscopic wave functions, sets it apart from other alternatives.

Pasquier and Haldane investigate the dipole picture of a system of
bosons in an isolated Landau level at filling factor one~\citep{pasquier1998}.
Their work, along with subsequent developments by other researchers~\citep{read1998,dong2020,gocanin2021},
sheds light on the construction of physical wave functions. They propose
interpreting vortices as auxiliary degrees of freedom that extend
the physical Hilbert space to a larger Hilbert space of composite
fermions. To obtain a physical state from a state of composite fermions
in the enlarged Hilbert space, one must eliminate the auxiliary degrees
of freedom by projecting the state into a physical subspace defined
by a pure state of the vortices. In this context, the Bijl-Jastrow
factor is interpreted as the complex conjugate of the wave function
of the vortex state, rather than the numerator of the singular gauge
factor in Eq.~(\ref{eq:CFCS}). The interpretation naturally leads
to a wave function ansatz alternative to Eq.~(\ref{eq:Psistd}),
and avoids the difficulties associating with the singular CS transformation~\citep{read1998}.

In this paper, we present a theory of quantum mechanics for composite
fermions based on the dipole picture and Pasquier-Haldane's interpretation.
The theory comprises three complimentary components: a wave equation
for determining the wave functions of a composite fermion in ideal
fractional quantum Hall states and when subjected to external perturbations,
a wave function ansatz for mapping a many-body wave function of composite
fermions to a physical wave function of electrons, and a microscopic
approach for determining the effective Hamiltonian of the composite
fermion. In our theory, the state of a composite fermion is represented
by a bivariate wave function that is holomorphic (anti-holomorphic)
in the coordinate of its constituent electron (vortex), and is defined
in a Bergman space with a weight determined by the spatial profiles
of the physical and the emergent CS magnetic fields. The wave equation
is derived by applying the rules of quantization in the Bergman space
to the phenomenological dipole model proposed in Ref.~\onlinecite{ji2021}.
It resembles the ordinary Schr\"{o}dinger equation but has drift
velocity corrections which are not present in the HLR theory. The
wave-function ansatz constructs a physical wave function of electrons
by projecting a state of composite fermions onto a half-filled bosonic
Laughlin state of vortices. Remarkably, Jain's wave function ansatz,
which underlies the success of the theory of composite fermions, can
be recast to the form of the new ansatz using an alternative wave-function
representation for composite fermions. The phenomenological dipole
model can be derived from the microscopic model of interacting electrons
confined in a Landau level by applying a Hartree-like approximation,
with its parameters determined from first principles. In this framework,
we can construct the physical wave function of a fractional quantum
Hall state deductively by solving the wave equation and applying the
wave function ansatz, rather than relying on intuition or educated
guesses. We further demonstrate that the reformulated theory of composite
fermions can be easily generalized for flat Chern bands, which are
also predicted to host the fractional quantum Hall states~\citep{parameswaran2013,bergholtz2013}.

The remainder of the paper is organized as follows. In Sec.~\ref{sec:Overview},
we introduce the dipole model which is the basis of our discussions,
and give an overview of the main results of this work. In Sec.~\ref{sec:Ideal-systems-of},
we develop the theory for ideal fractional quantum Hall systems which
are subjected only to uniform magnetic fields. In Sec.~\ref{sec:General-theory},
the theory is established for general systems which could be subjected
to spatially and temporarily fluctuating external perturbations and
have inhomogeneous densities. In Sec.~\ref{sec:Derivation-of-the},
we provide a microscopic underpinning for our theory by deriving the
phenomenological dipole model from the microscopic Hamiltonian of
interacting electrons confined in a Landau level. In Sec.~\ref{sec:Generalization-for-flat},
as an application of our reformulation, we generalize the theory of
composite fermions for flat Chern bands. In Sec.~\ref{sec:Summary-and-discussion},
we summarize and discuss our results. Certain details of derivations
are presented in Appendices.

\section{Overview \label{sec:Overview}}

\subsection{Dipole model \label{subsec:Dipole-model}}

Our theory is based on the dipole picture which was originally proposed
by Read for the half-filled Landau level~\citep{read1994}. The picture
can be generalized to a dipole model of composite fermions which can
be applied to arbitrary filling factors~\citep{shi2018,ji2020a,ji2021}.

\begin{figure}
\includegraphics{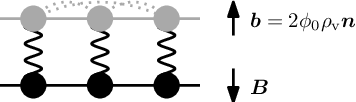}

\caption{\label{fig:CFdipolemodel_FQH}Dipole model of composite fermions.
A composite fermion consists of an electron (black) and a vortex (gray).
The electron is confined in the Landau level induced by the physical
magnetic field $\bm{B}$, while the vortex belongs to a bosonic liquid
of vortices in the $\nu=1/2$ Laughlin state. Under the mean-field
approximation, the vortex is considered as an independent particle
confined in the Landau level induced by an emergent CS magnetic field
$\bm{b}$. The electron and the vortex are bound together by a binding
potential which can be modeled as the harmonic potential Eq.~(\ref{eq:Vbound})
in a lowest Landau level.}
\end{figure}

The model is illustrated in Fig.~\ref{fig:CFdipolemodel_FQH}. According
to the model, a composite fermion consists of an electron and a vortex
confined in two separate Landau levels, one for the electron is the
Landau level induced by the physical magnetic field $\bm{B}=-B\bm{n}$,
another for the vortex is the fictitious Landau level induced by an
emergent CS magnetic field $\bm{b}=b\bm{n}$ with its strength determined
by the CS self-consistent condition $b=(2h/e)\rho_{\mathrm{v}}$,
where $\rho_{\mathrm{v}}$ is the density of vortices, and $\bm{n}$
denotes the normal vector of the two-dimensional plane of the system.
In general, both the physical magnetic field and the CS magnetic field
can be non-uniform. The two particles are bounded by a binding potential,
which can be shown to be well approximated by a harmonic potential
for the lowest-Landau-level (see Sec.~\ref{subsec:Energy}).

The non-interacting dipole model is actually a mean-field approximation
for an underlying correlated system of composite fermions. First of
all, electrons confined in the physical Landau level are interacting.
On the other hand, in Pasquier-Haldane's interpretation~\citep{pasquier1998,read1998},
vortices are auxiliary degrees of freedom introduced to extend the
physical Hilbert space of electrons to a larger Hilbert space of composite
fermions. They are assumed to form a collective half-filled bosonic
Laughlin state (see Sec.~\ref{subsec:Wave-function-ansatz-1}). In
Sec.~\ref{sec:Derivation-of-the}, we will show how the non-interacting
dipole model emerges after applying a Hartree-like approximation in
the enlarged Hilbert space. We note that the standard interpretation
of the vortex, namely, an entity consisting of two quantized microscopic
vortices, is actually a property derived from the particular collective
state assumed for the vortices (see Sec.~\ref{subsec:Wave-function-ansatz}).

It is also possible to interpret the composite fermion as a point-particle
by defining its momentum $\bm{p}$ and coordinate $\bm{x}$. A definition
of the momentum, as pointed out by Read~\citep{read1994}, could
be
\begin{equation}
\bm{p}=\frac{\hbar}{l_{B}^{2}}\bm{n}\times(\bm{z}-\bm{\eta}),\label{eq:momentum}
\end{equation}
where $\bm{z}$ and $\bm{\eta}$ are the coordinates of the electron
and the vortex, respectively, and $l_{B}\equiv\sqrt{\hbar/eB}$ is
the magnetic length of the $B$-field. We can define $\bm{x=\bm{\eta}}$,
as suggested in Refs.~\onlinecite{shi2018,ji2021}. The composite
fermion can then be interpreted as a particle that is subjected to
a uniform momentum-space Berry-curvature~\citep{shi2018,ji2020b,ji2021}
and obeys the Sundaram-Niu dynamics~\citep{sundaram1999}. It is
notable that such a particle has a modified phase space measure~\citep{xiao2005}.
Consequently, for a Landau level with the particle-hole symmetry,
the kinetic energy of the composite fermion should be modeled as~\citep{ji2021}
\begin{equation}
T=D\frac{p^{2}}{2m^{\ast}},\label{eq:kineticenergy}
\end{equation}
where $m^{\ast}$ denotes the effective mass of the composite fermion,
and $D=b/B$ is the density-of-states correction factor due to the
modified phase space measure~\citep{ji2021}. In Sec.~\ref{subsec:Energy},
we will show that the microscopic derivation of the dipole model does
give rise to the peculiar form of the kinetic energy.

Finally, we note that one is actually free to choose the definition
of $(\bm{x},\bm{p})$ and have a different interpretation. The physical
results do not depend on the interpretation. This is demonstrated
in Ref.~\onlinecite{ji2021} where two different choices of the definition
and their interpretations are compared. 

\subsection{Summary of results}

We summarize the main results of this work as follows:
\begin{enumerate}[label=\Alph{enumi})]
\item The state of the composite fermion can be represented by a bivariate
wave-function that is holomorphic (anti-holomorphic) in the coordinate
of its constituent electron (vortex) (see Sec.~\ref{subsec:Segal-Bargmann-space}).
The Hilbert space is the tensor product of two Bergman spaces, one
for the electron and one for the vortex, with weights determined by
the spatial profiles of the physical and CS magnetic fields, respectively
(see Sec.~\ref{subsec:Bergman_space_generalized}).
\item A new wave function ansatz can be logically inferred from the dipole
model (see Sec.~\ref{subsec:Wave-function-ansatz-1}):
\begin{equation}
\Psi(\{z_{i}\})=\hat{P}_{\mathrm{v}}\Psi_{\mathrm{CF}}(\{z_{i},\bar{\eta}_{i}\}),
\end{equation}
where $\hat{P}_{\mathrm{v}}$ denotes the projection onto the collective
state assumed for vortices. Remarkably, the new ansatz and the standard
ansatz Eq.~(\ref{eq:Psistd}) are equivalent, although they use two
different wave-function representations for composite fermions. The
two representations can be related by a transformation shown in Eq.~(\ref{eq:transformPsiCF})
for ideal states and Eq.~(\ref{eq:CFtransform}) generally.
\item A general wave equation can be established for composite fermions,
valid not only for ideal systems, but also when external perturbations
are present. The wave equation has a biorthogonal form, shown in Eqs.~(\ref{eq:TDWEQ1},
\ref{eq:TDWEQ2}), and its Hamiltonian in the long-wavelength limit
has corrections from the drift velocities of the electron and the
vortex, shown in Eq.~(\ref{eq:Hdipole}). For ideal systems, the
wave equation yields wave functions identical to those prescribed
by the standard theory. However, the responses to external perturbations
predicted by the wave equation will differ from those predicted by
the HLR theory because of the drift-velocity corrections. It has been
shown that the dipole model yields long-wavelength responses identical
to those predicted by the Dirac theory of composite fermions with
a dipole correction~\citep{ji2021}.
\item The non-interacting dipole model can be derived from the underlying
microscopic model of a set of interacting electrons confined in a
Landau level by applying a Hartree-like approximation in the enlarged
Hilbert space of composite fermions. The origin of the fictitious
Chern-Simons fields is clarified. They are introduced to impose the
requirement of consistency of orthonormalities between the physical
Hilbert space and the enlarged Hilbert space (see Sec.~\ref{subsec:Chern-Simons-constraints}).
The derivation also confirms that the kinetic energy, which is basically
the Coulomb attraction energy between the electron and the charge
void induced by the vortex, is indeed proportional to $D=2\nu$, where
$\nu$ is the filling factor of the system, and can be well approximated
by the parabolic form Eq.~(\ref{eq:kineticenergy}) for a lowest-Landau-level
(see Sec.~\ref{subsec:Energy}).
\item The reformulated theory of composite fermions can be generalized for
a flat Chern band with a Chern number $|C|=1$ by substituting the
Chern band in place of the physical Landau level in the dipole model
shown in Fig.~\ref{fig:CFdipolemodel-FCI}. We find that the effective
band dispersion experienced by a composite fermion is renormalized
by the combination of the quantum metric and the Berry-curvature of
the band that gives rise to the heuristic ``trace condition''~\citep{roy2014}.
The observation rationalizes the ``trace condition'' for the stability
of a fractional Chern insulator state. It also suggests the possibility
of stabilizing a fractional Chern insulator state in a non-flat Chern
band when the renormalization cancels the dispersion of the Chern
band (see Sec.~\ref{sec:Generalization-for-flat}).
\end{enumerate}

\section{Theory for ideal systems \label{sec:Ideal-systems-of}}

In this section, we develop the quantum mechanics of composite fermions
in systems that are subjected only to uniform external magnetic fields
and have homogeneous densities. We will establish a new wave function
ansatz and a set of wave equations for composite fermions. Remarkably,
our approach can be shown to reproduce the well-established results
of the standard theory. The principles established in this section
will serve as the foundation for developing a general theory.

\subsection{Hilbert space \label{subsec:Segal-Bargmann-space}}

The Hilbert space of a composite fermion shown in Fig.~\ref{fig:CFdipolemodel_FQH}~
is the tensor product of two Hilbert spaces with respect to the two
Landau levels. It differs from that of an ordinary quantum particle
in a free space as assumed in the HLR theory. 

The Hilbert space spanned by a Landau level is a weighted Bergman
space~\citep{girvin1984,rohringer2003}. For a disc geometry, the
space includes all holomorphic polynomials in the complex electron
coordinate $z=x+\mathrm{i}y$. The inner product between two states
$\psi_{1}(z)$ and $\psi_{2}(z)$ in the space is defined by $\braket{\psi_{1}|\psi_{2}}=\int\mathrm{d}\mu_{B}^{(0)}(\bm{z})\psi_{1}^{\ast}(z)\psi_{2}(z)$
with the integral measure
\begin{equation}
\mathrm{d}\mu_{B}^{(0)}(\bm{z})\equiv\frac{\mathrm{d}\bm{z}}{2\pi l_{B}^{2}}e^{-\left|z\right|^{2}/2l_{B}^{2}}.
\end{equation}
A Bergman space with the Gaussian weight is also known as the Segal-Bargmann
space~\citep{hall1999}.

The Hilbert space of a vortex is also a Segal-Bargmann space consisting
of all anti-holomorphic polynomials in the complex conjugated vortex
coordinate $\bar{\eta}=\eta_{x}-\mathrm{i}\eta_{y}$, where $\eta_{x}$
and $\eta_{y}$ are the Cartesian components of the vortex coordinate
$\bm{\eta}\equiv(\eta_{x},\eta_{y})$. Note that because the direction
of the $b$-field is opposite to that of the $B$-field, wave functions
for the vortex are anti-holomorphic functions. The corresponding integral
measure is
\begin{equation}
\mathrm{d}\mu_{b}^{(0)}(\bm{\eta})\equiv\frac{\mathrm{d}\bm{\eta}}{2\pi l_{b}^{2}}e^{-\left|\eta\right|^{2}/2l_{b}^{2}},
\end{equation}
where $l_{b}=\sqrt{eb/\hbar}$ is magnetic length of the $b$-field.

The Hilbert space of a composite fermion is the tensor product of
the two Segal-Bargmann spaces for the electron and the vortex, respectively.
The state of a composite fermion can thus be naturally represented
by a bivariate function:
\begin{equation}
\psi(z,\bar{\eta}),\label{eq:bvhwf}
\end{equation}
which is holomorphic (anti-holomorphic) in the complex coordinate
$z$ ($\bar{\eta}$) of the electron (vortex). Unlike the wave function
$\psi(\bm{z})\equiv\psi(\bar{z},z)$ for an ordinary particle, the
two coordinates of the wave function Eq.~(\ref{eq:bvhwf}) belong
to different particles.

For a Bergman space, we can define a reproducing kernel, which is
basically the coordinate representation of the identity operator of
the space~\citep{hall1999}. For the spaces of the electron and the
vortex, their reproducing kernels are
\begin{align}
K_{B}^{(0)}(z,\bar{z}^{\prime}) & =e^{z\bar{z}^{\prime}/2l_{B}^{2}},\label{eq:KB0}\\
K_{b}^{(0)}(\bar{\eta},\eta^{\prime}) & =e^{\bar{\eta}\eta^{\prime}/2l_{b}^{2}},\label{eq:Kb0}
\end{align}
respectively. The kernels transform wave functions in the respective
Bergman spaces back to themselves:
\begin{align}
\psi(z) & =\int\mathrm{d}\mu_{B}^{(0)}(\bm{z}^{\prime})K_{B}^{(0)}(z,\bar{z}^{\prime})\psi(z^{\prime}),\label{eq:reproduceB}\\
\varphi(\bar{\eta}) & =\int\mathrm{d}\mu_{b}^{(0)}(\bm{\eta}^{\prime})K_{b}^{(0)}(\bar{\eta},\eta^{\prime})\varphi(\bar{\eta}^{\prime}).\label{eq:reproduceb}
\end{align}

The kernels can also be used to project non-holomorphic functions
into the Segal-Bergmann spaces~\citep{hall1999}. Actually, $\hat{P}_{\mathrm{LLL}}$
in Eq.~(\ref{eq:Psistd}), the projection operator to the lowest
Landau level, can be written as an integral transform using the reproducing
kernel:
\begin{equation}
\hat{P}_{\mathrm{LLL}}f\left(\bm{z}\right)\equiv\int\mathrm{d}\mu_{B}^{(0)}(\bm{\xi})K_{B}^{(0)}(z,\bar{\xi})f\left(\bm{\xi}\right),\label{eq:projection}
\end{equation}
where $f(\bm{z})$ is shorthand notation of a non-holomorphic function
$f(\bar{z},z)$. We will use the notations interchangeably in this
paper. The projection into the $\eta$-space can be defined similarly
using the reproducing kernel $K_{b}^{(0)}(\bar{\eta},\eta^{\prime})$.

\subsection{Wave function ansatz \label{subsec:Wave-function-ansatz-1}}

The wave function ansatz Eq.~(\ref{eq:Psistd}) maps a many-body
wave function in the fictitious world of composite fermions to a physical
wave function of interacting electrons in the real world. Although
the ansatz is customarily expressed in a form that suggests its connection
with the singular CS transformation Eq.~(\ref{eq:CFCS}), it can
actually be more naturally inferred from the dipole model, as we will
demonstrate in this subsection. 

Pasquier and Haldane presented an alternative approach of constructing
the many-body wave functions of fractional quantum Hall states~\citep{pasquier1998}.
The approach was further developed by Read~\citep{read1998} and
Dong and Senthil~\citep{dong2020}. They investigate a system of
bosons at filling factor one. Vortices of one flux quantum, which
are fermions, are introduced as auxiliary degrees of freedom for extending
the physical Hilbert space of bosons to a Hilbert space of composite
fermions. It is envisioned that in the enlarged Hilbert space, it
may become feasible to apply mean-field approximations for the composite
fermions. To obtain physical wave functions, on the other hand, one
needs to eliminate auxiliary degrees of freedom by projecting states
of composite fermions into a physical subspace. This leads to a relation
between a wave function of composite fermions $\Psi_{\mathrm{CF}}(\{z_{i},\bar{\eta}_{i}\})$
and its physical counterpart $\Psi(\{z_{i}\})$~\citep{read1998}:
\begin{align}
\Psi(\{z_{i}\}) & =\hat{P}_{\mathrm{v}}\Psi_{\mathrm{CF}}(\{z_{i},\bar{\eta}_{i}\})\label{eq:PsiCFtoPsi-0}\\
 & \equiv\int\prod_{i}\mathrm{d}\mu_{b}^{(0)}(\bm{\eta}_{i})\Psi_{\mathrm{v}}^{\ast}(\{\bar{\eta}_{i}\})\Psi_{\mathrm{CF}}(\{z_{i},\bar{\eta}_{i}\}),\label{eq:PsiCFtoPsi}
\end{align}
where $\Psi_{\mathrm{v}}(\{\bar{\eta}_{i}\})$ is the wave function
of a vortex state that defines the physical subspace in the enlarged
Hilbert space, and $\hat{P}_{\mathrm{v}}$ denotes the projection
into the subspace. For a system of bosons, the vortex state is assumed
to be a $\nu=1$ incompressible state of fermions with $\Psi_{\mathrm{v}}(\{\bar{\eta}_{i}\})=\prod_{i<j}(\bar{\eta}_{i}-\bar{\eta}_{j})$.
The corresponding physical wave function describes a Fermi-liquid
like state of bosons.

The general idea of Pasquier-Haldane-Read's approach can be adapted
for a system of electrons. We can introduce bosonic vortices as the
auxiliary degrees of freedom. We assume that the vortices form a $\nu=1/2$
bosonic Laughlin state with the wave function
\begin{equation}
\Psi_{\mathrm{v}}\left(\left\{ \bar{\eta}_{i}\right\} \right)=J^{\ast}\left(\left\{ \eta_{i}\right\} \right).\label{eq:PsiB}
\end{equation}
By substituting the vortex wave function into Eq.~(\ref{eq:PsiCFtoPsi}),
we obtain an ansatz for constructing physical wave functions of electrons:
\begin{equation}
\Psi\left(\left\{ z_{i}\right\} \right)=\int\prod_{i}\mathrm{d}\mu_{b}^{(0)}(\bm{\eta}_{i})J\left(\left\{ \eta_{i}\right\} \right)\Psi_{\mathrm{CF}}\left(\left\{ z_{i},\bar{\eta}_{i}\right\} \right).\label{eq:wfansatzdp}
\end{equation}

Remarkably, the ansatz can be shown to be equivalent to the standard
(Jain's) ansatz Eq.~(\ref{eq:Psistd}). To see this, we express Eq.~(\ref{eq:Psistd})
in an integral form by using Eq.~(\ref{eq:projection}):
\begin{align}
\Psi(\{z_{i}\})= & \int\prod_{i}\mathrm{d}\mu_{B}^{(0)}(\bm{\xi}_{i})e^{\sum_{i}z_{i}\bar{\xi}_{i}/2l_{B}^{2}}J(\{\xi_{i}\})\tilde{\Psi}_{\mathrm{CF}}(\{\bm{\xi}_{i}\})\nonumber \\
= & \int\prod_{i}\mathrm{d}\mu_{b}^{(0)}(\bm{\eta}_{i})J(\{\eta_{i}\})\int\prod_{i}\mathrm{d}\mu_{B}^{(0)}(\bm{\xi}_{i})\nonumber \\
 & \times e^{\sum_{i}(z_{i}\bar{\xi}_{i}/2l_{B}^{2}+\bar{\eta}_{i}\xi_{i}/2l_{b}^{2})}\tilde{\Psi}_{\mathrm{CF}}(\{\bm{\xi}_{i}\}).\label{eq:ansatzderiv}
\end{align}
where we insert the reproducing kernel Eq.~(\ref{eq:Kb0}) for each
of the composite fermions. Comparing it with the new ansatz Eq.~(\ref{eq:wfansatzdp}),
we have
\begin{multline}
\Psi_{\mathrm{CF}}\left(\left\{ z_{i},\bar{\eta}_{i}\right\} \right)=\int\prod_{i}\mathrm{d}\mu_{B}^{(0)}\left(\bm{\xi}_{i}\right)\\
\times e^{\sum_{i}\left(z_{i}\bar{\xi}_{i}/2l_{B}^{2}+\xi_{i}\bar{\eta}_{i}/2l_{b}^{2}\right)}\tilde{\Psi}_{\mathrm{CF}}\left(\left\{ \bm{\xi}_{i}\right\} \right).\label{eq:transformPsiCF}
\end{multline}
We see that the two ansatze are equivalent but use different wave-function
representations for a state of composite fermions. In the following,
we will refer to the two representations as the dipole representation
($\Psi_{\mathrm{CF}}$) and the standard representation ($\tilde{\Psi}_{\mathrm{CF}}$),
respectively.

\subsection{Wave equation: the dipole representation \label{subsec:Wave-equation}}

The theory of composite fermions often relies on intuition or educated
guesses when selecting wave functions for composite fermions. The
resulting ansatz wave functions are then justified a posteriori by
showing high overlaps with wave functions obtained from exact diagonalizations~\citep{jain1997}.
Is it possible to determine appropriate wave functions for composite
fermions a priori, as we do for ordinary electrons? In this subsection,
we take the first step towards demonstrating the possibility by developing
a wave equation for composite fermions in ideal systems.

The wave equation can be derived from the variational principle $\delta L=0$,
with the Lagrangian $L$ defined by
\begin{equation}
L=\int\mathrm{d}\mu_{B}^{(0)}(\bm{z})\mathrm{d}\mu_{b}^{(0)}(\bm{\eta})\left[\epsilon-T(\bm{z},\bm{\eta})\right]\left|\psi(z,\bar{\eta})\right|^{2},
\end{equation}
where $\epsilon$ is the Lagrange multiplier for the normalization
constraint of the wave function
\begin{equation}
\int\mathrm{d}\mu_{B}^{(0)}(\bm{z})\int\mathrm{d}\mu_{b}^{(0)}(\bm{\eta})\left|\psi\left(z,\bar{\eta}\right)\right|^{2}=1,\label{eq:normalization}
\end{equation}
and $T$ is the binding energy of a composite fermion modeled as a
harmonic potential 
\begin{equation}
T=\frac{\hbar^{2}}{2m^{\ast}l_{B}^{2}l_{b}^{2}}\left|\bm{z}-\bm{\eta}\right|^{2}.\label{eq:Vbound}
\end{equation}
It becomes the kinetic energy Eq.~(\ref{eq:kineticenergy}) in the
point-particle interpretation discussed in Sec.~\ref{subsec:Dipole-model}. 

Differentiating the Lagrangian with respect to $\psi^{\ast}(z,\bar{\eta})$
gives rise to the wave equation $\epsilon\psi=\hat{H}\psi$, with
the Hamiltonian defined by
\begin{equation}
\left[\hat{H}\psi\right]\left(z,\bar{\eta}\right)\equiv\hat{P}T\left(\bm{z},\bm{\eta}\right)\psi\left(z,\bar{\eta}\right),\label{eq:Hpsi}
\end{equation}
where $\hat{P}$ denoting the projection into the Hilbert space of
the composite fermion defined in Sec.~\ref{subsec:Segal-Bargmann-space}.
Applying the rule of the projection into Landau levels, we map $\bar{z}$
and $\eta$ to the operators $\hat{\bar{z}}\equiv2l_{B}^{2}\partial_{z}$
and $\hat{\eta}\equiv2l_{b}^{2}\partial_{\bar{\eta}}$, respectively~\citep{jain2007}.
The stationary-state wave equation of the composite fermion can then
be written as:
\begin{equation}
\epsilon\psi(z,\bar{\eta})=-\frac{\hbar^{2}}{2m^{\ast}}\left(2\partial_{z}-\frac{\bar{\eta}}{l_{B}^{2}}\right)\left(2\partial_{\bar{\eta}}-\frac{z}{l_{b}^{2}}\right)\psi(z,\bar{\eta}),\label{eq:cfweq}
\end{equation}
where an unimportant constant term in the Hamiltonian due to the ordering
of operators is ignored.

We can transform the wave equation to an ordinary Schrödinger equation
for a charged particle subjected to a uniform magnetic field by applying
the transformation 
\begin{equation}
\psi(z,\bar{\eta})=\sqrt{2\pi}l_{B}\exp\left[\frac{1}{4}\left(\frac{1}{l_{B}^{2}}+\frac{1}{l_{b}^{2}}\right)z\bar{\eta}\right]\varphi(z,\bar{\eta}).\label{eq:psivarphi}
\end{equation}
 For $\varphi(\bm{\xi})\equiv\varphi(z,\bar{\eta})|_{z\rightarrow\xi,\bar{\eta}\rightarrow\bar{\xi}}$,
we have 
\begin{equation}
\epsilon\varphi(\bm{\xi})=-\frac{\hbar^{2}}{2m^{\ast}}\left(2\partial_{\xi}-\frac{\sigma\bar{\xi}}{2l^{2}}\right)\left(2\partial_{\bar{\xi}}+\frac{\sigma\xi}{2l^{2}}\right)\varphi(\bm{\xi}),\label{eq:cfweq1}
\end{equation}
with $l\equiv\sqrt{\hbar/e|\mathcal{B}|}$ being the magnetic length
of the effective magnetic field $\mathcal{B}=B-b$, and $\sigma\equiv\mathrm{sgn}(\mathcal{B})$
indicating its direction.

\subsection{Wave equation: the standard representation \label{subsec:Wave-equation:-the-1}}

We can also have a wave equation for the standard representation.
In the case of non-interacting composite fermions, both $\Psi_{\mathrm{CF}}$
and $\tilde{\Psi}_{\mathrm{CF}}$ are Slater determinants of single-particle
wave functions. The single-particle counterpart of the transformation
Eq.~(\ref{eq:transformPsiCF}) is
\begin{equation}
\psi\left(z,\bar{\eta}\right)=\int\mathrm{d}\mu_{B}^{(0)}\left(\bm{\xi}\right)e^{z\bar{\xi}/2l_{B}^{2}+\xi\bar{\eta}/2l_{b}^{2}}\tilde{\psi}\left(\bm{\xi}\right),\label{eq:transformpsiCF}
\end{equation}
where $\tilde{\psi}(\bm{\xi})$ denotes a single-particle wave function
in the standard representation. Substituting it into Eq.~(\ref{eq:cfweq}),
we obtain the wave equation (see Appendix \ref{subsec:Ideal-systems-operators})

\begin{equation}
\epsilon\tilde{\varphi}(\bm{\xi})=-\frac{\hbar^{2}}{2m^{\ast}}\left(2\partial_{\xi}-\frac{\sigma\bar{\xi}}{2l^{2}}\right)\left(2\partial_{\bar{\xi}}+\frac{\sigma\xi}{2l^{2}}\right)\tilde{\varphi}(\bm{\xi}),\label{eq:cfweqstd}
\end{equation}
and 
\begin{equation}
\tilde{\psi}(\bm{\xi})=\sqrt{2\pi}l_{B}\exp\left(\sigma\frac{\left|\xi\right|^{2}}{2l^{2}}\right)\tilde{\varphi}(\bm{\xi}).\label{eq:psitildevarphitilde}
\end{equation}
We see that the wave equation for $\tilde{\varphi}(\bm{\xi})$ is
just the ordinary Schrödinger equation for a charge particle in the
uniform effective magnetic field. 

Our theory reproduces the well-established results of the standard
theory for ideal states. The eigen-solutions of the wave equation
(see Appendix \ref{appendix:-levels-of-the}) are exactly the $\Lambda$
orbits of the standard theory of composite fermions~\citep{jain2007}.
The interpretation of the fractional series is also the same: a fractional
state of electrons corresponds to an integer filling state of composite
fermions. The filling factor $\nu=n/(2n+1)<1/2$, $n\in Z$ corresponds
to $n$ filled $\Lambda$-levels, and $\nu=(n+1)/(2n+1)>1/2$ corresponds
to $n+1$ filled $\Lambda$-levels with $\sigma=-1$. For the special
case of $\nu=1/2$, the effective magnetic field vanishes. The eigen-solutions
of the wave equation become the plane-wave states, and composite fermions
will form a Fermi sea. The resulting physical wave function is exactly
the Rezayi-Read wave function of the composite Fermi-liquid~\citep{rezayi1994}.

We note that the states $\{\varphi_{i}\}$and $\{\tilde{\varphi}_{i}\}$
are dual to each other, and form a biorthogonal system. This can be
seen by applying Eq.~(\ref{eq:transformpsiCF}) to rewrite the orthonormal
condition $\int\mathrm{d}\mu_{B}^{(0)}(\bm{z})\int\mathrm{d}\mu_{b}^{(0)}(\bm{\eta})\psi_{i}^{\ast}(z,\bar{\eta})\psi_{j}(z,\bar{\eta})=\delta_{ij}$
as
\begin{equation}
\int\mathrm{d}\bm{\xi}~\tilde{\varphi}_{i}^{\ast}(\bm{\xi})\varphi_{j}(\bm{\xi})=\delta_{ij}.
\end{equation}

\section{General theory \label{sec:General-theory}}

In this section, we generalize our theory to systems that are subjected
not only to strong uniform magnetic fields, but also to spatial and
temporal fluctuations of electromagnetic fields, and generally have
inhomogeneous densities. In the HLR theory, this can be done trivially
by assuming that the composite fermion obeys the ordinary Schrödinger
equation. In the dipole model, however, the composite fermion is far
from being an ordinary particle, as we see in Sec.~\ref{subsec:Dipole-model}.
We need to derive the general quantum theory of composite fermions
in a logical way, as we demonstrate in the last section for the ideal
systems. 

\subsection{Bergman space \label{subsec:Bergman_space_generalized}}

In this subsection, we show that the Hilbert space of a particle confined
in a Landau level by a non-uniform magnetic field is generally a Bergman
space with its weight determined by the spatial profile of the magnetic
field. Consequently, the Hilbert space of a composite fermion is the
tensor product of two Bergman spaces with their weights determined
by the spatial profiles of the physical and the CS magnetic fields,
respectively.

We consider a non-relativistic electron confined in the lowest Landau-level
by a non-uniform magnetic field $\bm{B}(\bm{z})=-B(\bm{z})\bm{n}$,
and assume $B(\bm{z})=B_{0}+B_{1}(\bm{z})>0$, $|B_{1}(\bm{z})|/B_{0}\ll1$.
The Hamiltonian of the system, in complex coordinates, is given by~\citep{jain2007}
\begin{equation}
\hat{H}=-\frac{\hbar^{2}}{2m_{\mathrm{e}}}\left(2\partial_{z}+\mathrm{i}\frac{e}{\hbar}\bar{A}\right)\left(2\partial_{\bar{z}}+\mathrm{i}\frac{e}{\hbar}A\right)+\frac{e\hbar B(\bm{z})}{2m_{\mathrm{e}}},
\end{equation}
with $A\equiv A_{x}(\bm{z})+\mathrm{i}A_{y}(\bm{z})$ and $\bar{A}\equiv A^{\ast}$
being the complex components of the vector potential of the magnetic
field. The first term of the Hamiltonian yields zero-energy for a
state with the wave function $\varphi(\bm{z})$ satisfying the constraint
\begin{equation}
\left[2\partial_{\bar{z}}+\mathrm{i}\frac{e}{\hbar}A(\bm{z})\right]\varphi\left(\bm{z}\right)=0.\label{eq:LL-constraint}
\end{equation}
All such states form the lowest Landau level in the non-uniform magnetic
field~\citep{spodyneiko2023}, and define the physical Hilbert space
of the electron in the zero-electron-mass limit $m_{\mathrm{e}}\rightarrow0$.
The second term of the Hamiltonian, on the other hand, can be interpreted
as the orbital magnetization energy of the electron, and will become
a part of the scalar potential experienced by composite fermions~\citep{simon1996}.
We note that for a two-dimensional massless Dirac particle, Equation~(\ref{eq:LL-constraint})
is an exact constraint for the zero-energy Landau level, and there
is no orbital magnetization energy.

To fulfill the constraint, a wave function in the Hilbert space must
have the form
\begin{equation}
\varphi(\bm{z})=\psi(z)\exp\left[-\frac{1}{2}f_{B}(\bar{z},z)\right],\label{eq:phicomplete}
\end{equation}
where $\psi(z)$ is a holomorphic function in $z$, and $f_{B}(\bar{z},z)$
is determined by the equation
\begin{equation}
\partial_{\bar{z}}f_{B}(\bar{z},z)=\mathrm{i}\frac{e}{\hbar}A(\bar{z},z).\label{eq:pzfb}
\end{equation}
Fixing the vector potential in the Coulomb gauge, we have $\partial_{z}A=-\partial_{\bar{z}}\bar{A}=-\mathrm{i}B(\bm{z})/2,$
and
\begin{equation}
\partial_{z}\partial_{\bar{z}}f_{B}(\bar{z},z)=\frac{e}{2\hbar}B(\bm{z}).\label{eq:feq}
\end{equation}
We can then choose $f_{B}(\bar{z},z)$ to be a real solution of the
equation. 

The Hilbert space of the electron is therefore a weighted Bergman
space consisting all holomorphic polynomials that are normalized by
the condition $\int\mathrm{d}\mu_{B}(z)|\psi(z)|^{2}=1$, where the
integral measure is modified to
\begin{equation}
\mathrm{d}\mu_{B}(\bm{z})=w_{B}\left(\bm{z}\right)\mathrm{d}\bm{z}\equiv\frac{\mathrm{d}\bm{z}}{2\pi l_{B}^{2}}\exp\left[-f_{B}(\bar{z},z)\right],
\end{equation}
with $w_{B}(\bm{z})$ being the weight of the Bergman space, and $l_{B}\equiv\sqrt{\hbar/eB_{0}}$.
We can choose the constant of integration for $f_{B}$ to normalize
the measure: $\int\mathrm{d}\mu_{B}(\bm{z})=1$.

Similarly, for a vortex in a non-uniform CS magnetic field $\bm{b}(\bm{\eta})=b(\bm{\eta})\bm{n}$,
$b(\bm{\eta})>0$, its Hilbert space is a Bergman space consisting
all anti-holomorphic polynomials in $\bar{\eta}$ with the modified
integral measure
\begin{equation}
\mathrm{d}\mu_{b}(\bm{\eta})=w_{b}\left(\bm{\eta}\right)\mathrm{d}\bm{\eta}\equiv\frac{\mathrm{d}\bm{\eta}}{2\pi l_{b}^{2}}\exp\left[-f_{b}(\bar{\eta},\eta)\right],\label{eq:dmub}
\end{equation}
where $f_{b}(\bar{\eta},\eta)$ is a real solution of the equation
\begin{equation}
\partial_{\bar{\eta}}\partial_{\eta}f_{b}(\bar{\eta},\eta)=\frac{e}{2\hbar}b(\bm{\eta}).\label{eq:ppfb}
\end{equation}
The counterpart of Eq.~(\ref{eq:pzfb}) for the vortex is
\begin{equation}
\partial_{\eta}f_{b}(\bar{\eta},\eta)=\mathrm{i}\frac{e}{\hbar}\bar{a}(\bar{\eta},\eta).\label{eq:pfb}
\end{equation}
where $\bar{a}\equiv a_{x}-\mathrm{i}a_{y}$ denotes the complex-conjugated
component of the vector potential $(a_{x},a_{y})$ of the CS magnetic
field. 

As in the ideal case, the state of a composite fermion is represented
by a bi-variate wave function that is holomorphic in the coordinate
of the electron and anti-holomorphic in the coordinate of the vortex,
defined in the Hilbert space that is the tensor product of the two
Bergman spaces.

We can also define the reproducing kernels $K_{B}(z,\bar{z}^{\prime})$
and $K_{b}(\bar{\eta},\eta^{\prime})$ for the weighted Bergman spaces
of electrons and vortices, respectively~\citep{hall1999}. $K_{B}$
transforms a wave function defined in the $B$-Bergman spaces back
to itself:
\begin{align}
\psi(z) & =\int\mathrm{d}\mu_{B}(\bm{z}^{\prime})K_{B}(z,\bar{z}^{\prime})\psi(z^{\prime}).\label{eq:reprodkern}
\end{align}
It also defines the projection into the space:
\begin{equation}
\hat{P}_{\mathrm{LLL}}f(\bm{z})\equiv\int\mathrm{d}\mu_{B}(\bm{\xi})K_{B}(z,\bar{\xi})f(\bm{\xi}),
\end{equation}
In general, we do not have a closed form of the reproducing kernel
like Eq.~(\ref{eq:KB0}). We formally express the reproducing kernels
as
\begin{equation}
K_{B}(z,\bar{\xi})\equiv e^{F_{B}(\bar{\xi},z)}\label{eq:KBFB}
\end{equation}
by introducing the function $F_{B}(\bar{\xi},z)$. In the long-wavelength
limit, $F_{B}$ and $f_{B}$ can be related approximately, see Appendix~\ref{subsec:Long-wavelength-approximations}.
The reproducing kernel of the $b$-space has a similar set of properties.

\subsection{Wave function ansatz \label{subsec:Wave-function-ansatz}}

Using the modified integral measure Eq.~(\ref{eq:dmub}), we can
generalize the wave function ansatz Eq.~(\ref{eq:PsiCFtoPsi}) straightforwardly:
\begin{equation}
\Psi\left(\left\{ z_{i}\right\} \right)=\int\prod_{i}\mathrm{d}\mu_{b}\left(\bm{\eta}_{i}\right)J\left(\left\{ \eta_{i}\right\} \right)\Psi_{\mathrm{CF}}\left(\left\{ z_{i},\bar{\eta}_{i}\right\} \right),\label{eq:PsiCFtoPsiGen}
\end{equation}
where we only change the integral measures for $\{\bm{\eta}_{i}\}$,
and assume that the wave function of the vortices defining the physical
subspace remains the same as in the ideal case.

Due to the change of the weight of the $\bm{\eta}$\textendash Bergman
space, vortices are actually in a deformed bosonic Laughlin state
with an inhomogeneous density. The joint density distribution of vortices
is proportional to
\begin{equation}
e^{\sum_{i<j}4\ln\left|\eta_{i}-\eta_{j}\right|-\sum_{i}f_{b}\left(\bar{\eta}_{i},\eta_{i}\right)}.
\end{equation}
Using Laughlin's plasma analogy~\citep{laughlin1983a}, we can interpret
it as the distribution function of a set of classical particles, each
of which carries two unit ``charges'', on a non-uniform neutralizing
background with the ``charge'' density $\partial_{\eta}\partial_{\bar{\eta}}f_{b}(\bar{\eta},\eta)/\pi$.
Such a system is expected to be nearly ``charge-neutral'' everywhere.
It implies that the single-particle density of vortices should be
\begin{equation}
2\rho_{\mathrm{v}}\left(\bm{\eta}\right)\simeq\frac{1}{\pi}\partial_{\eta}\partial_{\bar{\eta}}f_{b}\left(\bar{\eta},\eta\right)=\frac{e}{h}b\left(\bm{\eta}\right),
\end{equation}
where we make use of Eq.~(\ref{eq:ppfb}). We see that the CS self-consistent
condition, which relates the vortex density to the strength of the
CS magnetic field, arises as a result of the constraint of the physical
subspace.

The standard ansatz can be generalized and shown to be equivalent
to the new ansatz. By using the reproducing kernel of the electron
Bergman space, the standard ansatz Eq.~(\ref{eq:Psistd}) can be
written as
\begin{multline}
\Psi\left(\left\{ z_{i}\right\} \right)=\int\prod_{i}\mathrm{d}\mu_{B}\left(\bm{\xi}_{i}\right)\prod_{i}K_{B}\left(z_{i},\bar{\xi}_{i}\right)\\
\times J\left(\left\{ \xi_{i}\right\} \right)\tilde{\Psi}_{\mathrm{CF}}\left(\left\{ \bm{\xi}_{i}\right\} \right).\label{eq:wfansatzstdgen}
\end{multline}
The general relation between the dipole representation and the standard
representation reads
\begin{multline}
\Psi_{\mathrm{CF}}\left(\left\{ z_{i},\bar{\eta}_{i}\right\} \right)=\int\prod_{i}\mathrm{d}\mu_{B}\left(\bm{\xi}_{i}\right)\\
\times\left[\prod_{i}K_{B}\left(z_{i},\bar{\xi}_{i}\right)K_{b}\left(\bar{\eta}_{i},\xi_{i}\right)\right]\tilde{\Psi}_{\mathrm{CF}}\left(\left\{ \bm{\xi}_{i}\right\} \right).\label{eq:CFtransform}
\end{multline}

\subsection{General wave equation \label{subsec:General-wave-equation} }

In this subsection, we generalize the wave equation Eq.~(\ref{eq:cfweq})
for systems that are subjected to external perturbations. We assume
that the external magnetic field has a strong uniform component $B_{0}$
and a small fluctuating component $B_{1}(\bm{z})$ which varies slowly
over space with $|B_{1}(\bm{z})|/B_{0}\ll1$, $|\nabla B(\bm{z})|l_{B}/B_{0}\ll1$,
and the strength of the external electric field is weak and does not
induce inter-Landau-level transitions. The resulting theory will be
adequate for predicting long-wavelength responses to electromagnetic
fields~\citep{ji2021}. In this limit, we can establish a general
wave equation while not obscured by excessive microscopic details.
A more general theory would require taking into account microscopic
details which will be elucidated in Sec.~\ref{sec:Derivation-of-the}.

The Lagrangian of the dipole model for a set of composite fermions,
in terms of the single-particle wave-functions $\{\psi_{i}(z,\bar{\eta})\}$,
can be generally written as:
\begin{multline}
L=\sum_{i}\int\mathrm{d}\mu_{B}\left(\bm{z}\right)\mathrm{d}\mu_{b}(\bm{\eta})\biggr\{\epsilon_{i}\left|\psi_{i}\left(z,\bar{\eta}\right)\right|^{2}-\frac{\hbar^{2}}{2m^{\ast}}\\
\times\int\mathrm{d}\mu_{b}(\bm{\eta}^{\prime})\psi_{i}^{\ast}(z,\bar{\eta})\frac{\left(\bar{z}-\bar{\eta}\right)\left(z-\eta^{\prime}\right)}{l_{b}^{2}(\bm{z})l_{B}^{2}(\bm{z})}K_{b}(\bar{\eta},\eta^{\prime})\psi_{i}(z,\bar{\eta}^{\prime})\biggr\}\\
-\int\mathrm{d}\bm{z}~\Phi\left(\bm{z}\right)\rho_{\mathrm{e}}(\bm{z})-E_{\mathrm{xc}}\left[\rho_{\mathrm{e}}\right]\\
-\frac{e^{2}}{8\pi\varepsilon}\int\mathrm{d}\bm{z}\mathrm{d}\bm{z}^{\prime}\frac{\left[\rho_{\mathrm{e}}(\bm{z})-\rho_{0}\right]\left[\rho_{\mathrm{e}}(\bm{z}^{\prime})-\rho_{0}\right]}{\left|\bm{z}-\bm{z}^{\prime}\right|}\\
-\int\mathrm{d}\bm{\eta}\phi\left(\bm{\eta}\right)\left[\rho_{\mathrm{v}}\left(\bm{\eta}\right)-\frac{e}{2h}b\left(\bm{\eta}\right)\right],\label{eq:Lagrangian}
\end{multline}
where the summation is over the occupied states of composite fermions,
and $\epsilon_{i}$ is the Lagrange multiplier for the normalization
constraint of the wave functions. The second term in the braces is
the kinetic energy, which is basically the harmonic binding potential
Eq.~(\ref{eq:Vbound}) written in a form that implies anti-normal
ordering when quantizing $\bm{\eta}$ (see Appendix \ref{sec:quantization}),
with a space-dependent coefficient parametrized in the local magnetic
lengths of the external field $l_{B}(\bm{z})\equiv\sqrt{\hbar/eB(\bm{z})}$
and the CS field $l_{b}(\bm{z})\equiv\sqrt{\hbar/eb(\bm{z})}$. The
third term is the energy due to the single-body scalar potential $\Phi(\bm{z})$
experienced by electrons, which includes the scalar potential of the
external electromagnetic field as well as the orbital magnetization
energy discussed in Sec.~\ref{subsec:Bergman_space_generalized},
and
\begin{equation}
\rho_{\mathrm{e}}(\bm{z})=w_{B}\left(\bm{z}\right)\sum_{i}\int\mathrm{d}\mu_{b}\left(\bm{\eta}\right)\left|\psi_{i}\left(z,\bar{\eta}\right)\right|^{2}\label{eq:rhoe}
\end{equation}
is the local density of electrons. The next two terms are the Coulomb
energy and an exchange-correlation energy functional $E_{\mathrm{xc}}[\rho_{\mathrm{e}}]$
which accounts for the exchange and correlation effects of composite
fermions. The last term imposes the CS constraint which relates the
local density of vortices
\begin{equation}
\rho_{\mathrm{v}}(\bm{\eta})=w_{b}\left(\bm{\eta}\right)\sum_{i}\int\mathrm{d}\mu_{B}\left(\bm{z}\right)\left|\psi_{i}\left(z,\bar{\eta}\right)\right|^{2}\label{eq:rhov}
\end{equation}
to the local strength of the CS magnetic field $b(\bm{\eta})$, with
$\phi(\bm{\eta})$ serving as the Lagrange multiplier. The effects
of the non-uniform physical and CS magnetic fields are included implicitly
in the integral measures $\mathrm{d}\mu_{B}$ and $\mathrm{d}\mu_{b}$,
respectively. In Sec.~\ref{sec:Derivation-of-the}, we will derive
the Lagrangian from first principles.

Differentiating the Lagrangian with respect to $\psi_{i}^{\ast}$,
we obtain a generalized wave equation for the stationary state of
a composite fermion:
\begin{align}
\epsilon\psi(z,\bar{\eta})= & \int\mathrm{d}\mu_{B}(\bm{\xi})\mathrm{d}\mu_{b}(\bm{\eta}^{\prime})K_{B}(z,\bar{\xi})K_{b}(\bar{\eta},\eta^{\prime})\nonumber \\
 & \times\mathcal{E}(\bm{\xi};\bar{\eta},\eta^{\prime})\psi(\xi,\bar{\eta}^{\prime}),\\
\mathcal{E}(\bm{\xi};\bar{\eta},\eta^{\prime})= & \frac{\hbar^{2}}{2m^{\ast}}\frac{\left(\bar{z}-\bar{\eta}\right)\left(z-\eta^{\prime}\right)}{l_{b}^{2}(\bm{z})l_{B}^{2}(\bm{z})}+\Phi_{\mathrm{eff}}(\bm{z})+\phi(\bm{\eta}),
\end{align}
where we drop the state index subscripts for brevity, $\Phi_{\mathrm{eff}}$
is the effective scalar potential experienced by electrons, defined
by
\begin{equation}
\Phi_{\mathrm{eff}}(\bm{z})=\Phi(\bm{z})+\frac{e^{2}}{4\pi\varepsilon}\int\mathrm{d}\bm{z}^{\prime}\frac{\rho_{\mathrm{e}}(\bm{z}^{\prime})-\rho_{0}}{\left|\bm{z}-\bm{z}^{\prime}\right|}+v_{\mathrm{xc}}[\rho_{\mathrm{e}}](\bm{z}),\label{eq:Phieff}
\end{equation}
with the exchange-correlation potential $v_{\mathrm{xc}}[\rho_{\mathrm{e}}]\equiv\delta E_{\mathrm{xc}}[\rho_{\mathrm{e}}]/\delta\rho_{\mathrm{e}}+\tau_{\mathrm{xc}}(\bm{z})$,
and
\begin{equation}
\tau_{\mathrm{xc}}(\bm{z})=\frac{2\pi\hbar^{2}}{m^{\ast}}\int\mathrm{d}\mu_{b}(\bm{\eta})\sum_{i}\left|\frac{z-\hat{\eta}}{l_{B}(\bm{z})}\psi_{i}(z,\bar{\eta})\right|^{2},\label{eq:tauxc}
\end{equation}
which is obtained by differentiating the kinetic energy with respect
to $\rho_{\mathrm{e}}(\bm{z})\approx\rho_{\mathrm{v}}(\bm{z})=1/4\pi l_{b}^{2}(\bm{z})$
and applying the quantization (see below), and $\phi(\bm{\eta})$
can be interpreted as the scalar potential experienced by vortices.
The orthonormal condition between two eigen-states is:
\begin{equation}
\int\mathrm{d}\mu_{B}(\bm{z})\int\mathrm{d}\mu_{b}(\bm{\eta})\psi_{i}^{\ast}\left(z,\bar{\eta}\right)\psi_{j}\left(z,\bar{\eta}\right)=\delta_{ij}.\label{eq:orthonormal-1}
\end{equation}

Applying the rules of quantization defined in Appendix~\ref{sec:quantization},
we can write the wave equation as

\begin{equation}
\epsilon\psi\left(z,\bar{\eta}\right)=\hat{H}_{\psi}\psi\left(z,\bar{\eta}\right),\label{eq:weqdipole}
\end{equation}
with the effective Hamiltonian operator
\begin{multline}
\hat{H}_{\psi}=\frac{\hbar^{2}}{2m^{\ast}}\left(\hat{\bar{z}}-\bar{\eta}\right)\frac{1}{l_{b}^{2}(\bar{\eta},z)l_{B}^{2}(\bar{\eta},z)}\left(z-\hat{\eta}\right)\\
+N_{+}\left[\Phi_{\mathrm{eff}}\left(\hat{\bar{z}},z\right)+\phi\left(\bar{\eta},\hat{\eta}\right)\right],\label{eq:Hpsi-1}
\end{multline}
where $\hat{\bar{z}}$ and $\hat{\eta}$ are defined by (see Appendix
\ref{subsec:Quantization-in-the})
\begin{align}
\left[\hat{\bar{z}}\psi\right](z,\bar{\eta}) & \equiv\int\mathrm{d}\mu_{B}(\bm{\xi})K_{B}(z,\bar{\xi})\bar{\xi}\psi(\xi,\bar{\eta}),\label{eq:zhatdef}\\
\left[\hat{\eta}\psi\right](z,\bar{\eta}) & \equiv\int\mathrm{d}\mu_{b}(\bm{\zeta})K_{b}(\bar{\eta},\zeta)\zeta\psi(z,\bar{\zeta}),\label{eq:etahatdef}
\end{align}
and $N_{+}[\cdots]$ denotes the normal ordering that places $\hat{\bar{z}}$
and $\hat{\eta}$ on the left of all $z$'s and $\eta$'s. We apply
the approximations $l_{B}^{2}(\bm{z})\approx l_{B}^{2}(\bar{\eta},z)$
and $l_{b}^{2}(\bm{z})\approx l_{b}^{2}(\bar{\eta},z)$ for the coefficient
of the kinetic energy.

The wave equation is complemented by a set of CS self-consistent conditions,
which are obtained by differentiating the Lagrangian Eq.~(\ref{eq:Lagrangian})
with respect to $\bm{a}$ and $\phi$. We have
\begin{align}
b(\bm{\eta}) & =\frac{2h}{e}\rho_{\mathrm{v}}(\bm{\eta}),\label{eq:csb-1}\\
\bm{E}_{\mathrm{v}}(\bm{\eta}) & =\frac{2h}{e}\bm{n}\times\bm{j}_{\mathrm{v}}(\bm{\eta}),\label{eq:cse-1}
\end{align}
where $\bm{E}_{\mathrm{v}}$ and $b$ are the CS electric and magnetic
fields, respectively, $\bm{j}_{\mathrm{v}}(\bm{\eta})$ denotes the
current density of vortices, which can be written as (see Appendix
\ref{sec:Current-densities})
\begin{multline}
\bm{j}_{\mathrm{v}}(\bm{\eta})=\frac{\rho_{\mathrm{v}}(\bm{\eta})}{b(\bm{\eta})}\bm{E}_{\mathrm{v}}(\bm{\eta})\times\bm{n}+\frac{\hbar}{m^{\ast}}w_{b}(\bm{\eta})\sum_{i}\\
\times\int\mathrm{d}\mu_{B}\left(\bm{z}\right)\psi_{i}^{\ast}\left(z,\bar{\eta}\right)\frac{\bm{n}\times(\bm{z}-\bm{\eta})}{l_{B}^{2}(\bm{z})}\psi_{i}\left(z,\bar{\eta}\right).\label{eq:jv}
\end{multline}

\subsection{Biorthogonal quantum mechanics \label{subsec:Wave-equation:-the}}

As in the theory for ideal systems, we can determine the wave equation
for the standard representation, and define a biorthogonal system
of wave functions. In general, the single-particle wave functions
of the dipole representation and the standard representation are related
by the transformation
\begin{equation}
\psi\left(z,\bar{\eta}\right)=\int\mathrm{d}\mu_{B}\left(\bm{\xi}\right)K_{B}\left(z,\bar{\xi}\right)K_{b}\left(\bar{\eta},\xi\right)\tilde{\psi}\left(\bm{\xi}\right).\label{eq:transforCFgeneral}
\end{equation}
The operators in the dipole representation can be mapped to their
counterparts in the standard representation accordingly, see Appendix~\ref{subsec:General-systems-operators}.

We further introduce the transformations

\begin{align}
\psi\left(\bm{\xi}\right) & =\sqrt{2\pi}l_{B}\exp\left[\frac{f_{B}(\bm{\xi})+f_{b}(\bm{\xi})}{2}\right]\varphi\left(\bm{\xi}\right),\label{eq:psipsi1}\\
\tilde{\psi}(\bm{\xi}) & =\sqrt{2\pi}l_{B}\exp\left[\frac{f_{B}(\bm{\xi})-f_{b}(\bm{\xi})}{2}\right]\tilde{\varphi}\left(\bm{\xi}\right),\label{eq:cftransform}
\end{align}
which are the counterparts of the transformations Eq.~(\ref{eq:psivarphi})
and Eq.~(\ref{eq:psitildevarphitilde}), respectively. The orthonormal
condition Eq.~(\ref{eq:orthonormal-1}) can then be rewritten as
\begin{equation}
\int\mathrm{d}\bm{\xi}\tilde{\varphi}_{i}^{\ast}(\bm{\xi})\varphi_{j}(\bm{\xi})=\delta_{ij}.\label{eq:orthonormal-2}
\end{equation}
We see that as in ideal systems, $\{\varphi_{i}\}$ and $\{\tilde{\varphi}_{i}\}$
are dual to each other and form a biorthogonal system. 

The wave equations for $\varphi$ and $\tilde{\varphi}$ have the
form of the biorthogonal quantum mechanics~\citep{brody2013}. In
general, we can show that the Hamiltonian for $\tilde{\varphi}(\bm{\xi})$
is the complex conjugate of that for $\varphi(\bm{\xi})$(see Appendix
\ref{subsec:Hamiltonian-in-the}). Therefore, we have 
\begin{align}
\epsilon\varphi(\bm{\xi}) & =\hat{H}\varphi(\bm{\xi}),\label{eq:weqdipole-1}\\
\epsilon\tilde{\varphi}\left(\bm{\xi}\right) & =\hat{H}^{\dagger}\tilde{\varphi}\left(\bm{\xi}\right),\label{eq:weqstd}
\end{align}
where the first equation is transformed from Eq.~(\ref{eq:weqdipole})
with
\begin{equation}
\hat{H}\equiv e^{-\frac{f_{B}+f_{b}}{2}}\hat{H}_{\psi}e^{\frac{f_{B}+f_{b}}{2}}.\label{eq:Hdef}
\end{equation}
Note that $\hat{H}$ is non-Hermitian in general.

In the long-wavelength limit, the effective Hamiltonian of a composite
fermion can be written as (see Appendix \ref{subsec:Long-wavelength-approximations}):
\begin{multline}
\hat{H}=-\frac{\hbar^{2}}{2m^{\ast}}\left[2\partial_{\xi}+\mathrm{i}\frac{e}{\hbar}\bar{\mathcal{A}}\left(\bm{\xi}\right)+\mathrm{i}\frac{m^{\ast}}{\hbar}\bar{v}(\bm{\xi})\right]\\
\times\left[2\partial_{\bar{\xi}}+\mathrm{i}\frac{e}{\hbar}\mathcal{A}\left(\bm{\xi}\right)+\mathrm{i}\frac{m^{\ast}}{\hbar}V(\bm{\xi})\right]+\Phi_{\mathrm{eff}}\left(\bm{\xi}\right)+\phi\left(\bm{\xi}\right),\label{eq:Hdipole}
\end{multline}
where $(\bar{\mathcal{A}},\mathcal{A})$ denotes the effective vector
potential experienced by composite fermions:
\begin{equation}
\bm{\mathcal{A}}=\bm{a}+\bm{A},
\end{equation}
and $V(\bar{\eta},z)\equiv2\mathrm{i}\partial_{\bar{\eta}}\Phi_{\mathrm{eff}}(\bar{\eta},z)/eB(\bar{\eta},z)$
and $\bar{v}(\bar{\eta},z)\equiv2\mathrm{i}\partial_{z}\phi(\bar{\eta},z)/eb(\bar{\eta},z)$
are the complex components of the drift velocities $\bm{V}=\bm{E}\times\bm{B}/B^{2}$
and $\bm{v}=\bm{E}_{\mathrm{v}}\bm{\times\bm{b}}/b^{2}$ in the presence
of the electric fields $\bm{E}\equiv e^{-1}\bm{\nabla}\Phi_{\mathrm{eff}}$
and $\bm{E}_{\mathrm{v}}\equiv e^{-1}\bm{\nabla}\phi$ for electrons
and vortices, respectively.

The set of wave equations can be further generalized for time-dependent
systems. We have (see Appendix \ref{subsec:Time-dependent-systems}):

\begin{align}
\mathrm{i}\hbar\frac{\partial\varphi(\bm{\xi},t)}{\partial t} & =\hat{H}\varphi(\bm{\xi},t),\label{eq:TDWEQ1}\\
\mathrm{i}\hbar\frac{\partial\tilde{\varphi}(\bm{\xi},t)}{\partial t} & =\hat{H}^{\dagger}\tilde{\varphi}(\bm{\xi},t),\label{eq:TDWEQ2}
\end{align}
where $\hat{H}$ is formally identical to the stationary state Hamiltonian
Eq.~(\ref{eq:Hdipole}) in the long-wavelength limit, but the electric
fields $\bm{E}$ and $\bm{E}_{\mathrm{v}}$, which determine the drift
velocities, are replaced by their gauge-invariant forms $\bm{E}=e^{-1}\bm{\nabla}\Phi_{\mathrm{eff}}-\partial_{t}\bm{A}$
and $\bm{E}_{\mathrm{v}}=e^{-1}\bm{\nabla}\phi-\partial_{t}\bm{a}$. 

The wave equations (\ref{eq:TDWEQ2}, \ref{eq:TDWEQ1}), together
with the CS self-consistent conditions Eqs.~(\ref{eq:csb-1},~\ref{eq:cse-1})
and the self-consistent equation for the effective potential Eq.~(\ref{eq:Phieff}),
define the effective theory of composite fermions in the presence
of long-wavelength external perturbations. It is evident that the
effective theory differs from the heuristic HLR theory because of
the corrections from the drift velocities in Eq.~(\ref{eq:Hdipole}).
The corrections have previously been identified as either anomalous
velocity corrections~\citep{ji2021} or side-jump corrections~\citep{wang2017}
in the context of the semi-classical theory of composite fermions.
Equation (\ref{eq:Hdipole}) shows how these corrections are manifested
in the quantum mechanics of composite fermions.

\section{Microscopic underpinning \label{sec:Derivation-of-the}}

In this section, we derive the phenomenological dipole model, which
underlies our derivation of the quantum mechanics of composite fermions,
from the microscopic model of interacting electrons confined in a
Landau level in the zero-electron-mass limit. The microscopic Lagrangian
of such a system can be written as
\begin{equation}
L_{\mathrm{M}}=\Braket{\Psi|E-V_{\mathrm{ee}}-\Phi|\Psi},\label{eq:L}
\end{equation}
where $\Psi$ denotes the many-body wave function of electrons, $E$
is a Lagrange multiplier for the normalization constraint  $\braket{\Psi|\Psi}=1$,
$V_{\mathrm{ee}}=(e^{2}/4\pi\varepsilon)\sum_{i<j}|z_{i}-z_{j}|^{-1}+V_{\mathrm{B}}$
denotes the Coulomb interaction between electrons with $V_{\mathrm{B}}$
being the potential from a uniform neutralizing positive charge background,
and $\Phi\equiv\sum_{i}\Phi(\bm{z}_{i})$ denotes the energy of an
externally applied scalar potential. The kinetic energy of electrons
is ignored since it is completely quenched in a Landau level. 

Our derivation is based on the general variational principle of quantum
mechanics. By introducing the fictitious degrees of freedom of the
vortices, we basically embed the physical Hilbert space into a larger
Hilbert space of composite fermions, in the hope that the strongly
correlated state of electrons can be viewed as the projection of a
non-correlated state of composite fermions onto a lower-dimensional
subspace. Therefore, we choose the trial electron wave functions for
$\ket{\Psi}$ to be the ansatz form Eq.~(\ref{eq:PsiCFtoPsiGen})
with $\Psi_{\mathrm{CF}}$ being the Slater determinant of a set of
single-body trial wave-functions $\{\psi_{i}\}$ of composite fermions.
We will show that the Lagrangian Eq.~(\ref{eq:Lagrangian}) in terms
of $\{\psi_{i}\}$ can be derived from the microscopic Lagrangian
Eq.~(\ref{eq:L}). The set of the single-body trial wave-functions
should be determined by applying the variational principle
\begin{equation}
\delta L=0,
\end{equation}
which gives rise to the wave-equations and the CS self-consistent
conditions.

\subsection{Chern-Simons constraints \label{subsec:Chern-Simons-constraints}}

A notable feature of the theory of composite fermions is the presence
of the fictitious CS fields which are determined self-consistently
by Eqs.~(\ref{eq:csb-1}, \ref{eq:cse-1}). In this subsection, we
show how the CS fields and the self-consistent conditions emerge in
a microscopic theory.

It is easy to see that for the Slater determinant wave function
\begin{equation}
\Psi_{\mathrm{CF}}\left(\left\{ z_{i},\bar{\eta}_{i}\right\} \right)=\frac{1}{\sqrt{N!}}\mathrm{det}\left[\psi_{j}(z_{i},\bar{\eta}_{i})\right],\label{eq:PsiCFDet}
\end{equation}
two sets of single particle trial wave functions that are related
by a non-singular linear transformation yield the same physical wave
function after applying Eq.~(\ref{eq:PsiCFtoPsi-0}) or Eq.~(\ref{eq:PsiCFtoPsiGen}).
To eliminate the redundancy, it is necessary to impose the orthonormal
condition:
\begin{equation}
\int\mathrm{d}\mu_{B}(\bm{\xi})\mathrm{d}\mu_{b}(\bm{\eta})\psi_{i}^{\ast}(\xi,\bar{\eta})\psi_{j}(\xi,\bar{\eta})=\delta_{ij}.\label{eq:orthonormal}
\end{equation}
We note that the orthonormality depends on the weight in $\mathrm{d}\mu_{b}$,
which is not yet defined at this point.

To proceed, we adopt an approximation analogue to the Hartree approximation.
Basically, we determine the state of a composite fermion in an effective
medium formed by other composite fermions. In the spirit of the Hartree
approximation~\citep{robertg.parr1994}, we introduce a test particle
that is distinguishable from other composite fermions but interacts
and correlates just like them. The physical wave function of a system
with $N$ composite fermions plus such a test particle can be written
as
\begin{align}
\Psi^{\mathrm{t}}\left(z,\left\{ z_{i}\right\} \right)= & \int\mathrm{d}\mu_{b}\left(\bm{\eta}\right)\Psi_{\eta}^{\mathrm{v}}\left(\left\{ z_{i}\right\} \right)\psi\left(z,\bar{\eta}\right),\label{eq:wfCFtest}\\
\Psi_{\eta}^{\mathrm{v}}\left(\left\{ z_{i}\right\} \right)= & \int\prod_{i=1}^{N}\mathrm{d}\mu_{b}(\bm{\eta}_{i})\prod_{i=1}^{N}(\eta-\eta_{i})^{2}\nonumber \\
 & \times J\left(\left\{ \eta_{i}\right\} \right)\Psi_{\mathrm{CF}}\left(\left\{ z_{i},\bar{\eta}_{i}\right\} \right),\label{eq:Psiv}
\end{align}
where the test particle has the wave function $\psi(z,\bar{\eta}$),
and correlates with other composite fermions via the Bijl-Jastrow
factor. Because the test particle has no exchange symmetry with other
composite fermions, it can occupy any state, including those already
occupied in $\Psi_{\mathrm{CF}}$. Our approximation is to assume
that the set of single particle trial wave functions for constructing
$\Psi_{\mathrm{CF}}$ can be chosen from eigen-wave-functions of the
test particle. 

With the approximation, we can determine the weight of $\mathrm{d}\mu_{b}$,
self-consistently, by requiring that the orthonormality Eq.~(\ref{eq:orthonormal})
in the Hilbert space of composite fermions is consistent with that
of the physical Hilbert space. This is to require
\begin{equation}
\Braket{\Psi_{i}^{\mathrm{t}}|\Psi_{j}^{\mathrm{t}}}=\delta_{ij},\label{eq:Psiortho}
\end{equation}
where $\ket{\Psi_{i}^{\mathrm{t}}}$ and $\ket{\Psi_{j}^{\mathrm{t}}}$
denote two physical states obtained by setting $\psi=\psi_{i}$ and
$\psi=\psi_{j}$ in Eq.~(\ref{eq:wfCFtest}), respectively, and $\psi_{i}$
and $\psi_{j}$ satisfy the orthonormal condition Eq.~(\ref{eq:orthonormal}).
Equation (\ref{eq:Psiortho}) can be rewritten as
\begin{multline}
\int\mathrm{d}\mu_{B}(\bm{z})\mathrm{d}\mu_{b}(\bm{\eta})\mathrm{d}\mu_{b}(\bm{\eta}^{\prime})\psi_{i}^{\ast}(z,\bar{\eta})\\
\times K_{b}(\bar{\eta},\eta^{\prime})\psi_{j}(z,\bar{\eta}^{\prime})=\delta_{ij},\label{eq:orthK}
\end{multline}
with
\begin{equation}
K_{b}(\bar{\eta},\eta^{\prime})\equiv\Braket{\Psi_{\eta}^{\mathrm{v}}|\Psi_{\eta^{\prime}}^{\mathrm{v}}}.\label{eq:Kb}
\end{equation}
To make Eq.~(\ref{eq:orthK}) consistent with Eq.~(\ref{eq:orthonormal}),
we can adjust the weight of $\mathrm{d}\mu_{b}$ so that $K_{b}(\bar{\eta},\eta^{\prime})$
is the corresponding reproducing kernel. Equation (\ref{eq:orthK})
can then be reduced to Eq.~(\ref{eq:orthonormal}) by integrating
out $\bm{\eta}^{\prime}$.

The requirement that $K_{b}(\bar{\eta},\eta^{\prime})$ is the reproducing
kernel of the $\bm{\eta}$\textendash space gives rise to the CS constraint
Eq.~(\ref{eq:csb-1}) in the long-wavelength limit. To see this,
we rewrite Eq.~(\ref{eq:Kb}) as $K_{b}(\bar{\eta},\eta^{\prime})=\langle e^{\mathcal{F}}\rangle$,
with
\begin{multline}
\langle e^{\mathcal{F}}\rangle\equiv\int\prod_{i=1}^{N}\mathrm{d}\mu_{B}(\bm{z}_{i})\mathrm{d}\mu_{b}(\bm{\eta}_{i})\mathrm{d}\mu_{b}(\bm{\eta}_{i}^{\prime})e^{\mathcal{F}(\bar{\eta},\eta^{\prime},\{\bar{\eta}_{i},\eta_{i}^{\prime}\})}\\
\times J^{\ast}(\{\eta_{i}\})J(\{\eta_{i}^{\prime}\})\Psi_{\mathrm{CF}}^{\ast}(\{z_{i},\bar{\eta}_{i}\})\Psi_{\mathrm{CF}}(\{z_{i},\bar{\eta}_{i}^{\prime}\})\label{eq:avgF}
\end{multline}
and $\mathcal{F}(\bar{\eta},\eta^{\prime},\{\bar{\eta}_{i},\eta_{i}^{\prime}\})\equiv2\sum_{i}\ln(\bar{\eta}-\bar{\eta}_{i})(\eta^{\prime}-\eta_{i}^{\prime})$.
Using the cumulant expansion, we can approximate $F_{b}(\bar{\eta},\eta^{\prime})\equiv\ln K_{b}(\bar{\eta},\eta^{\prime})$
as:
\begin{equation}
F_{b}\left(\bar{\eta},\eta^{\prime}\right)\approx\left\langle \mathcal{F}\right\rangle +\frac{1}{2}\left\langle \left(\mathcal{F}-\left\langle \mathcal{F}\right\rangle \right)^{2}\right\rangle +\cdots,
\end{equation}
To the lowest order, we ignore the fluctuation and higher order corrections,
and have $F_{b}\approx\langle\mathcal{F}\rangle=2\sum_{i}\langle\ln(\bar{\eta}-\bar{\eta}_{i})(\eta^{\prime}-\eta_{i}^{\prime})\rangle$.
To evaluate the $i$-th term of the summation, we expand the Slater
determinant Eq.~(\ref{eq:PsiCFDet}) along its $i$-th row, substitute
the expansion into Eq.~(\ref{eq:avgF}), and ignore contributions
involving particle exchanges. We obtain
\begin{multline}
F_{b}\left(\bar{\eta},\eta^{\prime}\right)\approx\sum_{i}\int\mathrm{d}\mu_{B}\left(\bm{z}_{i}\right)\mathrm{d}\mu_{b}\left(\bm{\eta}_{i}\right)\mathrm{d}\mu_{b}\left(\bm{\eta}_{i}^{\prime}\right)\\
\times2\left[\ln(\bar{\eta}-\bar{\eta}_{i})+\ln(\eta^{\prime}-\eta_{i}^{\prime})\right]\\
\times\frac{1}{N}\sum_{a}\psi_{a}^{\ast}\left(z_{i},\bar{\eta}_{i}\right)K_{b}^{(a)}\left(\bar{\eta}_{i},\eta_{i}^{\prime}\right)\psi_{a}\left(z_{i},\bar{\eta}_{i}^{\prime}\right),\label{eq:Fbapprox}
\end{multline}
where $K_{b}^{(a)}$ is defined by Eq.~(\ref{eq:Kb}) but with one
composite fermion in the state $\psi_{a}$ being removed from Eq.~(\ref{eq:PsiCFDet}).
We assume that the effect of removing a composite fermion from the
effective medium of $N$ composite fermions is negligible, thus have
\begin{equation}
K_{b}^{(a)}(\bar{\eta}_{i},\eta_{i}^{\prime})\approx K_{b}(\bar{\eta}_{i},\eta_{i}^{\prime}).\label{eq:Kbapprox}
\end{equation}
 After integrating out $\bm{\eta}_{i}^{\prime}$, we obtain
\begin{equation}
F_{b}\left(\bar{\eta},\eta\right)\approx2\int\mathrm{d}^{2}\bm{\eta}_{1}\ln\left(\left|\eta-\eta_{1}\right|^{2}\right)\rho_{\mathrm{v}}\left(\bm{\eta}_{1}\right).
\end{equation}
where $\rho_{\mathrm{v}}(\bm{\eta}_{1})$ is the vortex density defined
in Eq.~(\ref{eq:rhov}). Applying the identity $\partial_{\eta}\partial_{\bar{\eta}}\ln(|\eta-\eta_{1}|^{2})=\pi\delta(\bm{\eta}-\bm{\eta}_{1})$,
we have
\begin{equation}
\partial_{\eta}\partial_{\bar{\eta}}F_{b}\left(\bar{\eta},\eta\right)=2\pi\rho_{\mathrm{v}}\left(\bm{\eta}\right).\label{eq:Fbsc}
\end{equation}
In the long-wavelength limit, we have $F_{b}(\bar{\eta},\eta)\approx f_{b}(\bm{\eta})-\ln[l_{b}^{2}(\bm{\eta})/l_{b}^{2}]$
(see Appendix~\ref{subsec:Long-wavelength-approximations}). Substituting
the relation into Eq.~(\ref{eq:Fbsc}), ignoring the spatial gradient
of the magnetic length, and applying Eq.~(\ref{eq:ppfb}), we obtain
the CS constraint Eq.~(\ref{eq:csb-1}).

We can then replace the normalization constraint $\braket{\Psi|\Psi}=1$
in Eq.~(\ref{eq:L}) with normalization constraints of the single-body
wave-functions as well as the CS constraint Eq.~(\ref{eq:csb-1}),
and introduce $\epsilon_{i}$ and $\phi(\bm{\eta})$ as the respective
Lagrange multipliers. The Lagrangian becomes
\begin{multline}
L=\int\mathrm{d}\mu_{B}(\bm{z})\mathrm{d}\mu_{b}(\bm{\eta})\sum_{i}\epsilon_{i}\left|\psi_{i}(z,\bar{\eta})\right|^{2}-\int\mathrm{d}\bm{\eta}\phi(\bm{\eta})\\
\times\left[\rho_{\mathrm{v}}\left(\bm{\eta}\right)-\frac{e}{2h}b\left(\bm{\eta}\right)\right]-\Braket{\Psi|V_{\mathrm{ee}}+\Phi|\Psi},
\end{multline}

\subsection{Energy \label{subsec:Energy}}

In this subsection, we determine the expectation value $\Braket{\Psi|V_{\mathrm{ee}}+\Phi|\Psi}.$
We shall show how the kinetic energy of a composite fermion, i.e.,
the electron-vortex binding potential, as well as its peculiar density-of-states
correction factor (see Sec.~\ref{subsec:Wave-equation}), would emerge. 

We first determine the expectation value of the scalar potential $\braket{\Psi|\Phi|\Psi}$.
Similar to Eq.~(\ref{eq:Fbapprox}), we have
\begin{multline}
\Braket{\Psi|\Phi|\Psi}\approx\sum_{i}\int\mathrm{d}\mu_{B}\left(\bm{z}_{i}\right)\mathrm{d}\mu_{b}\left(\bm{\eta}_{i}\right)\mathrm{d}\mu_{b}\left(\bm{\eta}_{i}^{\prime}\right)\Phi(\bm{z}_{i})\\
\times\frac{1}{N}\sum_{a}\psi_{a}^{\ast}\left(z_{i},\bar{\eta}_{i}\right)K_{b}^{(a)}\left(\bar{\eta}_{i},\eta_{i}^{\prime}\right)\psi_{a}\left(z_{i},\bar{\eta}_{i}^{\prime}\right).\label{eq:Phiapprox}
\end{multline}
Applying the approximation Eq.~(\ref{eq:Kbapprox}), we obtain
\begin{equation}
\Braket{\Psi|\Phi|\Psi}\approx\int\mathrm{d}\bm{z}\Phi(\bm{z})\rho_{\mathrm{e}}(\bm{z}).
\end{equation}

Next, we determine the expectation value of the Coulomb interaction
energy. It can be written as
\begin{equation}
\left\langle V_{\mathrm{ee}}\right\rangle =\frac{e^{2}}{8\pi\varepsilon}\int\mathrm{d}\bm{z}\mathrm{d}\bm{z}^{\prime}\frac{\rho_{2}(\bm{z},\bm{z}^{\prime})-2\rho_{\mathrm{e}}(\bm{z})\rho_{0}+\rho_{0}^{2}}{\left|z-z^{\prime}\right|},
\end{equation}
where $\rho_{2}(\bm{z},\bm{z}^{\prime})=\braket{\Psi|\sum_{i\ne j}\delta(\bm{z}-\bm{z}_{i})\delta(\bm{z}^{\prime}-\bm{z}_{j})|\Psi}$
is the two-particle reduced density matrix of electrons. We decompose
$\langle V_{\mathrm{ee}}\rangle$ into two parts. The first part is
the mean-field contribution of the Coulomb interaction
\begin{equation}
\bar{V}_{\mathrm{ee}}=\frac{e^{2}}{8\pi\varepsilon}\int\mathrm{d}\bm{z}\mathrm{d}\bm{z}^{\prime}\frac{\left[\rho_{\mathrm{e}}(\bm{z})-\rho_{0}\right]\left[\rho_{\mathrm{e}}(\bm{z}^{\prime})-\rho_{0}\right]}{\left|\bm{z}-\bm{z}^{\prime}\right|},
\end{equation}
which gives rise to the Coulomb energy term of the Lagrangian Eq.~(\ref{eq:Lagrangian}).
The second part is the correlation contribution
\begin{equation}
T=\frac{e^{2}}{8\pi\varepsilon}\int\mathrm{d}\bm{z}\mathrm{d}\bm{z}^{\prime}\frac{\rho_{2}(\bm{z},\bm{z}^{\prime})-\rho_{\mathrm{e}}(\bm{z})\rho_{\mathrm{e}}(\bm{z}^{\prime})}{\left|\bm{z}-\bm{z}^{\prime}\right|},\label{eq:Vbdef}
\end{equation}
which gives rise to the binding energy between electrons and vortices.

We determine the two-particle reduced density matrix by applying the
Hartree-like approximation introduced in the last subsection. We have
$\rho_{2}(\bm{z},\bm{z}^{\prime})=N(N-1)w_{B}(\bm{z})w_{B}(\bm{z}^{\prime})\int\prod_{i=3}^{N}\mathrm{d}\mu_{B}(\bm{z}_{i})|\Psi(\{z_{i}\}|^{2}$
with $z_{1}=z$ and $z_{2}=z^{\prime}$. We treat the first particle
($\bm{z}_{1}$) as a test particle, and the ensemble of other $N-1$
particles as an effective medium. By expanding the Slater determinant
Eq.~(\ref{eq:PsiCFDet}) along its first row, ignoring exchange terms
in $|\Psi(\{z_{i}\})|^{2}$, and replacing the $N-1$ particle effective
medium with the $N$-particle one as in Eq.~(\ref{eq:wfCFtest}),
we can approximate $\rho_{2}$ as
\begin{align}
\rho_{2}(\bm{z},\bm{z}_{1})\approx & ~w_{B}(\bm{z})\int\mathrm{d}\mu_{b}(\bm{\eta})\mathrm{d}\mu_{b}(\bm{\eta}^{\prime})K_{b}(\bar{\eta},\eta^{\prime})\nonumber \\
 & \times\sum_{a}\psi_{a}^{\ast}(z,\bar{\eta})\psi_{a}(z,\bar{\eta}^{\prime})\rho_{\mathrm{c}}(\bm{z}_{1};\bar{\eta},\eta^{\prime}),\label{eq:rho2}\\
\rho_{\mathrm{c}}(\bm{z}_{1};\bm{\eta})= & w_{B}(\bm{z}_{1})N\int\prod_{i=2}^{N}\mathrm{d}\mu_{B}(\bm{z}_{i})\frac{\left|\Psi_{\eta}^{\mathrm{v}}\left(\left\{ z_{i}\right\} \right)\right|^{2}}{\Braket{\Psi_{\eta}^{\mathrm{v}}|\Psi_{\eta}^{\mathrm{v}}}},\label{eq:rhoc}
\end{align}
and $\rho_{\mathrm{c}}(\bm{z}_{1};\bar{\eta},\eta^{\prime})\equiv\rho_{\mathrm{c}}(\bm{z}_{1},\bm{\eta})|_{\eta\rightarrow\eta^{\prime}}$.
$\rho_{\mathrm{c}}(\bm{z}_{1};\bm{\eta})$ is the density profile
of electrons surrounding a vortex at $\bm{\eta}$, which suppresses
the electron density in its vicinity, resulting in a void of electrons.

The Coulomb attraction between the test (first) electron and the void
gives rise to the binding energy of a composite fermion. Substituting
Eq.~(\ref{eq:rho2}) and (\ref{eq:rhoe}) into Eq.~(\ref{eq:Vbdef}),
we obtain

\begin{multline}
T\approx\int\mathrm{d}\mu_{B}(\bm{z})\mathrm{d}\mu_{b}(\eta)\mathrm{d}\mu_{b}(\eta^{\prime})K_{b}(\bar{\eta},\eta^{\prime})\\
\times\epsilon_{\mathrm{b}}^{\bigstar}(\bm{z};\bar{\eta},\eta^{\prime})\sum_{a}\psi_{a}^{\ast}(z,\bar{\eta})\psi_{a}(z,\bar{\eta}^{\prime}),\label{eq:T}
\end{multline}
with $\epsilon_{\mathrm{b}}^{\bigstar}(\bm{z};\bar{\eta},\eta^{\prime})\equiv\epsilon_{\mathrm{b}}(\bm{z};\bm{\eta})|_{\eta\rightarrow\eta^{\prime}}$,
\begin{equation}
\epsilon_{\mathrm{b}}(\bm{z};\bm{\eta})=\frac{e^{2}}{8\pi\varepsilon}\int\mathrm{d}\bm{z}_{1}\frac{\Delta\rho_{\mathrm{e}}(\bm{z}_{1};\bm{\eta})}{|z-z_{1}|},\label{eq:ebstar}
\end{equation}
and $\Delta\rho_{\mathrm{e}}(\bm{z}_{1};\bm{\eta})\equiv\rho_{\mathrm{c}}(\bm{z}_{1};\bm{\eta})-\rho_{\mathrm{e}}(\bm{z}_{1})$. 

\begin{figure}
\includegraphics{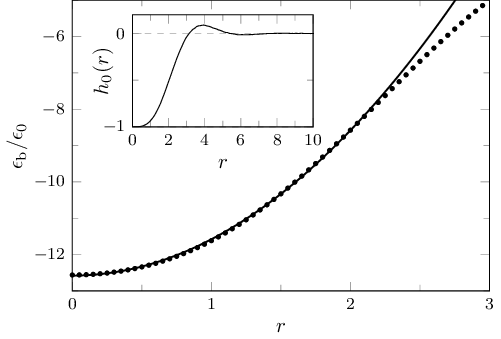}

\caption{\label{fig:Electron-vortex-binding-potentia}Electron-vortex binding
potential vs. $r\equiv|z-\eta|/l_{B}$ for the $\nu=1/3$ Laughlin
state. Dots are numerical results, and the solid line shows the fitting
$\epsilon_{\mathrm{b}}(r)/\epsilon_{0}=-12.6+r^{2}$, $\epsilon_{0}\equiv\nu e^{2}/16\pi^{2}\varepsilon l_{B}$.
Inset: the electron-vortex pair correlation function $h_{0}(r)$.
Calculated by Hao Jin. }
\end{figure}

The form of $\Delta\rho_{\mathrm{e}}(\bm{z}_{1};\bm{\eta})$ is constrained~\citep{tafelmayer1993}.
The electron density is suppressed near the center of the vortex,
and recovers in a length scale $\sim l_{B}$ (see the inset of Fig.~\ref{fig:Electron-vortex-binding-potentia}).
Thus we have $\Delta\rho_{\mathrm{e}}(\bm{\bm{z}_{1}};\bm{\eta})<0$
for $\bm{z}_{1}\rightarrow\bm{\eta}$ and $\Delta\rho_{\mathrm{e}}(\bm{z}_{1};\bm{\eta})\rightarrow0$
for $|\bm{z}_{1}-\bm{\eta}|\gg l_{B}$. Moreover, because the insertion
of a $2h/e$ vortex should induce a charge void with a total charge
of $2\nu e$, where $\nu$ is the filling factor, we have the sum
rule

\begin{equation}
\int\Delta\rho_{\mathrm{e}}(\bm{z}_{1};\bm{\eta})\mathrm{d}\bm{z}_{1}=-2\nu(\bm{\eta}),\label{eq:sumrule}
\end{equation}
where $\nu(\bm{\eta})$ denotes the local value of the filling factor,
and we assume that the electron density is nearly homogeneous.

The fact that the binding energy of a composite fermion originates
from the Coulomb attraction between an electron and a void with a
total charge of $2\nu e$ suggests that it should be proportional
to $2\nu$, which is exactly the density-of-states correction factor
$D\equiv b/B=2\nu$ appeared in Eq.~(\ref{eq:kineticenergy}). The
peculiar factor in the kinetic energy turns out to be a natural consequence
of the interaction origin of the binding energy.

\begin{comment}
In the long-wavelength limit, the density varies slowly over space.
We can approximate $h(\bm{z}_{1};\bm{\eta})$ as the pair correlation
function of a homogeneous system: $h(\bm{z}_{1};\bm{\eta})\approx h_{0}(|\bm{z}_{1}-\bm{\eta}|/l_{B}(\bm{\eta}))$.
For $h_{0}(r)$, the sum rule can be written as
\begin{equation}
\int_{0}^{\infty}\mathrm{d}rh_{0}(r)r=-2.
\end{equation}
It is notable that the sum rule is independent of the density. It
is thus reasonable to expect that $h_{0}(r)$ only weakly depends
on the density (or the filling fraction) since it has similar asymptotic
behavior for different densities while constrained by a sum rule independent
of the density.

The binding energy of a composite fermion can then be written as
\begin{align}
\epsilon_{\mathrm{b}}^{\bigstar}(\bm{z};\bm{\eta}) & \approx\frac{e^{2}l_{B}(\bm{z})}{8\pi\varepsilon}\rho_{\mathrm{e}}(\bm{z})u\left(\frac{|z-\eta|}{l_{B}(\bm{z})}\right),\label{eq:eb}\\
u(|x|) & =\int\mathrm{d}\bm{y}\frac{h_{0}(|\bm{y}|)}{|\bm{x+\bm{y}}|}.\label{eq:ur}
\end{align}
We see that the binding energy is proportional to the local electron
density $\rho_{\mathrm{e}}(\bm{z})\approx\rho_{\mathrm{v}}(\bm{z})\equiv1/4\pi l_{b}^{2}(\bm{z})$.
It gives rise to the density-of-sates correction of the kinetic energy
discussed in Sec.~\ref{subsec:Wave-equation}. 
\end{comment}

In the long-wavelength limit, we can approximate $\Delta\rho_{\mathrm{e}}(\bm{z}_{1};\bm{\eta})$
as $h(\bm{z}_{1};\bm{\eta})\approx\rho_{\mathrm{e}}(\bm{\eta})h_{0}(|\bm{z}_{1}-\bm{\eta}|/l_{B}(\bm{\eta}))$,
where $h_{0}(r)\equiv\Delta\rho_{\mathrm{e}}(r)/\rho_{0}$ is the
electron-vortex correlation function of a homogeneous system, with
$\Delta\rho_{\mathrm{e}}(r)$ being the change of electron density
relative to the average density $\rho_{0}$ in the vicinity of a vortex
at the origin of space. We expect that the binding energy, apart from
the density-of-states correction factor, should depend only weakly
on the density since $h_{0}(r)$ is constrained by the density-independent
sum rule $\int_{0}^{\infty}\mathrm{d}rh_{0}(r)r=-2$ and the overall
form. We thus estimate the binding energy using the Laughlin state
at $\nu=1/3$, for which we can complete the integrals with respect
to $\{\bm{\eta}_{i}\}$ in Eq.~(\ref{eq:Psiv}), and obtain
\begin{equation}
\Psi_{\eta=0}^{\mathrm{v}}(\{z_{i}\})=\prod_{i}z_{i}^{2}\prod_{i<j}(z_{i}-z_{j})^{3}.
\end{equation}
The density profile of the electrons near the origin can be determined
numerically using the Monte-Carlo method. The result is shown in Fig.~\ref{fig:Electron-vortex-binding-potentia}.
We find that the binding energy can be well fitted by a quadratic
function for $|z-\eta|\lesssim2l_{B}$~\footnote{Hao Jin, private communication.},
and approximated as
\begin{equation}
\epsilon_{\mathrm{b}}(\bm{z};\bm{\eta})\approx-g_{0}\frac{e^{2}l_{B}(\bm{z})}{\varepsilon}\rho_{\mathrm{e}}(\bm{z})+\frac{\hbar^{2}}{2m^{\ast}}\frac{\left|\bm{z}-\bm{\eta}\right|^{2}}{l_{B}^{2}(\bm{z})l_{b}^{2}(\bm{z})},\label{eq:ebapprox}
\end{equation}
with $g_{0}\approx0.5$ and 
\begin{align}
\frac{\hbar^{2}}{m^{\ast}} & \approx0.08\frac{e^{2}l_{B}(\bm{z})}{4\pi\varepsilon}.
\end{align}
The estimated effective mass is about four times larger than the one
usually assumed in the literature ($\hbar^{2}/m^{\ast}\approx0.3e^{2}l_{B}/4\pi\varepsilon$)~\citep{halperin1993,hu2019}.
On the other hand, the effective masses determined in experiments
vary with the measurement methods~\citep{jain2007}. Our estimate
is actually close to the cyclotron effective mass measured by Kukushkin
et al.~\citep{kukushkin2002}. There is also a theoretical proposal
that the effective mass should be four times larger~\citep{predin2023a}.

We can define an exchange-correlation functional
\begin{equation}
E_{\mathrm{xc}}[\rho_{\mathrm{e}}]=-g_{0}\frac{e^{2}}{\varepsilon}\int\mathrm{d}\bm{z}l_{B}(\bm{z})\rho_{\mathrm{e}}^{2}(\bm{z})+\dots~,
\end{equation}
which includes the contribution of the first term of Eq.~(\ref{eq:ebapprox}),
as well as contributions that are ignored in our derivation, in particular
the effect of particle exchanges. In the spirit of the Kohn-Sham approach
of the density functional theory, we could define $E_{\mathrm{xc}}[\rho_{\mathrm{e}}]$
as the difference between the exact ground state energy of a system
with a uniform density $\rho_{\mathrm{e}}$ and the total kinetic
energy of non-interacting composite fermions at the same density~\citep{zhang2015,hu2019}.

Combining all, we obtain the Lagrangian Eq.~(\ref{eq:Lagrangian}).

\section{Generalization for flat Chern bands \label{sec:Generalization-for-flat}}

The fractional quantum Hall effect is also predicted to emerge in
flat Chern bands, i.e., Bloch bands which are nearly dispersion-less
and have non-zero Chern numbers~\citep{parameswaran2013,bergholtz2013}.
A flat Chern band is considered as a generalized ``Landau level''
which possesses essential properties for hosting the fractional quantum
Hall effect. Conversely, a Landau level could be interpreted as an
ideal flat Chern band with a Chern number $|C|=1$~\citep{zhang2016a}.
One expects that interacting electrons confined in a flat Chern band
behave similarly as in an ordinary Landau level. The expectation is
recently confirmed in experiments~\citep{cai2023,zeng2023,park2023,xu2023}.

\begin{figure}
\includegraphics{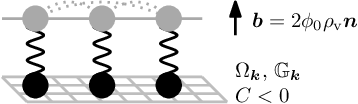}

\caption{\label{fig:CFdipolemodel-FCI}Dipole model of composite fermions for
a flat Chern band. Compared to the model presented in Fig.~\ref{fig:CFdipolemodel_FQH}
for a Landau level, the electron is now confined in a Bloch band characterized
by a Chern number $C$ and other parameters such as the Berry curvature
$\Omega_{\bm{k}}$ and the quantum metric $\mathbb{G}_{\bm{k}}$.
A Landau level can actually be interpreted as an ideal flat Chern
band with $C=-1$, a constant Berry curvature, and vanishing $\mathrm{Tr}\mathbb{G}_{\bm{k}}-|\Omega_{\bm{k}}|$.
The Landau level can be continuously evolved to a flat Chern band
with the same Chern number. One expects that the continuous evolution
should not induce a topological phase transition to the state of vortices.
The possibility that the vortices adopt other topological collective
states, in particular for flat Chern bands with $|C|\protect\ne1$,
is not yet considered in this work.}
\end{figure}

The generalization of our approach for flat Chern bands with $|C|=1$
is straightforward. A dipole model is shown in Fig.~\ref{fig:CFdipolemodel-FCI},
where we replace the electron Landau level in Fig.~\ref{fig:CFdipolemodel_FQH}
with a flat Chern band. The general idea presented in Sec.~\ref{subsec:Wave-function-ansatz-1}
for constructing many-body wave-functions of electrons is still applicable.
We introduce vortices as auxiliary degrees of freedom which should
be projected out in the end, and require that electrons always reside
in their original and physical Hilbert space. We thus have the wave
function ansatz for flat bands with $C=-1$~\footnote{For $C>0$, the topological flat band could be continuously connected
to Landau level(s) in a magnetic field with $B<0$. The definitions
of the complex coordinates $\eta$ and $\bar{\eta}$ should be exchanged. }:
\begin{equation}
\Psi\left(\left\{ \bm{r}_{i}\right\} \right)=\int\prod_{i}\mathrm{d}\mu_{b}(\bm{\eta}_{i})J\left(\left\{ \eta_{i}\right\} \right)\Psi_{\mathrm{CF}}\left(\left\{ \bm{r}_{i},\bar{\eta}_{i}\right\} \right),\label{eq:wfansatzdp-1}
\end{equation}
where $\{\bm{r}_{i}\}$ denotes the set of electron coordinates. For
a flat Chern band, unlike a Landau level, the wave functions $\Psi$
and $\Psi_{\mathrm{CF}}$ are generally not holomorphic in the electron
coordinates. Instead, they should be expanded in the Bloch states
of the flat band which span the physical Hilbert space. Thus, the
single-body wave function of a composite fermion can be written as
\begin{equation}
\psi(\bm{r},\bar{\eta})=\sum_{\bm{k}\in\mathrm{BZ}}\varphi_{\bm{k}}(\bar{\eta})e^{\mathrm{i}\bm{k}\cdot\bm{r}}u_{\bm{k}}(\bm{r}),\label{eq:psifcb}
\end{equation}
where $u_{\bm{k}}(\bm{r})$ denotes the periodic part of the Bloch
wave function at the quasi-wave-vector $\bm{k}$ of the flat band,
and the state of the composite fermion is represented by the wave
function $\varphi_{\bm{k}}(\bar{\eta})$.

We can introduce an effective Hamiltonian for composite fermions.
In the enlarged Hilbert space of composite fermions, each electron
in the flat band is bound to a vortex. While the binding potential
could be derived microscopically as we have demonstrated for a Landau
level in Sec.~\ref{subsec:Energy}, it is reasonable to assume that
the harmonic form Eq.~(\ref{eq:Vbound}) is a good first approximation.
Therefore, the effective Hamiltonian of a composite fermion can be
written as
\begin{equation}
\hat{H}_{\mathrm{CF}}=\hat{T}_{\mathrm{e}}+\frac{\hbar^{2}}{2m^{\ast}l_{B}^{2}l_{b}^{2}}\left|\bm{r}-\hat{\bm{\eta}}\right|^{2},
\end{equation}
where $\hat{T}_{\mathrm{e}}$ is the electron kinetic energy, and
we define the effective magnetic length
\begin{equation}
l_{B}^{2}\equiv\frac{V_{\mathrm{p}}}{2\pi},
\end{equation}
with $V_{\mathrm{p}}$ being the area of the primitive cell of the
system.

We can obtain the effective Hamiltonian for $\varphi_{\bm{k}}(\bar{\eta})$
by determining the expectation value $\braket{\psi|\hat{H}_{\mathrm{CF}}|\psi}$
for $\psi$ defined by Eq.~(\ref{eq:psifcb}). It is easy to prove
the identities:
\begin{align}
\Braket{\psi|\bm{r}|\psi} & =\sum_{\bm{k}}\varphi_{\bm{k}}^{\ast}(\bar{\eta})\left(\mathrm{i}\partial_{\bm{k}}+A_{\bm{k}}\right)\varphi_{\bm{k}}(\bar{\eta}),\\
\Braket{\psi|r^{2}|\psi} & =\sum_{\bm{k}}\varphi_{\bm{k}}^{\ast}(\bar{\eta})\left(\left|\mathrm{i}\partial_{\bm{k}}+A_{\bm{k}}\right|^{2}+\mathrm{Tr}\mathbb{G}_{\bm{k}}\right)\varphi_{\bm{k}}(\bar{\eta}),
\end{align}
where $\bm{A}_{\bm{k}}$ and $\mathbb{G}_{\bm{k}}$ are the Berry
connection and quantum metric tensor of the flat band, respectively,
defined by~\citep{roy2014}
\begin{align}
\bm{A}_{\bm{k}} & =\mathrm{i}\Braket{u_{\bm{k}}|\partial_{\bm{k}}u_{\bm{k}}},\\
\mathbb{G}_{\bm{k}}^{ab} & =\mathrm{Re}\Braket{\partial_{k_{a}}u_{\bm{k}}|\partial_{k_{b}}u_{\bm{k}}}-A_{\bm{k}}^{a}A_{\bm{k}}^{b}.
\end{align}
Applying the identities, we obtain:
\begin{equation}
\Braket{\psi|\hat{H}_{\mathrm{CF}}|\psi}=\sum_{\bm{k}}\varphi_{\bm{k}}^{\ast}(\bar{\eta})\hat{H}\varphi_{\bm{k}}(\bar{\eta}),
\end{equation}
and
\begin{multline}
\hat{H}=\epsilon_{\bm{k}}+\frac{\hbar^{2}}{2m^{\ast}l_{B}^{2}l_{b}^{2}}\left(\mathrm{Tr}\mathbb{G}_{\bm{k}}+\Omega_{\bm{k}}\right)\\
+\frac{\hbar^{2}}{2m^{\ast}l_{B}^{2}l_{b}^{2}}\left(2\mathrm{i}\partial_{k}+\bar{A}_{\bm{k}}-\bar{\eta}\right)\left(2\mathrm{i}\partial_{\bar{k}}+A_{\bm{k}}-\hat{\eta}\right),\label{eq:Hfcb}
\end{multline}
where $\epsilon_{\bm{k}}$ and $\Omega_{\bm{k}}$ are the dispersion
and Berry curvature of the flat band, respectively. The form of the
$\hat{\eta}$ operator depends on the spatial profile of the vortex
density. As a first approximation, one could assume a homogeneous
vortex density, thus $\hat{\eta}=2l_{b}^{2}\partial_{\bar{\eta}}$.

We could predict the stability of a fractional Chern insulator by
determining the eigen-spectrum of the single-body effective Hamiltonian
Eq.~(\ref{eq:Hfcb}). For an ideal flat band with a uniform Berry
curvature, only the last term remains, and it is easy to show that
the Hamiltonian gives rise to the ordinary $\Lambda$-levels (see
Appendix \ref{appendix:-levels-of-the}). For the more general cases,
however, we expect that the first two terms, which could be interpreted
as the renormalized band dispersion experienced by composite fermions,
will make $\Lambda$-levels non-degenerate and suppress excitation
gaps. When the gaps are closed, the fractional Chern insulator state
will be destroyed. The application of the effective Hamiltonian to
real materials is left for future investigation.

The form of the effective Hamiltonian seems to justify the heuristic
trace condition which requires $\mathrm{Tr}\mathbb{G}_{\bm{k}}-|\Omega_{\bm{k}}|\approx0$
everywhere in the Brillouin zone for the emergence of a fractional
Chern insulator~\citep{roy2014,wang2021a}. We see that the second
term, which is proportional to $\mathrm{Tr}\mathbb{G}_{\bm{k}}+\Omega_{\bm{k}}=\mathrm{Tr}\mathbb{G}_{\bm{k}}-|\Omega_{\bm{k}}|$
for $\Omega_{\bm{k}}<0$~\footnote{For $C>0$ and $\Omega_{\bm{k}}>0$, the definitions of $\eta$ and
$\bar{\eta}$, $k$ and $\bar{k}$ should be exchanged in Eq.~(\ref{eq:Hfcb}).
The second term of $\hat{H}$ will be proportional to $\mathrm{Tr}\mathbb{G}_{\bm{k}}-\Omega_{\bm{k}}=\mathrm{Tr}\mathbb{G}_{\bm{k}}-|\Omega_{\bm{k}}|$
for $\Omega_{\bm{k}}>0$.}, renormalizes the dispersion $\epsilon_{\bm{k}}$ of electrons. As
the renormalization tends to make a flat electron band non-flat, it
destabilizes a fractional Chern insulator. On the other hand, it could
be possible to engineer the correction to compensate the electron
dispersion of a non-flat electron band and make it flatter after the
renormalization. The latter suggests a novel possibility that fractional
Chern insulators could be stabilized in non-flat Chern bands.

\section{Summary and discussion \label{sec:Summary-and-discussion}}

In summary, we present a reformulation of the theory of composite
fermions based on the dipole model. Some new insights emerge.
\begin{enumerate}[label=\Alph{enumi})]
\item The states of composite fermions can be determined by solving a wave
equation, with an effective Hamiltonian that can be derived from first
principles. Such a deductive approach can reproduce the well-established
results of the standard theory of composite fermions, namely the wave
functions of the ideal fractional quantum Hall states in the lowest
Landau level. It may also provide an alternative to intuition and
educated guesses for understanding more complex states such as those
observed in higher Landau levels~\citep{baer2014}.
\item A wave-function ansatz $\Psi=\hat{P}_{\mathrm{v}}\Psi_{\mathrm{CF}}$,
or equivalently Jain's wave function ansatz in an alternative wave-function
representation of composite fermions, can be naturally inferred from
the dipole model. The Bijl-Jastrow factor in Jain's ansatz can be
interpreted the complex conjugate of the wave function of the collective
state of vortices, rather than the numerator of the singular CS transformation.
\item The effective theory specified by Eqs.~(\ref{eq:TDWEQ1}, \ref{eq:TDWEQ2})
differs from the HLR theory due to the drift-velocity corrections
in the effective Hamiltonian Eq.~(\ref{eq:Hdipole}). 
\item The wave function ansatz and the effective theory can be unified on
the common basis of the dipole model, and logically connected.
\item The Hilbert space of composite fermions has a simple structure, i.e.,
the tensor product of two separate Hilbert spaces for the physical
and fictitious degrees of freedom, respectively. The simple structure
makes it much easier and less prone to arbitrariness to generalize
the composite fermion theory, e.g., for the flat Chern bands.
\end{enumerate}
\begin{acknowledgments}
I acknowledges Hao Jin for the assistance of determining the binding
energy of composite fermion in $\nu=1/3$, Di Xiao for bringing to
my attention the recent developments of fractional Chern insulators
and sharing the manuscript of Ref.~\onlinecite{wang2024}, and Yinhan
Zhang, Guangyue Ji, Bo Yang, Yue Yu and Xi Lin for useful discussions.
I thank Osamu Sugino and Ryosuke Akashi for their hospitality during
the HISML workshop, during which the ideas for Sec.~\ref{sec:Generalization-for-flat}
were developed. The work is supported by the National Key R\&D Program
of China under Grand Nos.~ 2021YFA1401900 and 2018YFA0305603, and
the National Science Foundation of China under Grant No.~12174005. 
\end{acknowledgments}

\appendix

\section{$\bm{\Lambda}$-levels of the fractional quantum Hall states \label{appendix:-levels-of-the}}

The wave equation (\ref{eq:cfweq1}) is the same as that for an ordinary
charge particle in an effective magnetic field $\bm{\mathcal{B}}$
except for an unimportant constant. Therefore, the wave functions
of $\Lambda$-levels are just those for ordinary Landau levels, which
can be written as~\citep{jain2007}
\begin{equation}
\varphi_{n,m}(\bm{\xi})\propto\frac{e^{-|\xi|^{2}/4l^{2}}}{\sqrt{2\pi}l}f_{n,m}(\bm{\xi}),
\end{equation}
with
\begin{multline}
f_{n,m}(\bm{\xi})=c_{nm}l^{2n+m}e^{|\xi|^{2}/2l^{2}}\\
\times\begin{cases}
\partial_{\xi}^{n}\partial_{\bar{\xi}}^{m+n}e^{-|\xi|^{2}/2l^{2}} & \nu<\frac{1}{2}\\
\partial_{\bar{\xi}}^{n}\partial_{\xi}^{m+n}e^{-|\xi|^{2}/2l^{2}} & \nu>\frac{1}{2}
\end{cases}\label{eq:CFwf1}
\end{multline}
and $c_{nm}\equiv\sqrt{2^{2n+m}/n!(m+n)!}$, $m\ge-n$. $\psi_{mn}(z,\bar{\eta})$
is related to $\varphi_{n,m}(z,\bar{\eta})\equiv\varphi_{n,m}(\bm{\xi})|_{\xi\rightarrow z,\bar{\xi}\rightarrow\bar{\eta}}$
by Eq.~(\ref{eq:psivarphi}), and normalized by Eq.~(\ref{eq:normalization}).
We have

\begin{equation}
\psi_{n,m}(z,\bar{\eta})=f_{n,m}(z,\bar{\eta})\begin{cases}
\frac{l_{B}}{l}\left(\frac{l_{B}}{l_{b}}\right)^{n}e^{z\bar{\eta}/2l_{b}^{2}} & \nu<\frac{1}{2}\\
\frac{l_{b}}{l}\left(\frac{l_{b}}{l_{B}}\right)^{n}e^{z\bar{\eta}/2l_{B}^{2}} & \nu>\frac{1}{2}
\end{cases},\label{eq:CFwf}
\end{equation}
The corresponding eigen-energies are
\begin{equation}
\epsilon_{n,m}=\hbar\omega_{c}^{\ast}\begin{cases}
n & \nu<1/2\\
n+1 & \nu>1/2
\end{cases},
\end{equation}
with $\omega_{c}^{\ast}\equiv e|\mathcal{B}|/m^{\ast}$.

For the special case $\nu=1/2$, $B=b$, we have $\mathcal{B}=0$.
The wave function is plane-wave like:
\begin{equation}
\psi_{\bm{k}}(z,\bar{\eta})=\frac{l_{B}}{\sqrt{2\pi}}e^{\mathrm{i}\frac{\bar{k}z+k\bar{\eta}}{2}+\frac{z\bar{\eta}}{2l_{B}^{2}}-\frac{\left|k\right|^{2}l_{B}^{2}}{4}},\label{eq:psik}
\end{equation}
where $\bm{k}\equiv(k_{x},k_{y})$ denotes the wave-vector of the
state, and $k\equiv k_{x}+\mathrm{i}k_{y}$, $\bar{k}=k^{\ast}$.
The wave function is normalized by $\int\mathrm{d}\mu_{B}(\bm{z})\mathrm{d}\mu_{b}(\bm{\eta})\psi_{\bm{k}}^{\ast}(z,\bar{\eta})\psi_{\bm{k}^{\prime}}(z,\bar{\eta})=\delta(\bm{k}-\bm{k}^{\prime})$.

It is easy to show that the wave function does describe a bound state
of an electron and a vortex. Its spatial distribution can be written
as
\begin{multline}
\left|\psi_{n,m}\left(z,\bar{\eta}\right)\right|^{2}e^{-\left|z\right|^{2}/2l_{B}^{2}-\left|\eta\right|^{2}/2l_{b}^{2}}\\
\propto\begin{cases}
e^{-\left|z\right|^{2}/2l^{2}-\left|z-\eta\right|^{2}/2l_{b}^{2}} & \nu<1/2\\
e^{-\left|\eta\right|^{2}/2l^{2}-\left|z-\eta\right|^{2}/2l_{B}^{2}} & \nu>1/2
\end{cases}.
\end{multline}
We see that the electron and the vortex are bound by a Gaussian factor
with a length scale $l_{b}$ ($l_{B}$) for $\nu<1/2$ ($\nu>1/2$).

We can also solve Eq.~(\ref{eq:cfweqstd}), and obtain wave functions
in the standard representation:
\begin{equation}
\tilde{\varphi}_{n,m}\left(\bm{\xi}\right)=\frac{e^{-\frac{\left|\xi\right|^{2}}{4l^{2}}}}{\sqrt{2\pi}l}f_{n,m}\left(\bm{\xi}\right)\begin{cases}
\left(\frac{l_{b}}{l_{B}}\right)^{n} & \nu<\frac{1}{2}\\
\left(\frac{l_{B}}{l_{b}}\right)^{n+1} & \nu>\frac{1}{2}
\end{cases}.\label{eq:varphiprime}
\end{equation}
where the normalization constants are fixed using Eq.~(\ref{eq:orthonormal-2}).
$\tilde{\psi}_{n,m}(\bm{\xi})$ is related to $\tilde{\varphi}_{n,m}$
via Eq.~(\ref{eq:psitildevarphitilde}). It is straightforward to
verify that $\psi_{n,m}(z,\bar{\eta})$ and $\tilde{\psi}_{n,m}(\bm{\xi})$
are related by Eq.~(\ref{eq:transformpsiCF}).

An alternative way of solving the wave equation Eq.~(\ref{eq:cfweq})
is to define a set of ladder operators~\citep{jain2007}. For the
filing factor $\nu<1/2$, the ladder operators are
\begin{align}
\hat{a} & =\frac{1}{\sqrt{2}}\frac{l}{l_{B}l_{b}}\left(z-\hat{\eta}\right)\label{eq:a}\\
\hat{a}^{\dagger} & =\frac{1}{\sqrt{2}}\frac{l}{l_{B}l_{b}}\left(\hat{\bar{z}}-\bar{\eta}\right),\\
\hat{b} & =\frac{l}{\sqrt{2}}\left(\frac{\hat{\bar{z}}}{l_{B}^{2}}-\frac{\bar{\eta}}{l_{b}^{2}}\right),\\
\hat{b}^{\dagger} & =\frac{l}{\sqrt{2}}\left(\frac{z}{l_{B}^{2}}-\frac{\hat{\eta}}{l_{b}^{2}}\right).\label{eq:b+}
\end{align}
It is easy to verify the commutation relations $[\hat{a},\hat{a}^{\dagger}]=1$,
$[\hat{b},\hat{b}^{\dagger}]=1$, and $[\hat{a},\hat{b}]=0$. For
$\nu>1/2$, the ladder operators can be obtained by exchanging $l_{b}\leftrightarrow l_{B}$
and $z\leftrightarrow\bar{\eta}$ in the definitions.

\section{Quantization in Bergman spaces \label{sec:quantization}}

In this subsection, we discuss quantization in weighted Bergman spaces.
Two alternative forms of the quantization, corresponding to the normal
ordering and the anti-normal ordering of operators, respectively,
can be defined and related to each other.

\subsection{Quantization: normal ordering \label{subsec:Quantization-in-the}}

We can quantize an arbitrary function $H(\bar{z},z)$ (e.g., the energy
of a composite fermion) to an operator $\hat{H}$. In analogy to Eq.~(\ref{eq:Hpsi}),
the action of $\hat{H}$ on a wave function $\psi(z)$ can be defined
as: 
\begin{equation}
\left[\hat{H}\psi\right]\left(z\right)=\int\mathrm{d}\mu_{B}(\bm{\xi})K_{B}(z,\bar{\xi})H(\bar{\xi},\xi)\psi(\xi),\label{eq:Hhatpsi}
\end{equation}
where we use the reproducing kernel to project the non-holomorphic
function into the Bergman space. As we see in Sec.~\ref{subsec:Wave-equation},
varying the Lagrangian with respect to a wave function defined in
a Bergman space gives rise to a Hamiltonian operator of the form.

The operator can in general be written as
\begin{equation}
\hat{H}=N_{+}\left[H\left(\hat{\bar{z}},z\right)\right],
\end{equation}
where $\hat{\bar{z}}$ is defined in Eq.~(\ref{eq:zhatdef}). It
is sufficient to show the relation for $H(\bm{\xi})=\bar{\xi}^{m}\xi^{n}$,
$m,n\in Z$. Substituting the function into Eq.~(\ref{eq:Hhatpsi}),
we can combine $\xi^{n}$ with $\psi(\xi)$ (i.e., put it on the right),
and show that $\bar{\xi}^{m}$ is mapped to $\hat{\bar{z}}^{m}$.
For the latter, we have
\begin{align}
\hat{\bar{z}}^{2}\psi(z)\equiv & \int\mathrm{d}\mu_{B}(\bm{\xi}_{1})K_{B}(z,\bar{\xi}_{1})\bar{\xi}_{1}\nonumber \\
 & \times\int\mathrm{d}\mu_{B}(\bm{\xi})K_{B}(\xi_{1},\bar{\xi})\bar{\xi}\psi(\xi)\\
= & \int\mathrm{d}\mu_{B}(\bm{\xi})K_{B}(z,\bar{\xi})\bar{\xi}^{2}\psi(\xi),
\end{align}
where we apply the complex conjugate of Eq.~(\ref{eq:reprodkern})
when completing the integral with respect to $\bm{\xi}_{1}$.

We thus only need to determine the quantization of $\bar{z}$. In
general, the operator $\hat{\bar{z}}$ can be written as a function
of $\partial_{z}$ and $z$ satisfying
\begin{equation}
\hat{\bar{z}}\left(\partial_{z},z\right)K_{B}\left(z,\bar{\xi}\right)=\bar{\xi}K_{B}\left(z,\bar{\xi}\right).\label{eq:zhat}
\end{equation}
For the case of a uniform magnetic field with the reproducing kernel
Eq.~(\ref{eq:KB0}), it is easy to see that $\hat{\bar{z}}=2l_{B}^{2}\partial_{z}$
does satisfy the equation. For a general space, we substitute Eq.~(\ref{eq:KBFB})
into Eq.~(\ref{eq:zhat}), and obtain
\begin{equation}
\hat{\bar{z}}=2l_{B}^{2}\left(\partial_{z}+\frac{\mathrm{i}e}{\hbar}N_{-}\left[\bar{A}_{1}^{\bigstar}\left(\hat{\bar{z}},z\right)\right]\right),\label{eq:zbarhat}
\end{equation}
where we introduce the starred vector potential $\bar{A}^{\bigstar}$
defined by
\begin{equation}
\partial_{z}F_{B}(\bar{\xi},z)\equiv-\mathrm{i}\frac{e}{\hbar}\bar{A}^{\bigstar}(\bar{\xi},z),\label{eq:Abarstar}
\end{equation}
and decompose as $\bar{A}^{\bigstar}(\bar{\xi},z)=\mathrm{i}B_{0}\bar{\xi}/2+\bar{A}_{1}^{\bigstar}(\bar{\xi},z)$,
and $N_{-}[\dots]$ denotes the anti-normal ordering that puts $\hat{\bar{z}}$'s
to the right of all $z$'s.

The starred vector potential $\bar{A}^{\bigstar}$ can be related
to the physical vector potential $\bar{A}$, as we will show in the
next subsection. Consequently, the quantization can also be written
as
\begin{equation}
\hat{\bar{z}}=2l_{B}^{2}\left(\partial_{z}+\frac{\mathrm{i}e}{\hbar}N_{+}\left[\bar{A}_{1}\left(\hat{\bar{z}},z\right)\right]\right).\label{eq:zbarhat-1}
\end{equation}
These equations can be solved iteratively.

The vortex degree of freedom can be quantized similarly. We express
the reproducing kernel as $K_{b}(\bar{\eta},\eta^{\prime})\equiv\exp[F_{b}(\bar{\eta},\eta^{\prime})]$,
and define the corresponding starred vector potential with $\partial_{\bar{\eta}}F_{b}(\bar{\eta},\eta^{\prime})=-\mathrm{i}ea^{\bigstar}(\bar{\eta},\eta^{\prime})/\hbar$.
Making substitutions $z\rightarrow\bar{\eta}$, $\hat{\bar{z}}\rightarrow\hat{\eta}$,
$B\rightarrow b$ and $\bar{A}^{\bigstar}\rightarrow a^{\bigstar}$,
we have
\begin{align}
\hat{\eta} & =2l_{b}^{2}\left(\partial_{\bar{\eta}}+\frac{\mathrm{i}e}{\hbar}N_{-}\left[a_{1}^{\bigstar}\left(\bar{\eta},\hat{\eta}\right)\right]\right)\label{eq:etahat1}\\
 & =2l_{b}^{2}\left(\partial_{\bar{\eta}}+\frac{\mathrm{i}e}{\hbar}N_{+}\left[a_{1}\left(\bar{\eta},\hat{\eta}\right)\right]\right).\label{eq:etahat}
\end{align}
The normal (anti-normal) ordering should be re-interpreted accordingly
to put $\hat{\eta}$'s to the left (right) of all $\bar{\eta}$'s.

\subsection{Quantization: anti-normal ordering \label{subsec:Quantization:-anti-normal-orderi}}

Alternatively, we can define
\begin{equation}
\left[\hat{H}\psi\right]\left(z\right)=\int\mathrm{d}\mu_{B}(\bm{\xi})K_{B}\left(z,\bar{\xi}\right)H^{\bigstar}\left(z,\bar{\xi}\right)\psi\left(\xi\right),\label{eq:Hpsi1}
\end{equation}
where $H^{\bigstar}$ is defined by the transformation
\begin{multline}
H^{\bigstar}\left(z,\bar{\xi}\right)=\frac{1}{K_{B}\left(z,\bar{\xi}\right)}\\
\times\int\mathrm{d}\mu_{B}\left(\bm{\zeta}\right)K_{B}\left(z,\bar{\zeta}\right)H\left(\bar{\zeta},\zeta\right)K_{B}\left(\zeta,\bar{\xi}\right).\label{eq:Hstar}
\end{multline}
It is easy to verify that Eq.~(\ref{eq:Hpsi1}) reduces to Eq.~(\ref{eq:Hhatpsi})
after substituting Eq.~(\ref{eq:Hstar}).

It is not difficult to see that the operator can be written in the
form of the anti-normal ordering
\begin{equation}
\hat{H}=N_{-}\left[H^{\bigstar}(z,\hat{\bar{z}})\right].
\end{equation}
We thus have two alternative forms of the $\hat{H}$ operator in the
normal ordering and the anti-normal ordering, respectively, which
are related by the transformation Eq.~(\ref{eq:Hstar}).

The vector potential $\bar{A}$ and its starred counterpart $\bar{A}^{\bigstar}$
are also related by the transformation. To see that, we note that
$\partial_{z}\psi(z)$ can be written in two alternative forms:
\begin{align}
\partial_{z}\psi(z) & =\int\mathrm{d}\mu_{B}\left(\xi\right)K_{B}\left(z,\bar{\xi}\right)\partial_{\xi}\psi\left(\xi\right)\label{eq:ppsi1}\\
 & =\int\mathrm{d}\mu_{B}\left(\xi\right)\left[\partial_{z}K_{B}\left(z,\bar{\xi}\right)\right]\psi\left(\xi\right).
\end{align}
Applying integral by parts to the first form, we have
\begin{multline}
\int\mathrm{d}\mu_{B}\left(\xi\right)K_{B}\left(z,\bar{\xi}\right)\bar{A}\left(\bar{\xi},\xi\right)\psi\left(\xi\right)\\
=\int\mathrm{d}\mu_{B}\left(\xi\right)K_{B}\left(z,\bar{\xi}\right)\bar{A}^{\bigstar}\left(\bar{\xi},z\right)\psi\left(\xi\right).
\end{multline}
The two sides of the equation correspond to the two alternative quantization
forms of the vector potential. They are thus related by Eq.~(\ref{eq:Hstar}).

\section{Long-wavelength limit \label{subsec:Long-wavelength-approximations}}

We can related $f_{B}$ with $F_{B}$ when the non-uniform component
of the magnetic field is small and varies slowly over space. In this
limit, Equation~(\ref{eq:reprodkern}) can be approximated as
\begin{multline}
\psi(z,\bar{\eta})\approx\int\mathrm{d}\mu^{(0)}(\bm{\xi})K_{B}^{(0)}(z,\bar{\xi})\\
\times\left[1-f_{B}^{(1)}(\bm{\xi})+F_{B}^{(1)}(\bar{\xi},z)\right]\psi(\xi,\bar{\eta}),
\end{multline}
where $f_{B}^{(1)}$ and $F_{B}^{(1)}$ denote the corrections to
$f_{B}$ and $F_{B}$ due to the spatial fluctuation of the magnetic
field, respectively. Since an electron is always bound to a vortex
in a composite fermion with a length scale of the magnetic length
(see Appendix~\ref{appendix:-levels-of-the}), which is much smaller
than the wavelength of the fluctuating magnetic field, we expand $f_{B}^{(1)}$
and $F_{B}^{(1)}$ to the linear order of $\bar{\xi}$ around the
vortex coordinate $\bar{\eta}$, and complete the integral. We obtain
\begin{multline}
\psi(z,\bar{\eta})\approx\biggr[1-f_{B}^{(1)}(\bar{\eta},z)+F_{B}^{(1)}(\bar{\eta},z)\\
-2l_{B}^{2}\partial_{z}\partial_{\bar{\eta}}f_{B}^{(1)}(\bar{\eta},z)\biggr]\psi(z,\bar{\eta}).
\end{multline}
We thus have
\begin{equation}
F_{B}(z,\bar{\eta})\approx\left.\left[f_{B}(\bm{z})-\ln\frac{l_{B}^{2}(\bm{z})}{l_{B}^{2}}\right]\right|_{\bar{z}\rightarrow\bar{\eta}}.\label{eq:FBappr}
\end{equation}
Similarly, we can relate $f_{b}$ and $F_{b}$ in the long-wavelength
limit.

Differentiating the relation with respect to $z$, and ignoring the
gradient of the local magnetic length, we have
\begin{equation}
\bar{A}^{\bigstar}(\bar{\eta},z)\approx\left.\bar{A}(\bar{z},z)\right|{}_{\bar{z}\rightarrow\bar{\eta}}.
\end{equation}

We can have the approximate forms of $\hat{\bar{z}}$ and $\hat{\eta}$
in the long-wavelength limit as well. To do that, we apply the expansion
$N_{+}[\bar{A}_{1}(\hat{\bar{z}},z)]\approx\bar{A}_{1}(\bar{\eta},z)+(\hat{\bar{z}}-\bar{\eta})[\partial_{\bar{\eta}}\bar{A}_{1}(\bar{\eta},z)]$
and the similar one for $N_{+}[a_{1}(\bar{\eta},\hat{\eta})]$, substitute
them into Eq.~(\ref{eq:zbarhat-1}) and Eq.~(\ref{eq:etahat}),
respectively. We obtain
\begin{align}
\hat{\bar{z}}-\bar{\eta} & \approx2\left[\partial_{z}+\mathrm{i}\frac{e}{\hbar}\bar{A}\left(\bar{\eta},z\right)\right]l_{B}^{2}(\bar{\eta},z),\label{eq:deltazbar}\\
\hat{\eta}-z & \approx2\left[\partial_{\bar{\eta}}+\mathrm{i}\frac{e}{\hbar}a\left(\bar{\eta},z\right)\right]l_{b}^{2}\left(\bar{\eta},z\right),\label{eq:deltaeta}
\end{align}
with $l_{B}^{2}(z,\bar{\eta})\equiv l_{B}^{2}(\bm{z})|_{\bar{z}\rightarrow\bar{\eta}}$
and $l_{b}^{2}(z,\bar{\eta})\equiv l_{b}^{2}(\bm{z})|_{\bar{z}\rightarrow\bar{\eta}}$.

The scalar potentials can similarly be approximated as
\begin{align}
N_{+}[\Phi_{\mathrm{eff}}] & \approx\Phi_{\mathrm{eff}}(\bar{\eta},z)-V(\bar{\eta},z)\left[\mathrm{i}\hbar\partial_{z}-e\bar{A}(\bar{\eta},z)\right],\label{eq:Phiappr}\\
N_{+}[\phi] & \approx\phi(\bar{\eta},z)-\bar{v}(\bar{\eta},z)\left[-\mathrm{i}\hbar\partial_{\bar{\eta}}-ea(\bar{\eta},z)\right],\label{eq:phiappr}
\end{align}
where $V$ and $\bar{v}$ are the complex components of the drift
velocities defined in the main text.

We can then obtain the approximate form of $\hat{H}_{\psi}$ by applying
these approximations to Eq.~(\ref{eq:Hpsi-1}), and ignoring all
corrections proportional to the gradients of the magnetic and electric
fields as well as contributions beyond the linear order of $B_{1}(\bm{z})$,
$b_{1}(\bm{\eta})$ and the electric fields. Equation (\ref{eq:Hdipole})
can be obtained by applying Eq.~(\ref{eq:Hdef}).

\section{Time-dependent systems \label{subsec:Time-dependent-systems}}

We first note that the stationary state wave equation can also be
obtained from the variational principle $\delta L_{\mathrm{eff}}=0$,
where $L_{\mathrm{eff}}$ is the effective Lagrangian for a composite
fermion: 
\begin{equation}
L_{\mathrm{eff}}\equiv\int\mathrm{d}\mu_{B}(\bm{z})\mathrm{d}\mu_{b}(\bm{\eta})\psi^{\ast}(z,\bar{\eta})(\epsilon-\hat{H}_{\psi})\psi(z,\bar{\eta}),
\end{equation}

For time-dependent systems, the term proportional to $\epsilon$ in
the Lagrangian should be replaced by
\begin{multline}
\int\mathrm{d}\mu_{B}(\bm{z})\mathrm{d}\mu_{b}(\bm{\eta})\psi^{\ast}(z,\bar{\eta};t)\\
\times\left\{ \mathrm{i}\hbar\frac{\partial}{\partial t}-\frac{\mathrm{i}\hbar}{2}\left[\frac{\partial f_{B}(\bm{z},t)}{\partial t}+\frac{\partial f_{b}(\bm{\eta},t)}{\partial t}\right]\right\} \psi(z,\bar{\eta};t),
\end{multline}
where the second term in the braces originates from the exponential
factor in Eq.~(\ref{eq:phicomplete}) and its counterpart for vortices. 

After making the substitution, and applying Eqs.~(\ref{eq:transforCFgeneral}\textendash \ref{eq:cftransform}),
we obtain the Schrödinger action of a composite fermion:
\begin{equation}
S_{\mathrm{CF}}=\int\mathrm{d}t\int\mathrm{d}\bm{\xi}\tilde{\varphi}^{\ast}(\bm{\xi},t)\left(\mathrm{i}\hbar\frac{\partial}{\partial t}-\hat{H}_{1}\right)\varphi(\bm{\xi},t),\label{eq:SCF}
\end{equation}
with
\begin{multline}
\hat{H}_{1}\equiv\hat{H}+\frac{\mathrm{i}\hbar}{2}N_{+}\left[\partial_{t}f_{B}(\hat{z},\xi,t)-\partial_{t}f_{B}(\bm{\xi},t)\right]\\
+\frac{\mathrm{i}\hbar}{2}N_{+}\left[\partial_{t}f_{b}(\bar{\xi},\hat{\eta},t)-\partial_{t}f_{b}(\bm{\xi},t)\right].
\end{multline}

In the long-wavelength limit, we can approximate the correction terms
in the same way as in Eqs.~(\ref{eq:Phiappr}, \ref{eq:phiappr}).
It is not difficult to show that $\hat{H}_{1}$ has the same form
as $\hat{H}$ shown in Eq.~(\ref{eq:Hdipole}) but with the gauge-invariant
definitions of the electric fields.

\section{Operators in the standard representation}

\subsection{Ideal systems \label{subsec:Ideal-systems-operators}}

In ideal systems, wave functions of composite fermions in the dipole
representation and the standard representation are related by Eq.~(\ref{eq:transformpsiCF}).
Transforming to the standard representation, we have
\begin{align}
\bar{z}\psi\left(z,\bar{\eta}\right)\rightarrow & 2l_{B}^{2}\partial_{z}\psi\left(z,\bar{\eta}\right)\nonumber \\
= & \int\mathrm{d}\mu_{B}^{(0)}\left(\bm{\xi}\right)e^{z\bar{\xi}/2l_{B}^{2}+\xi\bar{\eta}/2l_{b}^{2}}\bar{\xi}\tilde{\psi}\left(\bm{\xi}\right),\\
z\psi\left(z,\bar{\eta}\right)= & \int\mathrm{d}\mu_{B}^{(0)}\left(\bm{\xi}\right)\left[2l_{B}^{2}\partial_{\bar{\xi}}e^{z\bar{\xi}/2l_{B}^{2}+\xi\bar{\eta}/2l_{b}^{2}}\right]\tilde{\psi}\left(\bm{\xi}\right)\nonumber \\
= & \int\mathrm{d}\mu_{B}^{(0)}\left(\bm{\xi}\right)e^{z\bar{\xi}/2l_{B}^{2}+\xi\bar{\eta}/2l_{b}^{2}}\nonumber \\
 & \times\left(-2l_{B}^{2}\partial_{\bar{\xi}}+\xi\right)\tilde{\psi}\left(\bm{\xi}\right),
\end{align}
and similar expressions for $\eta$ and $\bar{\eta}$. Therefore,
in the standard representation, the operators should be mapped to:
\begin{align}
z & \rightarrow-2l_{B}^{2}\partial_{\bar{\xi}}+\xi,\\
\bar{z} & \rightarrow\bar{\xi},\\
\eta & \rightarrow\xi,\\
\bar{\eta} & \rightarrow-2l_{b}^{2}\partial_{\xi}+\left(l_{b}^{2}/l_{B}^{2}\right)\bar{\xi}.
\end{align}

It is straightforward to apply the mappings to Eq.~(\ref{eq:Vbound})
and obtain a wave equation for $\tilde{\psi}(\bm{\xi})$. After applying
the transformation Eq.~(\ref{eq:psitildevarphitilde}), we obtain
Eq.~(\ref{eq:cfweqstd}).

\subsection{General systems \label{subsec:General-systems-operators}}

For general systems, wave functions in the dipole representation and
the standard representation are related by Eq.~(\ref{eq:transforCFgeneral}).
For the $\hat{\bar{z}}$ operator defined in Eq.~(\ref{eq:zhat}),
we have 
\begin{equation}
\hat{\bar{z}}\psi\left(z,\bar{\eta}\right)=\int\mathrm{d}\mu_{B}\left(\bm{\xi}\right)K_{B}\left(z,\bar{\xi}\right)K_{b}\left(\bar{\eta},\xi\right)\bar{\xi}\tilde{\psi}\left(\bm{\xi}\right).
\end{equation}

On the other hand, applying complex conjugation to Eq.~(\ref{eq:zhat})
and exchanging $\bm{z}$ and $\bm{\xi}$, we obtain
\begin{equation}
\hat{\bar{z}}^{\ast}\left(\partial_{\xi},\xi\right)K_{B}\left(z,\bar{\xi}\right)=zK_{B}\left(z,\bar{\xi}\right).
\end{equation}
Here, we make use of the relation $[K_{B}(z,\bar{\xi})]^{\ast}=K_{B}(\xi,\bar{z})$.
We thus have
\begin{align}
z\psi\left(z,\bar{\eta}\right)= & \int\mathrm{d}\mu_{B}\left(\xi\right)\left[\hat{\bar{z}}^{\ast}\left(\partial_{\xi},\xi\right)K_{B}\left(z,\bar{\xi}\right)\right]\nonumber \\
 & \times K_{b}\left(\bar{\eta},\xi\right)\tilde{\psi}\left(\bm{\xi}\right)\\
= & \int\mathrm{d}\mu_{B}\left(\xi\right)K_{B}\left(z,\bar{\xi}\right)K_{b}\left(\bar{\eta},\xi\right)\nonumber \\
 & \times\hat{\bar{z}}^{\dagger}\left(\partial_{\xi}-\frac{\mathrm{i}e}{\hbar}\bar{A}\left(\bm{\xi}\right),\xi\right)\tilde{\psi}\left(\bm{\xi}\right)
\end{align}
Here we apply integral by parts and make use of Eq.~(\ref{eq:pzfb}).
Note $(\partial_{\xi})^{\dagger}=-\partial_{\bar{\xi}}$. The mappings
of $\eta$ and $\bar{\eta}$ can be obtained similarly.

All summarized, we have the following mapping rules in the standard
representation:
\begin{align}
\bar{z} & \rightarrow\bar{\xi},\label{eq:mapzbar}\\
\eta & \rightarrow\xi,\\
z & \rightarrow\hat{\bar{z}}^{\dagger}\left(\partial_{\xi}-\frac{\mathrm{i}e}{\hbar}\bar{A}(\bm{\xi}),\xi\right),\\
\bar{\eta} & \rightarrow\hat{\eta}^{\dagger}\left(\partial_{\bar{\xi}}+\frac{\mathrm{i}e}{\hbar}A(\bm{\xi}),\xi\right).\label{eq:mapetabar}
\end{align}

\subsection{Transformations of Hamiltonians \label{subsec:Hamiltonian-in-the}}

We introduce an operator $\hat{K}$ to represent the transformation
Eq.~(\ref{eq:transforCFgeneral}):
\begin{equation}
\psi=\hat{K}\tilde{\psi}.
\end{equation}
 We further define
\begin{equation}
\hat{K}_{1}\equiv\hat{K}e^{f_{B}(\bm{\xi})}.
\end{equation}

Using the operator, the mapping rules Eqs~(\ref{eq:mapzbar}\textendash \ref{eq:mapetabar})
can be written as
\begin{align}
\hat{K}_{1}^{-1}\hat{\bar{z}}\hat{K}_{1} & =\bar{\xi},\\
\hat{K}_{1}^{-1}\hat{\eta}\hat{K}_{1} & =\xi,\\
\hat{K}_{1}^{-1}z\hat{K}_{1} & =\hat{\bar{z}}^{\dagger},\\
\hat{K}_{1}^{-1}\bar{\eta}\hat{K}_{1} & =\hat{\eta}^{\dagger}.
\end{align}

Applying the transformation to $\hat{H}_{\psi}$, we have
\begin{equation}
\hat{K}_{1}^{-1}\hat{H}_{\psi}\hat{K}_{1}=\hat{H}_{\psi}^{\dagger}.\label{eq:KHK}
\end{equation}

The Hamiltonians $\hat{H}$ and $\hat{\tilde{H}}$ that govern the
wave equations for $\varphi$ and $\tilde{\varphi}$. respectively,
can be identified to be
\begin{align}
\hat{H} & \equiv e^{-\frac{f_{B}+f_{b}}{2}}\hat{H}_{\psi}e^{\frac{f_{B}+f_{b}}{2}},\\
\hat{\tilde{H}} & \equiv e^{\frac{f_{B}+f_{b}}{2}}\hat{K}_{1}^{-1}\hat{H}_{\psi}\hat{K}_{1}e^{-\frac{f_{B}+f_{b}}{2}}.
\end{align}

Applying Eq.~(\ref{eq:KHK}), we obtain
\begin{equation}
\hat{\tilde{H}}=\hat{H}^{\dagger}.
\end{equation}

\section{Current densities \label{sec:Current-densities}}

\subsection{Current density in a Landau level}

In a Landau level, the particle density of a state $\psi$ can be
defined as
\begin{equation}
\rho(\bm{z},t)=w(\bm{z},t)\left|\psi(z,t)\right|^{2},
\end{equation}
where $w(\bm{z})$ denotes the weight of the Bergman space. The wave-equation
in the space can be generally written as
\begin{equation}
\mathrm{i}\hbar\frac{\partial\psi}{\partial t}=\int\mathrm{d}\mu(\bm{\xi})K(z,\bar{\xi})H^{\bigstar}(\bar{\xi},z)\psi(\xi,t).
\end{equation}

We can determine the current density by establishing a continuity
equation for $\rho(\bm{z})$. We have
\begin{multline}
\frac{\partial\rho(\bm{z},t)}{\partial t}=\frac{1}{\mathrm{i}\hbar}\int\mathrm{d}\mu(\bm{\xi}_{1})\mathrm{d}\mu(\bm{\xi}_{2})\psi^{\ast}(\xi_{1},t)K(\xi_{1},\bar{\xi}_{2})\\
\times H^{\bigstar}(\bar{\xi}_{2},\xi_{1})\psi(\xi_{2},t)\left[\delta\left(\bm{z}-\bm{\xi}_{1}\right)-\delta\left(\bm{z}-\bm{\xi}_{2}\right)\right].
\end{multline}
We then substitute the expansion 
\begin{multline}
\delta(\bm{z}-\bm{\xi}_{1})-\delta(\bm{z}-\bm{\xi_{2}})=-\frac{1}{2}\sideset{}{^{\prime}}\sum_{n,m=0}^{\infty}\biggr[\frac{1}{m!n!}(\xi_{1}-\xi_{2})^{m}\\
\times(\bar{\xi}_{1}-\bar{\xi}_{2})^{n}\partial_{z}^{m}\partial_{\bar{z}}^{n}\delta(\bm{z}-\bm{\xi}_{1})-\mathrm{(1\leftrightarrow2)}\biggr],
\end{multline}
where the summation excludes $(m,n)=(0,0)$. We obtain the continuity
equation
\begin{equation}
\frac{\partial\rho(\bm{z},t)}{\partial t}+\bm{\nabla}\cdot\bm{j}(\bm{z},t)=0,
\end{equation}
with the current density $j\equiv j_{x}+\mathrm{i}j_{y}$:
\begin{align}
j(\bm{z},t)= & ~j_{0}(\bm{z},t)-2\mathrm{i}\partial_{\bar{z}}m(\bm{z},t),\\
j_{0}(\bm{z},t)= & ~\frac{\mathrm{1}}{\hbar}w(\bm{z},t)\int\mathrm{d}\mu(\bm{\xi})(z-\xi)\nonumber \\
 & \times\mathrm{Im}\left[\psi^{\ast}(z,t)K(z,\bar{\xi})H^{\bigstar}(\bar{\xi},z)\psi(\xi,t)\right],\\
m(\bm{z},t)= & ~\frac{1}{4\hbar}\mathrm{Re}\sum_{m,n=1}^{\infty}\frac{\partial_{z}^{m-1}\partial_{\bar{z}}^{n-1}}{m!n!}w(\bm{z},t)\int\mathrm{d}\mu(\xi)\nonumber \\
 & \times\psi^{\ast}(z,t)K(z,\bar{\xi})H^{\bigstar}(\bar{\xi},z)\psi(\xi,t)\nonumber \\
 & \times(z-\xi)^{m}(\bar{z}-\bar{\xi})^{n}.
\end{align}
where $m(\bm{z},t)$ is the orbital magnetization density.

Applying the quantization rules shown in Appendix \ref{sec:quantization},
we can rewrite the equations in operator forms
\begin{align}
\bm{j}_{0}(\bm{z},t)= & ~w(\bm{z},t)\mathrm{Re}\left\{ \psi^{\ast}(z,t)\left[\hat{\bm{v}}\psi\right](z,t)\right\} ,\\
m(\bm{z},t)= & ~\mathrm{Re}\sum_{m,n=1}^{\infty}\frac{\partial_{z}^{m-1}\partial_{\bar{z}}^{n-1}}{m!n!}w(\bm{z},t)\psi^{\ast}(z,t)\nonumber \\
 & \times\left[\hat{m}_{mn}\psi\right](z,t),
\end{align}
with
\begin{align}
\hat{\bm{v}} & \equiv\frac{1}{\mathrm{i}\hbar}\left[\hat{\bm{z}},\hat{H}\right],\\
\hat{m}_{mn} & \equiv\frac{1}{4\hbar}[\underset{n}{\underbrace{\hat{\bar{z}}\dots,[\hat{\bar{z}}}},[\underset{m}{\underbrace{\hat{z},\dots[\hat{z}}},\hat{H}]\dots]]\dots].
\end{align}

In the long-wavelength limit, we can keep only $j_{0}(\bm{z},t)$,
and ignore the magnetization current.

\subsection{Current densities of a composite fermion system}

The result derived in the last subsection can be applied to composite
fermions with a straightforward generalization. The electron and vortex
current densities for a state $\psi(z,\bar{\eta})$ can be written
as
\begin{align}
\bm{j}_{\mathrm{e}}(\bm{z}) & \approx w_{B}(\bm{z})\mathrm{Re}\int\mathrm{d}\mu_{b}(\bm{\eta})\psi^{\ast}(z,\bar{\eta})\left[\hat{\bm{U}}\psi\right](z,\bar{\eta}),\label{eq:je}\\
\bm{j}_{\mathrm{v}}(\bm{\eta}) & \approx w_{b}(\bm{z})\mathrm{Re}\int\mathrm{d}\mu_{B}(\bm{z})\psi^{\ast}(z,\bar{\eta})\left[\hat{\bm{u}}\psi\right](z,\bar{\eta}),\label{eq:jv-1}
\end{align}
where $\hat{\bm{U}}\equiv[\hat{\bm{z}},\hat{H}_{\psi}]/\mathrm{i}\hbar$
and $\hat{\bm{u}}\equiv[\hat{\bm{\eta}},\hat{H_{\psi}}]/\mathrm{i}\hbar$
are the electron and vortex velocity operators, respectively, $\hat{H}_{\psi}$
is the effective Hamiltonian shown in Eq.~(\ref{eq:Hpsi-1}), and
we ignore the orbital magnetization contribution. 

In the long-wavelength limit, we can apply the approximate commutators
$\left[\hat{\bar{z}},\hat{z}\right]\approx2l_{B}^{2}(\bm{z})$, $\left[\hat{\eta},\hat{\bar{\eta}}\right]\approx2l_{b}^{2}(\bm{z})$.
The velocity operators are then approximated as
\begin{align}
\hat{\bm{U}} & \approx\frac{\hbar}{m^{\ast}}\frac{\bm{n}\times\left(\hat{\bm{z}}-\hat{\bm{\eta}}\right)}{l_{b}^{2}(\bm{z})}+\bm{V},\label{eq:U}\\
\hat{\bm{u}} & \approx\frac{\hbar}{m^{\ast}}\frac{\bm{n}\times\left(\hat{\bm{z}}-\hat{\bm{\eta}}\right)}{l_{B}^{2}(\bm{z})}+\bm{v}.\label{eq:ubar}
\end{align}
Substituting Eq.~(\ref{eq:ubar}) into Eq.~(\ref{eq:jv-1}) and
summing over occupied states of composite fermions, we obtain the
current density of vortices Eq.~(\ref{eq:jv}). The current density
of electrons can be obtained similarly using Eq.~(\ref{eq:U}).

\subsection{Dipole approximation}

We can also obtain approximate expressions for the particle and current
densities of electrons and vortices by differentiating the action
Eq.~(\ref{eq:SCF}) with respect to $\left(\Phi_{\mathrm{eff}},\bm{A}\right)$
and $(\phi,\bm{a})$, respectively. The approximation corresponds
to the multipole expansion discussed in Ref.~\onlinecite{ji2021}.
We have: 
\begin{align}
\rho_{\mathrm{e}}(\bm{\xi},t)\approx & \sum_{i}\tilde{\varphi}_{i}^{\ast}(\bm{\xi},t)\varphi_{i}(\bm{\xi},t)-\partial_{\bar{\xi}}\bar{P}(\bm{\xi},t),\\
\rho_{\mathrm{v}}(\bm{\xi},t)\approx & \sum_{i}\tilde{\varphi}_{i}^{\ast}(\bm{\xi},t)\varphi_{i}(\bm{\xi},t)+\partial_{\xi}P(\bm{\xi},t),
\end{align}
and
\begin{align}
\bar{j}_{\mathrm{e}}(\bm{\xi},t)\approx & \sum_{i}\left[\frac{-2\mathrm{i}\hbar\partial_{\bar{\xi}}+e\mathcal{A}+m^{\ast}v}{m^{\ast}}\tilde{\varphi}_{i}\right]^{\ast}\varphi_{i}+\partial_{t}\bar{P},\\
j_{\mathrm{v}}(\bm{\xi},t)\approx & \sum_{i}\tilde{\varphi}_{i}^{\ast}\left[\frac{-2\mathrm{i}\hbar\partial_{\bar{\xi}}+e\mathcal{A}+m^{\ast}V}{m^{\ast}}\varphi_{i}\right]-\partial_{t}P,
\end{align}
where $P$ and $\bar{P}$ are the complex components of the dipole
density, approximated as 
\begin{align}
P(\bm{\xi},t) & \approx-l_{b}^{2}(\bm{\xi})\sum_{i}\tilde{\varphi}_{i}^{\ast}\left(2\partial_{\bar{\xi}}+\mathrm{i}\frac{e}{\hbar}\mathcal{A}\right)\varphi_{i}.
\end{align}

\subsection{Vanishing dipole density \label{subsec:Vanishing-dipole-density}}

We can show that a system of composite fermions always has a vanishing
dipole density in the long-wavelength limit. To see this, we apply
the self-consistent condition Eq.~(\ref{eq:csb-1}), and find that
the first term of the current density Eq.~(\ref{eq:jv}) becomes
an anomalous Hall current with a half-quantized Hall conductance $\sigma_{xy}^{(\mathrm{v})}=-e^{2}/2h$~\citep{ji2021}.
Compare the current density to the second CS constraint Eq.~(\ref{eq:cse-1}),
we have
\begin{equation}
\bm{P}(\bm{\eta})\equiv w_{b}(\bm{\eta})\sum_{i}\int\mathrm{d}\mu_{B}(\bm{z})(\bm{z}-\bm{\eta})\left|\psi_{i}\left(z,\bar{\eta}\right)\right|^{2}\approx0,\label{eq:dpden}
\end{equation}
where we ignore the slow spatial variation of $1/l_{B}^{2}(\bm{z})$.

The vanishing dipole density suggests that, on average, the coordinate
of an electron, always coincides with the coordinate of the vortex
to which it is bound. The same identity has also been found in the
semi-classical theory~\citep{ji2021}. In Ref.~\onlinecite{read1998},
it was considered that this condition could replace the CS self-consistent
conditions and serves as the basis for a composite fermion theory
without the CS fields.

\bibliography{References,../physicsbooks}

%Control: production of eprint (-1) disabled
\begin{thebibliography}{50}%
\makeatletter
\providecommand \@ifxundefined [1]{%
 \@ifx{#1\undefined}
}%
\providecommand \@ifnum [1]{%
 \ifnum #1\expandafter \@firstoftwo
 \else \expandafter \@secondoftwo
 \fi
}%
\providecommand \@ifx [1]{%
 \ifx #1\expandafter \@firstoftwo
 \else \expandafter \@secondoftwo
 \fi
}%
\providecommand \natexlab [1]{#1}%
\providecommand \enquote  [1]{``#1''}%
\providecommand \bibnamefont  [1]{#1}%
\providecommand \bibfnamefont [1]{#1}%
\providecommand \citenamefont [1]{#1}%
\providecommand \href@noop [0]{\@secondoftwo}%
\providecommand \href [0]{\begingroup \@sanitize@url \@href}%
\providecommand \@href[1]{\@@startlink{#1}\@@href}%
\providecommand \@@href[1]{\endgroup#1\@@endlink}%
\providecommand \@sanitize@url [0]{\catcode `\\12\catcode `\$12\catcode
  `\&12\catcode `\#12\catcode `\^12\catcode `\_12\catcode `\%12\relax}%
\providecommand \@@startlink[1]{}%
\providecommand \@@endlink[0]{}%
\providecommand \url  [0]{\begingroup\@sanitize@url \@url }%
\providecommand \@url [1]{\endgroup\@href {#1}{\urlprefix }}%
\providecommand \urlprefix  [0]{URL }%
\providecommand \Eprint [0]{\href }%
\providecommand \doibase [0]{http://dx.doi.org/}%
\providecommand \selectlanguage [0]{\@gobble}%
\providecommand \bibinfo  [0]{\@secondoftwo}%
\providecommand \bibfield  [0]{\@secondoftwo}%
\providecommand \translation [1]{[#1]}%
\providecommand \BibitemOpen [0]{}%
\providecommand \bibitemStop [0]{}%
\providecommand \bibitemNoStop [0]{.\EOS\space}%
\providecommand \EOS [0]{\spacefactor3000\relax}%
\providecommand \BibitemShut  [1]{\csname bibitem#1\endcsname}%
\let\auto@bib@innerbib\@empty
%</preamble>
\bibitem [{\citenamefont {Tsui}\ \emph {et~al.}(1982)\citenamefont {Tsui},
  \citenamefont {Stormer},\ and\ \citenamefont {Gossard}}]{tsui1982}%
  \BibitemOpen
  \bibfield  {author} {\bibinfo {author} {\bibfnamefont {D.~C.}\ \bibnamefont
  {Tsui}}, \bibinfo {author} {\bibfnamefont {H.~L.}\ \bibnamefont {Stormer}}, \
  and\ \bibinfo {author} {\bibfnamefont {A.~C.}\ \bibnamefont {Gossard}},\
  }\bibfield  {title} {\enquote {\bibinfo {title} {Two-{{Dimensional
  Magnetotransport}} in the {{Extreme Quantum Limit}}},}\ }\href {\doibase
  10.1103/PhysRevLett.48.1559} {\bibfield  {journal} {\bibinfo  {journal}
  {Phys. Rev. Lett.}\ }\textbf {\bibinfo {volume} {48}},\ \bibinfo {pages}
  {1559} (\bibinfo {year} {1982})}\BibitemShut {NoStop}%
\bibitem [{\citenamefont {Jain}(2007)}]{jain2007}%
  \BibitemOpen
  \bibfield  {author} {\bibinfo {author} {\bibfnamefont {Jainendra~K.}\
  \bibnamefont {Jain}},\ }\href@noop {} {\emph {\bibinfo {title} {Composite
  Fermions}}}\ (\bibinfo  {publisher} {Cambridge University Press},\ \bibinfo
  {year} {2007})\BibitemShut {NoStop}%
\bibitem [{\citenamefont {Jain}\ and\ \citenamefont
  {Kamilla}(1997)}]{jain1997}%
  \BibitemOpen
  \bibfield  {author} {\bibinfo {author} {\bibfnamefont {J.~K.}\ \bibnamefont
  {Jain}}\ and\ \bibinfo {author} {\bibfnamefont {R.~K.}\ \bibnamefont
  {Kamilla}},\ }\bibfield  {title} {\enquote {\bibinfo {title} {Composite
  {{Fermions}} in the {{Hilbert Space}} of the {{Lowest Electronic Landau
  Level}}},}\ }\href {\doibase 10.1142/S0217979297001301} {\bibfield  {journal}
  {\bibinfo  {journal} {Int. J. Mod. Phys. B}\ }\textbf {\bibinfo {volume}
  {11}},\ \bibinfo {pages} {2621--2660} (\bibinfo {year} {1997})}\BibitemShut
  {NoStop}%
\bibitem [{\citenamefont {Halperin}\ \emph {et~al.}(1993)\citenamefont
  {Halperin}, \citenamefont {Lee},\ and\ \citenamefont {Read}}]{halperin1993}%
  \BibitemOpen
  \bibfield  {author} {\bibinfo {author} {\bibfnamefont {B.~I.}\ \bibnamefont
  {Halperin}}, \bibinfo {author} {\bibfnamefont {Patrick~A.}\ \bibnamefont
  {Lee}}, \ and\ \bibinfo {author} {\bibfnamefont {Nicholas}\ \bibnamefont
  {Read}},\ }\bibfield  {title} {\enquote {\bibinfo {title} {Theory of the
  half-filled {{Landau}} level},}\ }\href {\doibase 10.1103/PhysRevB.47.7312}
  {\bibfield  {journal} {\bibinfo  {journal} {Phys. Rev. B}\ }\textbf {\bibinfo
  {volume} {47}},\ \bibinfo {pages} {7312} (\bibinfo {year}
  {1993})}\BibitemShut {NoStop}%
\bibitem [{\citenamefont {Heinonen}(1997)}]{heinonen1997}%
  \BibitemOpen
  \bibfield  {author} {\bibinfo {author} {\bibfnamefont {O.}~\bibnamefont
  {Heinonen}},\ }\href@noop {} {\emph {\bibinfo {title} {Composite
  {{Fermions}}: {{A Unified View}} of the {{Quantum Hall Regime}}}}}\ (\bibinfo
   {publisher} {World Scientific},\ \bibinfo {year} {1997})\BibitemShut
  {NoStop}%
\bibitem [{\citenamefont {Lopez}\ and\ \citenamefont
  {Fradkin}(1991)}]{lopez1991}%
  \BibitemOpen
  \bibfield  {author} {\bibinfo {author} {\bibfnamefont {Ana}\ \bibnamefont
  {Lopez}}\ and\ \bibinfo {author} {\bibfnamefont {Eduardo}\ \bibnamefont
  {Fradkin}},\ }\bibfield  {title} {\enquote {\bibinfo {title} {Fractional
  quantum {{Hall}} effect and {{Chern-Simons}} gauge theories},}\ }\href
  {\doibase 10.1103/PhysRevB.44.5246} {\bibfield  {journal} {\bibinfo
  {journal} {Phys. Rev. B}\ }\textbf {\bibinfo {volume} {44}},\ \bibinfo
  {pages} {5246} (\bibinfo {year} {1991})}\BibitemShut {NoStop}%
\bibitem [{\citenamefont {Murthy}\ and\ \citenamefont
  {Shankar}(2003)}]{murthy2003}%
  \BibitemOpen
  \bibfield  {author} {\bibinfo {author} {\bibfnamefont {Ganpathy}\
  \bibnamefont {Murthy}}\ and\ \bibinfo {author} {\bibfnamefont
  {R.}~\bibnamefont {Shankar}},\ }\bibfield  {title} {\enquote {\bibinfo
  {title} {Hamiltonian theories of the fractional quantum {{Hall}} effect},}\
  }\href {\doibase 10.1103/RevModPhys.75.1101} {\bibfield  {journal} {\bibinfo
  {journal} {Rev. Mod. Phys.}\ }\textbf {\bibinfo {volume} {75}},\ \bibinfo
  {pages} {1101} (\bibinfo {year} {2003})}\BibitemShut {NoStop}%
\bibitem [{\citenamefont {Read}(1994)}]{read1994}%
  \BibitemOpen
  \bibfield  {author} {\bibinfo {author} {\bibfnamefont {N.}~\bibnamefont
  {Read}},\ }\bibfield  {title} {\enquote {\bibinfo {title} {Theory of the
  half-filled {{Landau}} level},}\ }\href {\doibase
  10.1088/0268-1242/9/11S/002} {\bibfield  {journal} {\bibinfo  {journal}
  {Semicond. Sci. Technol.}\ }\textbf {\bibinfo {volume} {9}},\ \bibinfo
  {pages} {1859} (\bibinfo {year} {1994})}\BibitemShut {NoStop}%
\bibitem [{\citenamefont {Son}(2015)}]{son2015}%
  \BibitemOpen
  \bibfield  {author} {\bibinfo {author} {\bibfnamefont {Dam~Thanh}\
  \bibnamefont {Son}},\ }\bibfield  {title} {\enquote {\bibinfo {title} {Is the
  {{Composite Fermion}} a {{Dirac Particle}}?}}\ }\href {\doibase
  10.1103/PhysRevX.5.031027} {\bibfield  {journal} {\bibinfo  {journal} {Phys.
  Rev. X}\ }\textbf {\bibinfo {volume} {5}},\ \bibinfo {pages} {031027}
  (\bibinfo {year} {2015})}\BibitemShut {NoStop}%
\bibitem [{\citenamefont {Balram}\ and\ \citenamefont
  {Jain}(2016)}]{balram2016}%
  \BibitemOpen
  \bibfield  {author} {\bibinfo {author} {\bibfnamefont {Ajit~C.}\ \bibnamefont
  {Balram}}\ and\ \bibinfo {author} {\bibfnamefont {J.~K.}\ \bibnamefont
  {Jain}},\ }\bibfield  {title} {\enquote {\bibinfo {title} {Nature of
  composite fermions and the role of particle-hole symmetry: {{A}} microscopic
  account},}\ }\href {\doibase 10.1103/PhysRevB.93.235152} {\bibfield
  {journal} {\bibinfo  {journal} {Phys. Rev. B}\ }\textbf {\bibinfo {volume}
  {93}},\ \bibinfo {pages} {235152} (\bibinfo {year} {2016})}\BibitemShut
  {NoStop}%
\bibitem [{\citenamefont {Rezayi}\ and\ \citenamefont
  {Read}(1994)}]{rezayi1994}%
  \BibitemOpen
  \bibfield  {author} {\bibinfo {author} {\bibfnamefont {E.}~\bibnamefont
  {Rezayi}}\ and\ \bibinfo {author} {\bibfnamefont {N.}~\bibnamefont {Read}},\
  }\bibfield  {title} {\enquote {\bibinfo {title} {Fermi-liquid-like state in a
  half-filled {{Landau}} level},}\ }\href {\doibase 10.1103/PhysRevLett.72.900}
  {\bibfield  {journal} {\bibinfo  {journal} {Phys. Rev. Lett.}\ }\textbf
  {\bibinfo {volume} {72}},\ \bibinfo {pages} {900} (\bibinfo {year}
  {1994})}\BibitemShut {NoStop}%
\bibitem [{\citenamefont {Shi}(2017)}]{shi2017}%
  \BibitemOpen
  \bibfield  {author} {\bibinfo {author} {\bibfnamefont {Junren}\ \bibnamefont
  {Shi}},\ }\bibfield  {title} {\enquote {\bibinfo {title} {Chern-{{Simons
  Theory}} and {{Dynamics}} of {{Composite Fermions}}},}\ }\href@noop {}
  {\bibfield  {journal} {\bibinfo  {journal} {arXiv:1704.07712}\ } (\bibinfo
  {year} {2017})}\BibitemShut {NoStop}%
\bibitem [{\citenamefont {Shi}\ and\ \citenamefont {Ji}(2018)}]{shi2018}%
  \BibitemOpen
  \bibfield  {author} {\bibinfo {author} {\bibfnamefont {Junren}\ \bibnamefont
  {Shi}}\ and\ \bibinfo {author} {\bibfnamefont {Wencheng}\ \bibnamefont
  {Ji}},\ }\bibfield  {title} {\enquote {\bibinfo {title} {Dynamics of the
  {{Wigner}} crystal of composite particles},}\ }\href {\doibase
  10.1103/PhysRevB.97.125133} {\bibfield  {journal} {\bibinfo  {journal} {Phys.
  Rev. B}\ }\textbf {\bibinfo {volume} {97}},\ \bibinfo {pages} {125133}
  (\bibinfo {year} {2018})}\BibitemShut {NoStop}%
\bibitem [{\citenamefont {Ji}\ and\ \citenamefont {Shi}(2021)}]{ji2021}%
  \BibitemOpen
  \bibfield  {author} {\bibinfo {author} {\bibfnamefont {Guangyue}\
  \bibnamefont {Ji}}\ and\ \bibinfo {author} {\bibfnamefont {Junren}\
  \bibnamefont {Shi}},\ }\bibfield  {title} {\enquote {\bibinfo {title}
  {Emergence of {{Dirac}} composite fermions: {{Dipole}} picture},}\ }\href
  {\doibase 10.1103/PhysRevResearch.3.043055} {\bibfield  {journal} {\bibinfo
  {journal} {Phys. Rev. Research}\ }\textbf {\bibinfo {volume} {3}},\ \bibinfo
  {pages} {043055} (\bibinfo {year} {2021})}\BibitemShut {NoStop}%
\bibitem [{\citenamefont {Pasquier}\ and\ \citenamefont
  {Haldane}(1998)}]{pasquier1998}%
  \BibitemOpen
  \bibfield  {author} {\bibinfo {author} {\bibfnamefont {V.}~\bibnamefont
  {Pasquier}}\ and\ \bibinfo {author} {\bibfnamefont {F.~D.~M.}\ \bibnamefont
  {Haldane}},\ }\bibfield  {title} {\enquote {\bibinfo {title} {A dipole
  interpretation of the $\nu=1/2$ state},}\ }\href {\doibase
  10.1016/S0550-3213(98)00069-8} {\bibfield  {journal} {\bibinfo  {journal}
  {Nuclear Physics B}\ }\textbf {\bibinfo {volume} {516}},\ \bibinfo {pages}
  {719} (\bibinfo {year} {1998})}\BibitemShut {NoStop}%
\bibitem [{\citenamefont {Read}(1998)}]{read1998}%
  \BibitemOpen
  \bibfield  {author} {\bibinfo {author} {\bibfnamefont {N.}~\bibnamefont
  {Read}},\ }\bibfield  {title} {\enquote {\bibinfo {title}
  {Lowest-{{Landau-level}} theory of the quantum {{Hall}} effect: {{The
  Fermi-liquid-like}} state of bosons at filling factor one},}\ }\href
  {\doibase 10.1103/PhysRevB.58.16262} {\bibfield  {journal} {\bibinfo
  {journal} {Phys. Rev. B}\ }\textbf {\bibinfo {volume} {58}},\ \bibinfo
  {pages} {16262} (\bibinfo {year} {1998})}\BibitemShut {NoStop}%
\bibitem [{\citenamefont {Dong}\ and\ \citenamefont
  {Senthil}(2020)}]{dong2020}%
  \BibitemOpen
  \bibfield  {author} {\bibinfo {author} {\bibfnamefont {Zhihuan}\ \bibnamefont
  {Dong}}\ and\ \bibinfo {author} {\bibfnamefont {T.}~\bibnamefont {Senthil}},\
  }\bibfield  {title} {\enquote {\bibinfo {title} {Noncommutative field theory
  and composite {{Fermi}} liquids in some quantum {{Hall}} systems},}\ }\href
  {\doibase 10.1103/PhysRevB.102.205126} {\bibfield  {journal} {\bibinfo
  {journal} {Phys. Rev. B}\ }\textbf {\bibinfo {volume} {102}},\ \bibinfo
  {pages} {205126} (\bibinfo {year} {2020})}\BibitemShut {NoStop}%
\bibitem [{\citenamefont {Go{\v c}anin}\ \emph {et~al.}(2021)\citenamefont
  {Go{\v c}anin}, \citenamefont {Predin}, \citenamefont {{\'C}iri{\'c}},
  \citenamefont {Radovanovi{\'c}},\ and\ \citenamefont
  {Milovanovi{\'c}}}]{gocanin2021}%
  \BibitemOpen
  \bibfield  {author} {\bibinfo {author} {\bibfnamefont {Dragoljub}\
  \bibnamefont {Go{\v c}anin}}, \bibinfo {author} {\bibfnamefont {Sonja}\
  \bibnamefont {Predin}}, \bibinfo {author} {\bibfnamefont
  {Marija~Dimitrijevi{\'c}}\ \bibnamefont {{\'C}iri{\'c}}}, \bibinfo {author}
  {\bibfnamefont {Voja}\ \bibnamefont {Radovanovi{\'c}}}, \ and\ \bibinfo
  {author} {\bibfnamefont {Milica}\ \bibnamefont {Milovanovi{\'c}}},\
  }\bibfield  {title} {\enquote {\bibinfo {title} {Microscopic derivation of
  {{Dirac}} composite fermion theory: {{Aspects}} of noncommutativity and
  pairing instabilities},}\ }\href {\doibase 10.1103/PhysRevB.104.115150}
  {\bibfield  {journal} {\bibinfo  {journal} {Phys. Rev. B}\ }\textbf {\bibinfo
  {volume} {104}},\ \bibinfo {pages} {115150} (\bibinfo {year}
  {2021})}\BibitemShut {NoStop}%
\bibitem [{\citenamefont {Parameswaran}\ \emph {et~al.}(2013)\citenamefont
  {Parameswaran}, \citenamefont {Roy},\ and\ \citenamefont
  {Sondhi}}]{parameswaran2013}%
  \BibitemOpen
  \bibfield  {author} {\bibinfo {author} {\bibfnamefont {Siddharth~A.}\
  \bibnamefont {Parameswaran}}, \bibinfo {author} {\bibfnamefont {Rahul}\
  \bibnamefont {Roy}}, \ and\ \bibinfo {author} {\bibfnamefont {Shivaji~L.}\
  \bibnamefont {Sondhi}},\ }\bibfield  {title} {\enquote {\bibinfo {title}
  {Fractional quantum {{Hall}} physics in topological flat bands},}\ }\href
  {\doibase 10.1016/j.crhy.2013.04.003} {\bibfield  {journal} {\bibinfo
  {journal} {Comptes Rendus Physique}\ }\bibinfo {series} {Topological
  Insulators / {{Isolants}} Topologiques},\ \textbf {\bibinfo {volume} {14}},\
  \bibinfo {pages} {816} (\bibinfo {year} {2013})}\BibitemShut {NoStop}%
\bibitem [{\citenamefont {Bergholtz}\ and\ \citenamefont
  {Liu}(2013)}]{bergholtz2013}%
  \BibitemOpen
  \bibfield  {author} {\bibinfo {author} {\bibfnamefont {Emil~J.}\ \bibnamefont
  {Bergholtz}}\ and\ \bibinfo {author} {\bibfnamefont {Zhao}\ \bibnamefont
  {Liu}},\ }\bibfield  {title} {\enquote {\bibinfo {title} {Topological {{Flat
  Band Models}} and {{Fractional Chern Insulators}}},}\ }\href {\doibase
  10.1142/S021797921330017X} {\bibfield  {journal} {\bibinfo  {journal} {Int.
  J. Mod. Phys. B}\ }\textbf {\bibinfo {volume} {27}},\ \bibinfo {pages}
  {1330017} (\bibinfo {year} {2013})}\BibitemShut {NoStop}%
\bibitem [{\citenamefont {Ji}\ and\ \citenamefont
  {Shi}(2020{\natexlab{a}})}]{ji2020a}%
  \BibitemOpen
  \bibfield  {author} {\bibinfo {author} {\bibfnamefont {Guangyue}\
  \bibnamefont {Ji}}\ and\ \bibinfo {author} {\bibfnamefont {Junren}\
  \bibnamefont {Shi}},\ }\bibfield  {title} {\enquote {\bibinfo {title}
  {Asymmetry of the geometrical resonances of composite fermions},}\ }\href
  {\doibase 10.1103/PhysRevB.101.235301} {\bibfield  {journal} {\bibinfo
  {journal} {Phys. Rev. B}\ }\textbf {\bibinfo {volume} {101}},\ \bibinfo
  {pages} {235301} (\bibinfo {year} {2020}{\natexlab{a}})}\BibitemShut
  {NoStop}%
\bibitem [{\citenamefont {Ji}\ and\ \citenamefont
  {Shi}(2020{\natexlab{b}})}]{ji2020b}%
  \BibitemOpen
  \bibfield  {author} {\bibinfo {author} {\bibfnamefont {Guangyue}\
  \bibnamefont {Ji}}\ and\ \bibinfo {author} {\bibfnamefont {Junren}\
  \bibnamefont {Shi}},\ }\bibfield  {title} {\enquote {\bibinfo {title} {Berry
  phase in the composite {{Fermi}} liquid},}\ }\href {\doibase
  10.1103/PhysRevResearch.2.033329} {\bibfield  {journal} {\bibinfo  {journal}
  {Phys. Rev. Research}\ }\textbf {\bibinfo {volume} {2}},\ \bibinfo {pages}
  {033329} (\bibinfo {year} {2020}{\natexlab{b}})}\BibitemShut {NoStop}%
\bibitem [{\citenamefont {Sundaram}\ and\ \citenamefont
  {Niu}(1999)}]{sundaram1999}%
  \BibitemOpen
  \bibfield  {author} {\bibinfo {author} {\bibfnamefont {Ganesh}\ \bibnamefont
  {Sundaram}}\ and\ \bibinfo {author} {\bibfnamefont {Qian}\ \bibnamefont
  {Niu}},\ }\bibfield  {title} {\enquote {\bibinfo {title} {Wave-packet
  dynamics in slowly perturbed crystals: {{Gradient}} corrections and
  {{Berry-phase}} effects},}\ }\href {\doibase 10.1103/PhysRevB.59.14915}
  {\bibfield  {journal} {\bibinfo  {journal} {Phys. Rev. B}\ }\textbf {\bibinfo
  {volume} {59}},\ \bibinfo {pages} {14915--14925} (\bibinfo {year}
  {1999})}\BibitemShut {NoStop}%
\bibitem [{\citenamefont {Xiao}\ \emph {et~al.}(2005)\citenamefont {Xiao},
  \citenamefont {Shi},\ and\ \citenamefont {Niu}}]{xiao2005}%
  \BibitemOpen
  \bibfield  {author} {\bibinfo {author} {\bibfnamefont {Di}~\bibnamefont
  {Xiao}}, \bibinfo {author} {\bibfnamefont {Junren}\ \bibnamefont {Shi}}, \
  and\ \bibinfo {author} {\bibfnamefont {Qian}\ \bibnamefont {Niu}},\
  }\bibfield  {title} {\enquote {\bibinfo {title} {Berry {{Phase Correction}}
  to {{Electron Density}} of {{States}} in {{Solids}}},}\ }\href {\doibase
  10.1103/PhysRevLett.95.137204} {\bibfield  {journal} {\bibinfo  {journal}
  {Phys. Rev. Lett.}\ }\textbf {\bibinfo {volume} {95}},\ \bibinfo {pages}
  {137204} (\bibinfo {year} {2005})}\BibitemShut {NoStop}%
\bibitem [{\citenamefont {Roy}(2014)}]{roy2014}%
  \BibitemOpen
  \bibfield  {author} {\bibinfo {author} {\bibfnamefont {Rahul}\ \bibnamefont
  {Roy}},\ }\bibfield  {title} {\enquote {\bibinfo {title} {Band geometry of
  fractional topological insulators},}\ }\href {\doibase
  10.1103/PhysRevB.90.165139} {\bibfield  {journal} {\bibinfo  {journal} {Phys.
  Rev. B}\ }\textbf {\bibinfo {volume} {90}},\ \bibinfo {pages} {165139}
  (\bibinfo {year} {2014})}\BibitemShut {NoStop}%
\bibitem [{\citenamefont {Girvin}(1984)}]{girvin1984}%
  \BibitemOpen
  \bibfield  {author} {\bibinfo {author} {\bibfnamefont {S.~M.}\ \bibnamefont
  {Girvin}},\ }\bibfield  {title} {\enquote {\bibinfo {title} {Anomalous
  quantum {{Hall}} effect and two-dimensional classical plasmas: {{Analytic}}
  approximations for correlation functions and ground-state energies},}\ }\href
  {\doibase 10.1103/PhysRevB.30.558} {\bibfield  {journal} {\bibinfo  {journal}
  {Phys. Rev. B}\ }\textbf {\bibinfo {volume} {30}},\ \bibinfo {pages}
  {558--560} (\bibinfo {year} {1984})}\BibitemShut {NoStop}%
\bibitem [{\citenamefont {Rohringer}\ \emph {et~al.}(2003)\citenamefont
  {Rohringer}, \citenamefont {Burgd{\"o}rfer},\ and\ \citenamefont
  {Macris}}]{rohringer2003}%
  \BibitemOpen
  \bibfield  {author} {\bibinfo {author} {\bibfnamefont {Nina}\ \bibnamefont
  {Rohringer}}, \bibinfo {author} {\bibfnamefont {Joachim}\ \bibnamefont
  {Burgd{\"o}rfer}}, \ and\ \bibinfo {author} {\bibfnamefont {Nicolas}\
  \bibnamefont {Macris}},\ }\bibfield  {title} {\enquote {\bibinfo {title}
  {Bargmann representation for {{Landau}} levels in two dimensions},}\ }\href
  {\doibase 10.1088/0305-4470/36/14/318} {\bibfield  {journal} {\bibinfo
  {journal} {J. Phys. A: Math. Gen.}\ }\textbf {\bibinfo {volume} {36}},\
  \bibinfo {pages} {4173} (\bibinfo {year} {2003})}\BibitemShut {NoStop}%
\bibitem [{\citenamefont {Hall}(1999)}]{hall1999}%
  \BibitemOpen
  \bibfield  {author} {\bibinfo {author} {\bibfnamefont {Brian~C.}\
  \bibnamefont {Hall}},\ }\bibfield  {title} {\enquote {\bibinfo {title}
  {Holomorphic {{Methods}} in {{Mathematical Physics}}},}\ }\href@noop {}
  {\bibfield  {journal} {\bibinfo  {journal} {Contemporary Mathematics}\
  }\textbf {\bibinfo {volume} {260}},\ \bibinfo {pages} {1} (\bibinfo {year}
  {1999})}\BibitemShut {NoStop}%
\bibitem [{\citenamefont {Spodyneiko}(2023)}]{spodyneiko2023}%
  \BibitemOpen
  \bibfield  {author} {\bibinfo {author} {\bibfnamefont {Lev}\ \bibnamefont
  {Spodyneiko}},\ }\bibfield  {title} {\enquote {\bibinfo {title} {A note on
  {{GMP}} algebra, dipole symmetry, and {{Hohenberg-Mermin-Wagner}} theorem in
  the lowest {{Landau}} level},}\ }\href@noop {} {\bibfield  {journal}
  {\bibinfo  {journal} {arXiv.2304.09927}\ } (\bibinfo {year}
  {2023})}\BibitemShut {NoStop}%
\bibitem [{\citenamefont {Simon}(1996)}]{simon1996}%
  \BibitemOpen
  \bibfield  {author} {\bibinfo {author} {\bibfnamefont {Steven~H.}\
  \bibnamefont {Simon}},\ }\bibfield  {title} {\enquote {\bibinfo {title}
  {Coupling of surface acoustic waves to a two-dimensional electron gas},}\
  }\href {\doibase 10.1103/PhysRevB.54.13878} {\bibfield  {journal} {\bibinfo
  {journal} {Phys. Rev. B}\ }\textbf {\bibinfo {volume} {54}},\ \bibinfo
  {pages} {13878--13884} (\bibinfo {year} {1996})}\BibitemShut {NoStop}%
\bibitem [{\citenamefont {Laughlin}(1983)}]{laughlin1983a}%
  \BibitemOpen
  \bibfield  {author} {\bibinfo {author} {\bibfnamefont {R.~B.}\ \bibnamefont
  {Laughlin}},\ }\bibfield  {title} {\enquote {\bibinfo {title} {Anomalous
  {{Quantum Hall Effect}}: {{An Incompressible Quantum Fluid}} with
  {{Fractionally Charged Excitations}}},}\ }\href {\doibase
  10.1103/PhysRevLett.50.1395} {\bibfield  {journal} {\bibinfo  {journal}
  {Phys. Rev. Lett.}\ }\textbf {\bibinfo {volume} {50}},\ \bibinfo {pages}
  {1395} (\bibinfo {year} {1983})}\BibitemShut {NoStop}%
\bibitem [{\citenamefont {Brody}(2013)}]{brody2013}%
  \BibitemOpen
  \bibfield  {author} {\bibinfo {author} {\bibfnamefont {Dorje~C.}\
  \bibnamefont {Brody}},\ }\bibfield  {title} {\enquote {\bibinfo {title}
  {Biorthogonal quantum mechanics},}\ }\href {\doibase
  10.1088/1751-8113/47/3/035305} {\bibfield  {journal} {\bibinfo  {journal} {J.
  Phys. A: Math. Theor.}\ }\textbf {\bibinfo {volume} {47}},\ \bibinfo {pages}
  {035305} (\bibinfo {year} {2013})}\BibitemShut {NoStop}%
\bibitem [{\citenamefont {Wang}\ \emph {et~al.}(2017)\citenamefont {Wang},
  \citenamefont {Cooper}, \citenamefont {Halperin},\ and\ \citenamefont
  {Stern}}]{wang2017}%
  \BibitemOpen
  \bibfield  {author} {\bibinfo {author} {\bibfnamefont {Chong}\ \bibnamefont
  {Wang}}, \bibinfo {author} {\bibfnamefont {Nigel~R.}\ \bibnamefont {Cooper}},
  \bibinfo {author} {\bibfnamefont {Bertrand~I.}\ \bibnamefont {Halperin}}, \
  and\ \bibinfo {author} {\bibfnamefont {Ady}\ \bibnamefont {Stern}},\
  }\bibfield  {title} {\enquote {\bibinfo {title} {Particle-{{Hole Symmetry}}
  in the {{Fermion-Chern-Simons}} and {{Dirac Descriptions}} of a {{Half-Filled
  Landau Level}}},}\ }\href {\doibase 10.1103/PhysRevX.7.031029} {\bibfield
  {journal} {\bibinfo  {journal} {Phys. Rev. X}\ }\textbf {\bibinfo {volume}
  {7}},\ \bibinfo {pages} {031029} (\bibinfo {year} {2017})}\BibitemShut
  {NoStop}%
\bibitem [{\citenamefont {{Robert G. Parr}}\ and\ \citenamefont {{Weitao
  Yang}}(1994)}]{robertg.parr1994}%
  \BibitemOpen
  \bibfield  {author} {\bibinfo {author} {\bibnamefont {{Robert G. Parr}}}\
  and\ \bibinfo {author} {\bibnamefont {{Weitao Yang}}},\ }\href@noop {} {\emph
  {\bibinfo {title} {Density-{{Functional Theory}} of {{Atoms}} and
  {{Molecules}}}}}\ (\bibinfo  {publisher} {Oxford University Press},\ \bibinfo
  {year} {1994})\BibitemShut {NoStop}%
\bibitem [{\citenamefont {Tafelmayer}(1993)}]{tafelmayer1993}%
  \BibitemOpen
  \bibfield  {author} {\bibinfo {author} {\bibfnamefont {R.}~\bibnamefont
  {Tafelmayer}},\ }\bibfield  {title} {\enquote {\bibinfo {title} {Topological
  vortex solitons in effective field theories for the fractional quantum
  {{Hall}} effect},}\ }\href {\doibase 10.1016/0550-3213(93)90657-B} {\bibfield
   {journal} {\bibinfo  {journal} {Nuclear Physics B}\ }\textbf {\bibinfo
  {volume} {396}},\ \bibinfo {pages} {386--410} (\bibinfo {year}
  {1993})}\BibitemShut {NoStop}%
\bibitem [{Note1()}]{Note1}%
  \BibitemOpen
  \bibinfo {note} {Hao Jin, private communication.}\BibitemShut {Stop}%
\bibitem [{\citenamefont {Hu}\ and\ \citenamefont {Jain}(2019)}]{hu2019}%
  \BibitemOpen
  \bibfield  {author} {\bibinfo {author} {\bibfnamefont {Yayun}\ \bibnamefont
  {Hu}}\ and\ \bibinfo {author} {\bibfnamefont {J.~K.}\ \bibnamefont {Jain}},\
  }\bibfield  {title} {\enquote {\bibinfo {title} {Kohn-{{Sham Theory}} of the
  {{Fractional Quantum Hall Effect}}},}\ }\href {\doibase
  10.1103/PhysRevLett.123.176802} {\bibfield  {journal} {\bibinfo  {journal}
  {Phys. Rev. Lett.}\ }\textbf {\bibinfo {volume} {123}},\ \bibinfo {pages}
  {176802} (\bibinfo {year} {2019})}\BibitemShut {NoStop}%
\bibitem [{\citenamefont {Kukushkin}\ \emph {et~al.}(2002)\citenamefont
  {Kukushkin}, \citenamefont {Smet}, \citenamefont {{von Klitzing}},\ and\
  \citenamefont {Wegscheider}}]{kukushkin2002}%
  \BibitemOpen
  \bibfield  {author} {\bibinfo {author} {\bibfnamefont {I.~V.}\ \bibnamefont
  {Kukushkin}}, \bibinfo {author} {\bibfnamefont {J.~H.}\ \bibnamefont {Smet}},
  \bibinfo {author} {\bibfnamefont {K.}~\bibnamefont {{von Klitzing}}}, \ and\
  \bibinfo {author} {\bibfnamefont {W.}~\bibnamefont {Wegscheider}},\
  }\bibfield  {title} {\enquote {\bibinfo {title} {Cyclotron resonance of
  composite fermions},}\ }\href {\doibase 10.1038/415409a} {\bibfield
  {journal} {\bibinfo  {journal} {Nature}\ }\textbf {\bibinfo {volume} {415}},\
  \bibinfo {pages} {409} (\bibinfo {year} {2002})}\BibitemShut {NoStop}%
\bibitem [{\citenamefont {Predin}\ \emph {et~al.}(2023)\citenamefont {Predin},
  \citenamefont {Kne{\v z}evi{\'c}},\ and\ \citenamefont
  {Milovanovi{\'c}}}]{predin2023a}%
  \BibitemOpen
  \bibfield  {author} {\bibinfo {author} {\bibfnamefont {S.}~\bibnamefont
  {Predin}}, \bibinfo {author} {\bibfnamefont {A.}~\bibnamefont {Kne{\v
  z}evi{\'c}}}, \ and\ \bibinfo {author} {\bibfnamefont {M.~V.}\ \bibnamefont
  {Milovanovi{\'c}}},\ }\bibfield  {title} {\enquote {\bibinfo {title} {Dipole
  representation of half-filled {{Landau}} level},}\ }\href {\doibase
  10.1103/PhysRevB.107.155132} {\bibfield  {journal} {\bibinfo  {journal}
  {Phys. Rev. B}\ }\textbf {\bibinfo {volume} {107}},\ \bibinfo {pages}
  {155132} (\bibinfo {year} {2023})}\BibitemShut {NoStop}%
\bibitem [{\citenamefont {Zhang}\ and\ \citenamefont {Shi}(2015)}]{zhang2015}%
  \BibitemOpen
  \bibfield  {author} {\bibinfo {author} {\bibfnamefont {Yin-Han}\ \bibnamefont
  {Zhang}}\ and\ \bibinfo {author} {\bibfnamefont {Jun-Ren}\ \bibnamefont
  {Shi}},\ }\bibfield  {title} {\enquote {\bibinfo {title} {Density
  {{Functional Theory}} of {{Composite Fermions}}},}\ }\href {\doibase
  10.1088/0256-307X/32/3/037101} {\bibfield  {journal} {\bibinfo  {journal}
  {Chinese Phys. Lett.}\ }\textbf {\bibinfo {volume} {32}},\ \bibinfo {pages}
  {037101} (\bibinfo {year} {2015})}\BibitemShut {NoStop}%
\bibitem [{\citenamefont {Zhang}\ and\ \citenamefont {Shi}(2016)}]{zhang2016a}%
  \BibitemOpen
  \bibfield  {author} {\bibinfo {author} {\bibfnamefont {Yinhan}\ \bibnamefont
  {Zhang}}\ and\ \bibinfo {author} {\bibfnamefont {Junren}\ \bibnamefont
  {Shi}},\ }\bibfield  {title} {\enquote {\bibinfo {title} {Mapping a
  fractional quantum {{Hall}} state to a fractional {{Chern}} insulator},}\
  }\href {\doibase 10.1103/PhysRevB.93.165129} {\bibfield  {journal} {\bibinfo
  {journal} {Phys. Rev. B}\ }\textbf {\bibinfo {volume} {93}},\ \bibinfo
  {pages} {165129} (\bibinfo {year} {2016})}\BibitemShut {NoStop}%
\bibitem [{\citenamefont {Cai}\ \emph {et~al.}(2023)\citenamefont {Cai},
  \citenamefont {Anderson}, \citenamefont {Wang}, \citenamefont {Zhang},
  \citenamefont {Liu}, \citenamefont {Holtzmann}, \citenamefont {Zhang},
  \citenamefont {Fan}, \citenamefont {Taniguchi}, \citenamefont {Watanabe},
  \citenamefont {Ran}, \citenamefont {Cao}, \citenamefont {Fu}, \citenamefont
  {Xiao}, \citenamefont {Yao},\ and\ \citenamefont {Xu}}]{cai2023}%
  \BibitemOpen
  \bibfield  {author} {\bibinfo {author} {\bibfnamefont {Jiaqi}\ \bibnamefont
  {Cai}}, \bibinfo {author} {\bibfnamefont {Eric}\ \bibnamefont {Anderson}},
  \bibinfo {author} {\bibfnamefont {Chong}\ \bibnamefont {Wang}}, \bibinfo
  {author} {\bibfnamefont {Xiaowei}\ \bibnamefont {Zhang}}, \bibinfo {author}
  {\bibfnamefont {Xiaoyu}\ \bibnamefont {Liu}}, \bibinfo {author}
  {\bibfnamefont {William}\ \bibnamefont {Holtzmann}}, \bibinfo {author}
  {\bibfnamefont {Yinong}\ \bibnamefont {Zhang}}, \bibinfo {author}
  {\bibfnamefont {Fengren}\ \bibnamefont {Fan}}, \bibinfo {author}
  {\bibfnamefont {Takashi}\ \bibnamefont {Taniguchi}}, \bibinfo {author}
  {\bibfnamefont {Kenji}\ \bibnamefont {Watanabe}}, \bibinfo {author}
  {\bibfnamefont {Ying}\ \bibnamefont {Ran}}, \bibinfo {author} {\bibfnamefont
  {Ting}\ \bibnamefont {Cao}}, \bibinfo {author} {\bibfnamefont {Liang}\
  \bibnamefont {Fu}}, \bibinfo {author} {\bibfnamefont {Di}~\bibnamefont
  {Xiao}}, \bibinfo {author} {\bibfnamefont {Wang}\ \bibnamefont {Yao}}, \ and\
  \bibinfo {author} {\bibfnamefont {Xiaodong}\ \bibnamefont {Xu}},\ }\bibfield
  {title} {\enquote {\bibinfo {title} {Signatures of fractional quantum
  anomalous hall states in twisted {{MoTe$_2$}}},}\ }\href {\doibase
  10.1038/s41586-023-06289-w} {\bibfield  {journal} {\bibinfo  {journal}
  {Nature}\ }\textbf {\bibinfo {volume} {622}},\ \bibinfo {pages} {63}
  (\bibinfo {year} {2023})}\BibitemShut {NoStop}%
\bibitem [{\citenamefont {Zeng}\ \emph {et~al.}(2023)\citenamefont {Zeng},
  \citenamefont {Xia}, \citenamefont {Kang}, \citenamefont {Zhu}, \citenamefont
  {Knüppel}, \citenamefont {Vaswani}, \citenamefont {Watanabe}, \citenamefont
  {Taniguchi}, \citenamefont {Mak},\ and\ \citenamefont {Shan}}]{zeng2023}%
  \BibitemOpen
  \bibfield  {author} {\bibinfo {author} {\bibfnamefont {Yihang}\ \bibnamefont
  {Zeng}}, \bibinfo {author} {\bibfnamefont {Zhengchao}\ \bibnamefont {Xia}},
  \bibinfo {author} {\bibfnamefont {Kaifei}\ \bibnamefont {Kang}}, \bibinfo
  {author} {\bibfnamefont {Jiacheng}\ \bibnamefont {Zhu}}, \bibinfo {author}
  {\bibfnamefont {Patrick}\ \bibnamefont {Knüppel}}, \bibinfo {author}
  {\bibfnamefont {Chirag}\ \bibnamefont {Vaswani}}, \bibinfo {author}
  {\bibfnamefont {Kenji}\ \bibnamefont {Watanabe}}, \bibinfo {author}
  {\bibfnamefont {Takashi}\ \bibnamefont {Taniguchi}}, \bibinfo {author}
  {\bibfnamefont {Kin~Fai}\ \bibnamefont {Mak}}, \ and\ \bibinfo {author}
  {\bibfnamefont {Jie}\ \bibnamefont {Shan}},\ }\bibfield  {title} {\enquote
  {\bibinfo {title} {Thermodynamic evidence of fractional chern insulator in
  moir\'{e} {{MoTe$_2$}}},}\ }\href {\doibase 10.1038/s41586-023-06452-3}
  {\bibfield  {journal} {\bibinfo  {journal} {Nature}\ }\textbf {\bibinfo
  {volume} {622}},\ \bibinfo {pages} {69} (\bibinfo {year} {2023})}\BibitemShut
  {NoStop}%
\bibitem [{\citenamefont {Park}\ \emph {et~al.}(2023)\citenamefont {Park},
  \citenamefont {Cai}, \citenamefont {Anderson}, \citenamefont {Zhang},
  \citenamefont {Zhu}, \citenamefont {Liu}, \citenamefont {Wang}, \citenamefont
  {Holtzmann}, \citenamefont {Hu}, \citenamefont {Liu}, \citenamefont
  {Taniguchi}, \citenamefont {Watanabe}, \citenamefont {Chu}, \citenamefont
  {Cao}, \citenamefont {Fu}, \citenamefont {Yao}, \citenamefont {Chang},
  \citenamefont {Cobden}, \citenamefont {Xiao},\ and\ \citenamefont
  {Xu}}]{park2023}%
  \BibitemOpen
  \bibfield  {author} {\bibinfo {author} {\bibfnamefont {Heonjoon}\
  \bibnamefont {Park}}, \bibinfo {author} {\bibfnamefont {Jiaqi}\ \bibnamefont
  {Cai}}, \bibinfo {author} {\bibfnamefont {Eric}\ \bibnamefont {Anderson}},
  \bibinfo {author} {\bibfnamefont {Yinong}\ \bibnamefont {Zhang}}, \bibinfo
  {author} {\bibfnamefont {Jiayi}\ \bibnamefont {Zhu}}, \bibinfo {author}
  {\bibfnamefont {Xiaoyu}\ \bibnamefont {Liu}}, \bibinfo {author}
  {\bibfnamefont {Chong}\ \bibnamefont {Wang}}, \bibinfo {author}
  {\bibfnamefont {William}\ \bibnamefont {Holtzmann}}, \bibinfo {author}
  {\bibfnamefont {Chaowei}\ \bibnamefont {Hu}}, \bibinfo {author}
  {\bibfnamefont {Zhaoyu}\ \bibnamefont {Liu}}, \bibinfo {author}
  {\bibfnamefont {Takashi}\ \bibnamefont {Taniguchi}}, \bibinfo {author}
  {\bibfnamefont {Kenji}\ \bibnamefont {Watanabe}}, \bibinfo {author}
  {\bibfnamefont {Jiun-Haw}\ \bibnamefont {Chu}}, \bibinfo {author}
  {\bibfnamefont {Ting}\ \bibnamefont {Cao}}, \bibinfo {author} {\bibfnamefont
  {Liang}\ \bibnamefont {Fu}}, \bibinfo {author} {\bibfnamefont {Wang}\
  \bibnamefont {Yao}}, \bibinfo {author} {\bibfnamefont {Cui-Zu}\ \bibnamefont
  {Chang}}, \bibinfo {author} {\bibfnamefont {David}\ \bibnamefont {Cobden}},
  \bibinfo {author} {\bibfnamefont {Di}~\bibnamefont {Xiao}}, \ and\ \bibinfo
  {author} {\bibfnamefont {Xiaodong}\ \bibnamefont {Xu}},\ }\bibfield  {title}
  {\enquote {\bibinfo {title} {Observation of fractionally quantized anomalous
  {{Hall}} effect},}\ }\href {\doibase 10.1038/s41586-023-06536-0} {\bibfield
  {journal} {\bibinfo  {journal} {Nature}\ }\textbf {\bibinfo {volume} {622}},\
  \bibinfo {pages} {74} (\bibinfo {year} {2023})}\BibitemShut {NoStop}%
\bibitem [{\citenamefont {Xu}\ \emph {et~al.}(2023)\citenamefont {Xu},
  \citenamefont {Sun}, \citenamefont {Jia}, \citenamefont {Liu}, \citenamefont
  {Xu}, \citenamefont {Li}, \citenamefont {Gu}, \citenamefont {Watanabe},
  \citenamefont {Taniguchi}, \citenamefont {Tong}, \citenamefont {Jia},
  \citenamefont {Shi}, \citenamefont {Jiang}, \citenamefont {Zhang},
  \citenamefont {Liu},\ and\ \citenamefont {Li}}]{xu2023}%
  \BibitemOpen
  \bibfield  {author} {\bibinfo {author} {\bibfnamefont {Fan}\ \bibnamefont
  {Xu}}, \bibinfo {author} {\bibfnamefont {Zheng}\ \bibnamefont {Sun}},
  \bibinfo {author} {\bibfnamefont {Tongtong}\ \bibnamefont {Jia}}, \bibinfo
  {author} {\bibfnamefont {Chang}\ \bibnamefont {Liu}}, \bibinfo {author}
  {\bibfnamefont {Cheng}\ \bibnamefont {Xu}}, \bibinfo {author} {\bibfnamefont
  {Chushan}\ \bibnamefont {Li}}, \bibinfo {author} {\bibfnamefont
  {Yu}~\bibnamefont {Gu}}, \bibinfo {author} {\bibfnamefont {Kenji}\
  \bibnamefont {Watanabe}}, \bibinfo {author} {\bibfnamefont {Takashi}\
  \bibnamefont {Taniguchi}}, \bibinfo {author} {\bibfnamefont {Bingbing}\
  \bibnamefont {Tong}}, \bibinfo {author} {\bibfnamefont {Jinfeng}\
  \bibnamefont {Jia}}, \bibinfo {author} {\bibfnamefont {Zhiwen}\ \bibnamefont
  {Shi}}, \bibinfo {author} {\bibfnamefont {Shengwei}\ \bibnamefont {Jiang}},
  \bibinfo {author} {\bibfnamefont {Yang}\ \bibnamefont {Zhang}}, \bibinfo
  {author} {\bibfnamefont {Xiaoxue}\ \bibnamefont {Liu}}, \ and\ \bibinfo
  {author} {\bibfnamefont {Tingxin}\ \bibnamefont {Li}},\ }\bibfield  {title}
  {\enquote {\bibinfo {title} {Observation of integer and fractional quantum
  anomalous hall effects in twisted bilayer ${{{\mathrm{MoTe}}_{2}}}$},}\
  }\href {\doibase 10.1103/PhysRevX.13.031037} {\bibfield  {journal} {\bibinfo
  {journal} {Phys. Rev. X}\ }\textbf {\bibinfo {volume} {13}},\ \bibinfo
  {pages} {031037} (\bibinfo {year} {2023})}\BibitemShut {NoStop}%
\bibitem [{Note2()}]{Note2}%
  \BibitemOpen
  \bibinfo {note} {For $C>0$, the topological flat band could be continuously
  connected to Landau level(s) in a magnetic field with $B<0$. The definitions
  of the complex coordinates $\eta $ and $\protect \bar {\eta }$ should be
  exchanged.}\BibitemShut {Stop}%
\bibitem [{\citenamefont {Wang}\ \emph {et~al.}(2021)\citenamefont {Wang},
  \citenamefont {Cano}, \citenamefont {Millis}, \citenamefont {Liu},\ and\
  \citenamefont {Yang}}]{wang2021a}%
  \BibitemOpen
  \bibfield  {author} {\bibinfo {author} {\bibfnamefont {Jie}\ \bibnamefont
  {Wang}}, \bibinfo {author} {\bibfnamefont {Jennifer}\ \bibnamefont {Cano}},
  \bibinfo {author} {\bibfnamefont {Andrew~J.}\ \bibnamefont {Millis}},
  \bibinfo {author} {\bibfnamefont {Zhao}\ \bibnamefont {Liu}}, \ and\ \bibinfo
  {author} {\bibfnamefont {Bo}~\bibnamefont {Yang}},\ }\bibfield  {title}
  {\enquote {\bibinfo {title} {Exact {{Landau Level Description}} of
  {{Geometry}} and {{Interaction}} in a {{Flatband}}},}\ }\href {\doibase
  10.1103/PhysRevLett.127.246403} {\bibfield  {journal} {\bibinfo  {journal}
  {Phys. Rev. Lett.}\ }\textbf {\bibinfo {volume} {127}},\ \bibinfo {pages}
  {246403} (\bibinfo {year} {2021})}\BibitemShut {NoStop}%
\bibitem [{Note3()}]{Note3}%
  \BibitemOpen
  \bibinfo {note} {For $C>0$ and $\Omega _{\protect \bm {k}}>0$, the
  definitions of $\eta $ and $\protect \bar {\eta }$, $k$ and $\protect \bar
  {k}$ should be exchanged in Eq.~(\ref {eq:Hfcb}). The second term of
  $\protect \hat {H}$ will be proportional to $\protect \mathrm {Tr}\protect
  \mathbb {G}_{\protect \bm {k}}-\Omega _{\protect \bm {k}}=\protect \mathrm
  {Tr}\protect \mathbb {G}_{\protect \bm {k}}-|\Omega _{\protect \bm {k}}|$ for
  $\Omega _{\protect \bm {k}}>0$.}\BibitemShut {Stop}%
\bibitem [{\citenamefont {Baer}\ \emph {et~al.}(2014)\citenamefont {Baer},
  \citenamefont {R{\"o}ssler}, \citenamefont {Ihn}, \citenamefont {Ensslin},
  \citenamefont {Reichl},\ and\ \citenamefont {Wegscheider}}]{baer2014}%
  \BibitemOpen
  \bibfield  {author} {\bibinfo {author} {\bibfnamefont {S.}~\bibnamefont
  {Baer}}, \bibinfo {author} {\bibfnamefont {C.}~\bibnamefont {R{\"o}ssler}},
  \bibinfo {author} {\bibfnamefont {T.}~\bibnamefont {Ihn}}, \bibinfo {author}
  {\bibfnamefont {K.}~\bibnamefont {Ensslin}}, \bibinfo {author} {\bibfnamefont
  {C.}~\bibnamefont {Reichl}}, \ and\ \bibinfo {author} {\bibfnamefont
  {W.}~\bibnamefont {Wegscheider}},\ }\bibfield  {title} {\enquote {\bibinfo
  {title} {Experimental probe of topological orders and edge excitations in the
  second {{Landau}} level},}\ }\href {\doibase 10.1103/PhysRevB.90.075403}
  {\bibfield  {journal} {\bibinfo  {journal} {Phys. Rev. B}\ }\textbf {\bibinfo
  {volume} {90}},\ \bibinfo {pages} {075403} (\bibinfo {year}
  {2014})}\BibitemShut {NoStop}%
\bibitem [{\citenamefont {Wang}\ \emph {et~al.}(2024)\citenamefont {Wang},
  \citenamefont {Zhang}, \citenamefont {Liu}, \citenamefont {He}, \citenamefont
  {Xu}, \citenamefont {Ran}, \citenamefont {Cao},\ and\ \citenamefont
  {Xiao}}]{wang2024}%
  \BibitemOpen
  \bibfield  {author} {\bibinfo {author} {\bibfnamefont {Chong}\ \bibnamefont
  {Wang}}, \bibinfo {author} {\bibfnamefont {Xiao-Wei}\ \bibnamefont {Zhang}},
  \bibinfo {author} {\bibfnamefont {Xiaoyu}\ \bibnamefont {Liu}}, \bibinfo
  {author} {\bibfnamefont {Yuchi}\ \bibnamefont {He}}, \bibinfo {author}
  {\bibfnamefont {Xiaodong}\ \bibnamefont {Xu}}, \bibinfo {author}
  {\bibfnamefont {Ying}\ \bibnamefont {Ran}}, \bibinfo {author} {\bibfnamefont
  {Ting}\ \bibnamefont {Cao}}, \ and\ \bibinfo {author} {\bibfnamefont
  {Di}~\bibnamefont {Xiao}},\ }\bibfield  {title} {\enquote {\bibinfo {title}
  {Fractional chern insulator in twisted bilayer ${{\mathrm{MoTe}_{2}}}$},}\
  }\href {\doibase 10.1103/PhysRevLett.132.036501} {\bibfield  {journal}
  {\bibinfo  {journal} {Phys. Rev. Lett.}\ }\textbf {\bibinfo {volume} {132}},\
  \bibinfo {pages} {036501} (\bibinfo {year} {2024})}\BibitemShut {NoStop}%
\end{thebibliography}%

\end{document}